\documentclass[%
 aip,
 amsmath,amssymb,
 reprint,%
]{revtex4-1}

\usepackage{graphicx}
\usepackage{dcolumn}
\usepackage{bm}

\usepackage[utf8]{inputenc}
\usepackage[T1]{fontenc}
\usepackage{mathptmx}
\usepackage{etoolbox}
\usepackage{bbding}
\makeatletter
\def\@email#1#2{%
 \endgroup
 \patchcmd{\titleblock@produce}
  {\frontmatter@RRAPformat}
  {\frontmatter@RRAPformat{\produce@RRAP{*#1\href{mailto:#2}{#2}}}\frontmatter@RRAPformat}
  {}{}
}%
\usepackage{subfig,float,epstopdf,epsfig,url,hyperref,diagbox,xcolor}
\usepackage{mathtools}
\DeclarePairedDelimiter\abs{\lvert}{\rvert}%
\DeclarePairedDelimiter\norm{\lVert}{\rVert}%
\makeatletter
\let\oldabs\abs
\def\abs{\@ifstar{\oldabs}{\oldabs*}}
\let\oldnorm\norm
\def\norm{\@ifstar{\oldnorm}{\oldnorm*}}
\usepackage[makeroom]{cancel}
\makeatother
\begin{document}

\title{Modeling the compressible flow field of an impulsively started circular cylinder with refined potential flow theory}
\author{Taofiq Omoniyi Amoloye}

\affiliation{Department of Aeronautical and Astronautical Engineering,
Faculty of Engineering and Technology, Kwara State University, Malete, Nigeria.}
\author{Leke Thaddeus Oladimeji}

\affiliation{Department of Aeronautical and Astronautical Engineering,
Faculty of Engineering and Technology, Kwara State University, Malete, Nigeria.}
\author{Mahmoud A. Hayajnh}
	
\affiliation{Department of Aeronautical Engineering,
Jordan University of Science and Technology, Irbid, Jordan 22110.}
\author{Olalekan Adebayo Olayemi}
\affiliation{Department of Aeronautical and Astronautical Engineering,
Faculty of Engineering and Technology, Kwara State University, Malete, Nigeria.}

\email{olalekan.olayemi@kwasu.edu.ng}
\email{mahayajnh@just.edu.jo}
\email{leke.oladimeji14@gmail.com}
\email{taofiq.amoloye@kwasu.edu.ng}

\date{\today}

\begin{abstract}
Understanding compressible flows over bluff bodies, such as circular cylinders, is critical for applications in aerospace engineering, space exploration, and astrophysics. However, the complexity of viscous and unsteady flows, compounded by compressibility effects, remains a challenge for experimental and computational methods. This study explores the compressible flow field of an impulsively started circular cylinder using Refined Potential Flow Theory (RPT), an analytical model that extends classical potential flow theory to include compressibility effects. The governing equations, boundary conditions, and refined stream function are developed to capture the density, velocity, and pressure fields in subsonic and transonic regimes. The results demonstrate that compressibility marginally increases vortex enstrophy at low Reynolds numbers, while its influence on wake flow stability diminishes at higher values. The study reveals that compressibility suppresses shear layer instability, reduces wake velocity deficits, and introduces smaller-scale structures that disrupt the inertial range of turbulence in the low subsonic regime. At higher Mach numbers ($M_\infty \ge 0.6$), local supersonic pockets and shock waves emerge, forming complex $\lambda$-shock systems and bow shocks. The predictions of the Strouhal number from RPT show marginal changes for $0.2 \le M_\infty \le 0.5$, consistent with experimental and computational trends. Although RPT aligns with existing methods to capture key flow features, discrepancies in wake stability and recirculation zone dimensions highlight areas for further refinement. This research underscores the potential of RPT \color{black}in developing \color{black} optimization tool\color{black}s \color{black} to complement experimental and numerical studies, offering valuable insights into compressible flow physics for engineering applications in high-altitude flight and space exploration. 
\end{abstract}

\maketitle

\section{\label{Intro}Introduction}
Knowledge of compressible flows over bluff bodies, such as circular cylinders and spheres, is important for many applications, including high-altitude unmanned aircraft design\cite{Nagataetal2020b}, astrophysics\cite{KleimannandRoken2024,Jelineketal2022}, and interplanetary space exploration \cite{Nagataetal2020b,Anyojietal2011}. However, the complexity of viscous and unsteady cylinder flows\cite{Amoloye2024,MatheswaranandMiller2024a,MatheswaranandMiller2024b} is usually compounded by compressibility effects\cite{Li2025}. This is not well understood \cite{Awasthietal2025}. Early studies of cylinder flows focused on the incompressible regime. This is because the operational envelope of the wind tunnel is usually limited to the low Reynold number ($Re$: based on the freestream velocity, $V_\infty$, the cylinder diameter, $D$, and the fluid kinematic viscosity, $\nu$, as $Re=V_{\infty}D/\nu$) and the low compressible subsonic flow range \cite{Anyojietal2011}. These studies, including those by Taneda \cite{Taneda1956}, Dennis and Chang \cite{DennisandChang1970}, Coutanceau and Bouard \cite{CoutandBouard1977a,CoutandBouard1977b}, Roshko \cite{Roshko1953,Roshko1961}, Parnadeau et al. \cite{Parnaudeauetal2008}, and Behara and Mittal \cite{BeharaandMittal2010)}, characterize the wake phenomenon with $Re$ with other unique properties for each condition. These confirm that the flow around a circular cylinder is strongly dependent on $Re$. 

Experiments show that the force coefficients in sub-critical cylinder flow depend on both $Re$ and $M_\infty$ (freestream Mach number, $M_\infty=V_\infty/a_\infty$ where $a_\infty$ is the speed of sound in the freestream) in the subsonic compressible regime \cite{Lindsey1938,GowenandPerkins1952}. However, the influence of $Re$ fades as the flow becomes supersonic for both the sub-critical and super-critical regimes \cite{GowenandPerkins1952}. Wake transitions in the supersonic sub-critical regime have been experimentally studied \cite{McCarrhyandKubota1964}. Although the Strouhal number was experimentally examined as $0.18$ for $M_\infty <0.9$, the wake periodicity or discernible vortex shedding breaks down for $M_\infty > 0.9$ in the range $Re$ between \color{black}$30,000$ \color{black} and \color{black}$500,000$\color{black} \cite{MurthyandRose1977}.  A strong coupling effect has been observed between $M_\infty$ and the vortex street in the near wake for subsonic compressible and transonic cylinder flows in the $Re$ range from $0.8$ to \color{black}$2,000,000$\color{black} \cite{Rodriguez1984}. The strength of the wake vortices changes at $M_\infty>0.6$, reducing the base pressure drag until $M_\infty>0.9$ where local sonic flows present increased drag \cite{Ackermanetal2008}.  
Experimental studies of subsonic compressible flow in the intermediate sub-critical regime $Re$ have characterized the compressibility effects on the location of the shedding of the von K\'{a}rm\'{a}n vortices, the frequency of shedding, and the maximum width of the recirculation region for circular cylinders \cite{Nagataetal2020,Nagataetal2020b} and spheres \cite{Sansicaetal2018}. 

Computational fluid dynamics (CFD) studies have contributed to the exploration of compressible flow over a circular cylinder\cite{BurbeauandSagaut2002,BobenriethMiserdaandLeal2006,Lysenkoetal2012,Xuetal2020,Mataretal2023,HoffmannandWeiss2023}. Canuto and Taira \cite{CanutoandTaira2015} observed that compressibility effects lower the dominant frequency and growth rate of flow instability in the linear growth stage for $M_\infty\le 0.5$ in the low $Re$ regime. In this regime, Liu et al. \cite{Liuetal2023b} also observed that compressibility effects enhance flow stability, presenting a reduction in vortex shedding strength, shedding frequency, and aerodynamic force fluctuations as $M_\infty$ increases. Shang \cite{Shang1982} observed rapid and extreme variations in the flow field at $M_\infty=0.6$ and \color{black}$Re= 167,000$\color{black}, with large-scale structures contributing significantly to turbulence in the wake. Qualitative agreement was observed for the general flow features. The results of Ishii and Kuwahara show that compressibility suppresses the width of the wake \cite{IshiiandKuwahara1984}.
Pandolfi and Larocca \cite{PandolfiandLarocca1989} show that half cylinders experience separation behind shocks in transonic flows, whereas full cylinders show asymmetric, unsteady, periodic flows. Botta's solution \cite{Botta1995} has periodic behavior at $M_\infty$ between $0.5$ and $0.6$, undergoing two transitions as $M_\infty$ increases from $0.6$ to $0.98$. Shirani \cite{Shirani2001} observed that for $M_\infty>0.6$, the flow parameters become independent of $Re$, while at $M_\infty>0.8$, they become independent of $M_\infty$. Xu et al. \cite{Xuetal2009} demonstrated that above a critical $M_\infty\approx 0.9$, the flow is a quasi-steady flow with virtually stationary shock waves created in the near wake. Below this critical $M_\infty$, there exists an unsteady flow with moving shock waves interacting with the turbulent flow in the cylinder near wake. These drastic changes in the flow field near high-subsonic and transonic regimes have also been reported by others\cite{Xiaetal2016}. At \color{black}$Re=10,000$\color{black}, compressibility significantly affects the location and mechanism of the transition. The wake transition region shifts towards the cylinder shoulder as $M_\infty$ increases, with an eventual loss of wake coherence and the vortex shedding frequency  \cite{Rodriguezetal2023}. Compressibility also affects the dimensionality of the flow at low $Re$, stabilizing the instability of mode B and delaying the three-dimensionalization of the wake as $M_\infty$ increases \cite{Rolandietal2023}. At $Re=3,900$, compressibility suppresses the Kelvin-Helmholtz instability but increases velocity fluctuations in the boundary layer and the pressure difference between the freestream and the recirculation zone \cite{Xueetal2024}. Consequently, the recirculation zone shortens with $M_\infty$ up to $0.5$.

From the foregoing review it can be seen that numerous amounts of work have been done to evaluate the effect of compressibility on a circular cylinder experimentally and numerically, both of which have documented limitations \cite{LiSetal2024,Nagataetal2020b}. Hence, there is a need to consider other approaches. The Janzen-Rayleigh expansions \cite{WallersteinandKeshet2022} of the flow variables as functions of $M_\infty$ are the earliest attempts to theorize compressible flow over a circular cylinder. These were applied to study the d'Alembert's paradox in the subsonic compressible flow regime, assuming a steady, inviscid flow of a polytropic fluid with no external forces or initial vorticity \cite{WallersteinandKeshet2022}. Consequently, this formulation cannot replicate the foregoing features of the full range of the unsteady, viscous, and compressible flow of an impulsively started circular cylinder. These features are physical solutions of the governing equations that are coupled and notoriously non-linear, resisting general analytical descriptions of the entire flow \cite{KeshetandNaor2016}. Cherry and Temple \cite{CherryandTemple1947} attempted an exact solution of a two-dimensional subsonic compressible flow around a cylinder. However, there were non-physical discontinuities within the flow, and the full physics of the phenomena was not captured or reported. \color{black}Joseph\cite{Joseph2003,Joseph2006} explored viscous potential flows (VPF) as solutions to the governing equations, focusing on the inclusion of viscous normal stresses in the definition of pressure \cite{Joseph2006} and extending the VPF into the compressible flow regime for sound waves in a perturbation analysis of the linearized governing equations in a state of rest \cite{Joseph2003}. However, the applications of VPF analysis \cite{Josephetal2007} do not include compressible flow over a circular cylinder. \color{black}Amoloye \cite{Amoloye2024} successfully refined the classical potential flow theory \color{black}for cylinder flows \color{black}to model the wake characteristics of an impulsively started circular cylinder in the sub-critical regime. The model focuses on the incompressible flow. However, there are unpublished extensions of the formulation to include compressibility effects \cite{Amoloye2018,Amoloye2020,Amoloye2021,Amoloye2022}.

Therefore, the present publication seeks to explore the refined potential flow theory (RPT) in the study of compressible flow over an impulsively started circular cylinder in the subcritical regime, attempting to theorize subsonic compressible and transonic flow physics. This analytical model can \color{black} aid the development of \color{black} optimization tool\color{black}s \color{black} to complement numerical and experimental studies for engineering applications in space exploration and high-altitude unmanned aerial vehicle missions. It can also offer insights into compressible flow over blunt objects with potential applications in space physics and astrophysics \cite{KeshetandNaor2016,Jelineketal2022}. 

Section~\ref{GEBC} discusses the governing equations and boundary conditions. Section~\ref{RPT} develops RPT for compressible \color{black}circular cylinder \color{black}flows. Instantaneous streamline patterns of RPT, time-averaged flow field velocity profiles, wake velocity energy spectra, and predicted Strouhal numbers are reported and discussed in comparison with published results from experiments and CFD in Section~\ref{RnD}. Section~\ref{C} concludes the paper.

\section{\label{GEBC}Governing Equations and Boundary Conditions}
\color{black} Assuming an isotropic and Newtonian fluid in an infinite domain\color{black}, the governing equations for the compressible cylinder crossflow considered here are the continuity equation 
\begin{equation}
\label{contdiff}
\dfrac{\partial \rho}{\partial t} + \nabla{}\cdot (\rho \mathbf{V})=0\hspace{10pt} (t\ge 0),
\end{equation}
the \color{black}compressible \color{black}Navier-Stokes equations \color{black}(NSE)\color{black}
\begin{equation}
\label{NSEDiff}
\begin{array}{l}
\dfrac{\partial{}\left(\rho{}\mathbf{V}\right)}{\partial{}t}+\mathbf{V}\nabla{}\cdot\left(\rho{}\mathbf{V}\right)+(\rho{}\mathbf{V})\cdot\nabla{}\mathbf{V}=-\nabla{}p+\nabla{}\left(\left[\lambda + 2\mu{}\right]\nabla{}\cdot \mathbf{V}\right)\\[10pt]~~~~~~~~~~~~~~~~~~~~~~~~~~~~~~~~~~~~~~~~~~+\nabla{}\times{}\left(\mu{}\omega{}\right) \hspace{10pt} (t\ge 0),
\end{array}
\end{equation}
and the energy equations\color{black}
\begin{widetext}
\begin{equation}
\label{EnergyDiff}
\begin{array}{l}
\dfrac{\partial{}\left(\rho{}E\right)}{\partial{}t}+\nabla{}\cdot\left(\rho{}\mathbf{V}E+P\mathbf{V}\right)=\nabla{}\cdot\left(k\nabla{}T\right)+\nabla{}\cdot\left(\mathbf{V}\left[\nabla{}\left(\left[\lambda + 2\mu{}\right]\nabla{}\cdot \mathbf{V}\right)+\nabla{}\times{}\left(\mu{}\omega{}\right)\right]\right)\hspace{10pt} (t\ge 0)
\end{array}
\end{equation}
\end{widetext}\color{black}
where $\mathbf{V}$ is the velocity, $p$ is the pressure, $\rho$ is the density, $\omega$ is the vorticity vector, $\lambda$ is the second coefficient of viscosity, $\mu$ is the dynamic viscosity, $E$ is the total energy, $k$ is the thermal conductivity, $T$ is the temperature, \color{black}and $t$ is the dimensional time\color{black} \cite{Anderson2011,White2006}. Together, Eqs.~\ref{contdiff} to \ref{EnergyDiff} constitute an initial boundary value problem. The boundary conditions on the velocity are the infinity boundary condition at the freestream and the no-slip boundary condition at the cylinder surface.

\section{Refined Potential Flow Theory for Compressible Flows Over Circular Cylinders}\label{RPT} 
\color{black} For an incompressible \color{black}circular cylinder \color{black}flow, Eqs.~\ref{contdiff} and~\ref{NSEDiff} reduce to 
\begin{equation}
\label{contdiffinc}
\nabla{}\cdot\mathbf{V}=0\hspace{10pt} (t\ge 0),
\end{equation}
and
\begin{equation}
\label{NSEDiffinc}
\begin{array}{l}
\dfrac{\partial{}\mathbf{V}}{\partial{}t}+\nabla{}\left(\dfrac{p}{\rho{}}+\dfrac{V^2}{2}-\left[\dfrac{\lambda}{\rho}+2\nu{}\right]\nabla{}\cdot \mathbf{V}\right)\\[10pt]=\mathbf{V}\times{}\omega{}+\nabla{}\times{}\left(\nu{}\omega{}\right)
\hspace{10pt} (t\ge 0)
\end{array}
\end{equation}
respectively, where $V=|\mathbf{V}|$ and $\nu={\mu}/{\rho}$. \color{black}These constitute the Navier-Stokes problem \cite{CMI,Farwig2021}.\color{black}

\begin{figure*}
	\centering
	\includegraphics[width=0.5\textwidth,trim={0cm 0cm 0cm 0cm},clip]{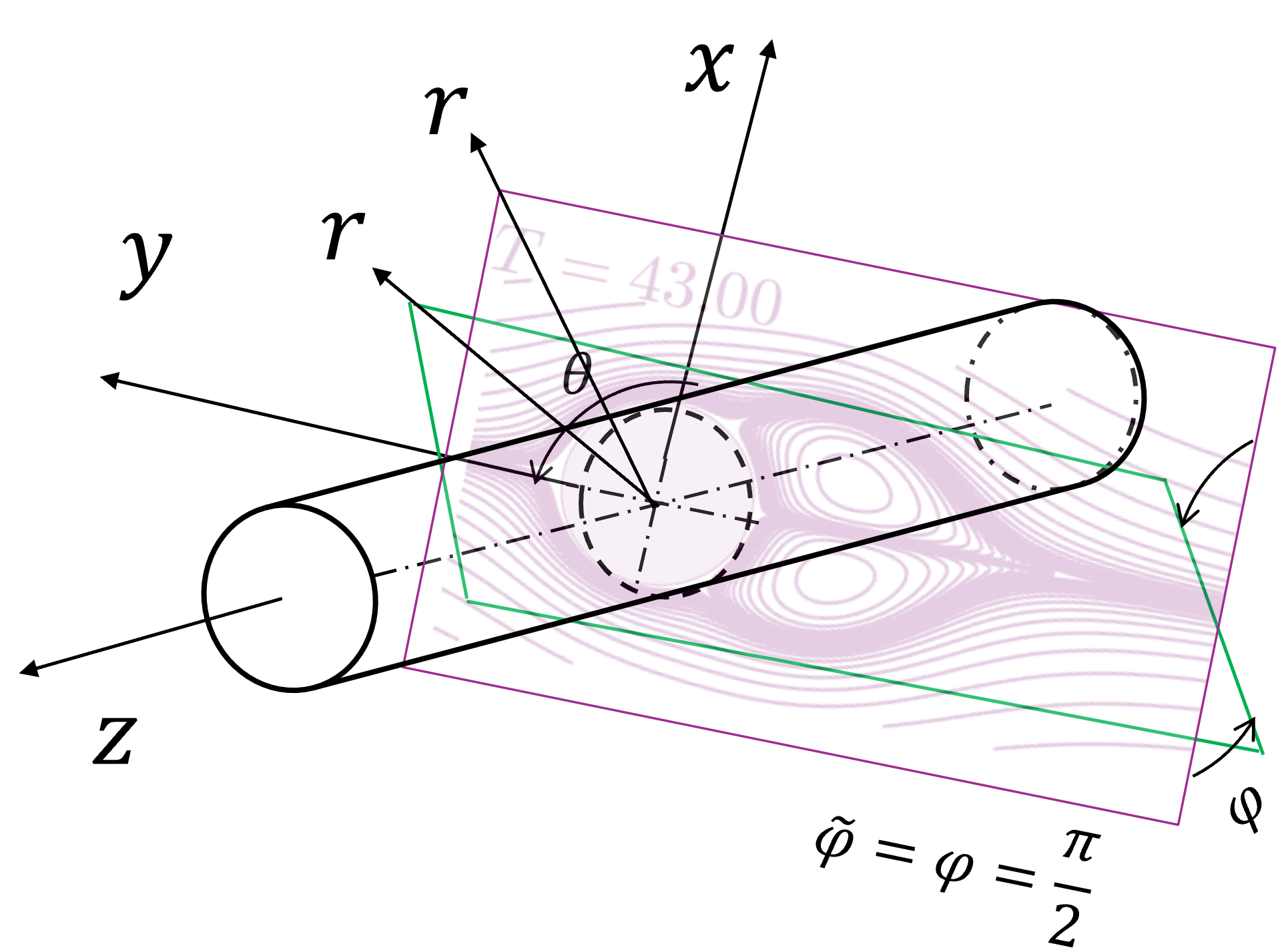}
	\caption{The geometric and wind coordinate axes \color{black}in relation to $\psi_{vis}$ \cite{Amoloye2024}. Reproduced with permission from Phys. Scr. $\mathbf{99}$ $085207$ $(2024)$. Copyright $2024$ IOP Publishing Ltd.\color{black}}\label{Fig1}
\end{figure*}

A rotational velocity field satisfying Eq.~\ref{contdiffinc} is obtained from a vector of unsteady, planar, and viscous stream function, $\overrightarrow{\psi}$, as
\begin{equation}
\label{SFVect}
\mathbf{V}=\nabla{}\times\overrightarrow{\psi{}},
\end{equation}
where $\overrightarrow{\psi}=\psi_{vis}\hat{\mathbf{e}}_z$\cite{Amoloye2024} \color{black}(vis: viscous)\color{black}. \color{black}However \color{black}, Eq.~\ref{SFVect} describes a planar flow whereas an actual cylinder flow is three-dimensional. For three-dimensional consideration in RPT, the distinction between Cartesian geometric coordinate axes of the flow domain and Cartesian wind coordinate axes (signified by the tilde) is introduced\cite{Amoloye2024}. The three-dimensional Cartesian geometric coordinate axes are 
\begin{equation}
\label{3Ddomain}
\begin{array}{l}
\hspace{10pt}x=r\cos{\theta}\sin{{\varphi}} \hspace{10pt}y=r\sin{\theta}\sin{{\varphi}} \hspace{10pt}z=r\cos{{\varphi}}\hspace{3pt}
\end{array}
\end{equation}
where $r$, $\theta$, and $\varphi$ are the radial, circumferential, and azimuthal coordinate variables of an associated spherical geometric coordinate system, respectively. However, the Cartesian wind coordinate axes are
\begin{equation}
\label{Vwind}
\begin{array}{l}
\tilde{x}=r\cos{\theta}\sin{\tilde{\varphi}} \hspace{10pt}\tilde{y}=r\sin{\theta}\sin{\tilde{\varphi}} \hspace{10pt}\tilde{z}=r\cos{\tilde{\varphi}}=0\\
\tilde{\varphi}=\dfrac{\pi}{2} \hspace{10pt}V_{\tilde{z}}=0 \hspace{10pt}
V_{\tilde{\varphi}}=0\hspace{3pt}.
\end{array}
\end{equation}
This system of axes is perpetually two-dimensional, that is, $\tilde{z}$ is always zero. For every $\varphi$ plane, there is a coincident $\tilde{\varphi}$ plane of the wind axes that has a constant value $\pi/2$ as depicted in Fig.~\ref{Fig1}. $V_{\tilde{z}}=V_{\tilde{\varphi}}=0$ for all these planes. The geometric axes of $\psi_{vis}$ are therefore transformed into the wind axes for a rotational symmetry of the flow in the three-dimensional space \cite{Amoloye2024}.

In terms of $\overrightarrow{\psi}$, Eq.~\ref{NSEDiffinc} becomes
\begin{equation}
\label{NSEPsi1a}
\begin{array}{l}
\nabla{}\times{}\dfrac{\partial{}\overrightarrow{\psi{}}}{\partial{}t}+\nabla{}\left(\dfrac{p}{\rho}+\dfrac{|\nabla{}\times\overrightarrow{\psi{}}|^2}{2}-\left[\dfrac{\lambda}{\rho}+2\nu{}\right]\nabla{}\cdot(\nabla{}\times\overrightarrow{\psi{}})\right)\\[10pt]=(\nabla{}\times\overrightarrow{\psi{}})\times{}\omega{}+\nabla{}\times{}\left(\nu{}\omega{}\right)\hspace{3pt}
\end{array}
\end{equation}
clearly presenting the difficulty of analytically obtaining a corresponding pressure field that conserves momentum.

A three-dimensional viscous potential function for which $\omega=0$ overcomes the difficulty presented in Eq.~\ref{NSEPsi1a} \cite{Amoloye2024,Joseph2006}. \color{black} Thus, t\color{black}o render \color{black}the velocity field from \color{black}$\psi_{vis}$ quasi-irrotational in the Cartesian wind coordinate system, \color{black}$\psi_{vis}$ \color{black}is defined on a principal axis of the flow about which $\omega$ is identically zero \cite{Amoloye2024}. The preceding transformations define $\tilde{\kappa}$, an incompressible \color{black} quasi-potential \color{black} stream function for the cylinder flow in RPT \cite{Amoloye2024}. $\tilde{\kappa}$ has the properties of a stream function in a cylindrical polar coordinate system that allow it to satisfy the conservation of mass (Eq.~\ref{contdiffinc}) \color{black}and produce a rotational velocity field \color{black}. Also, it has the properties of a potential function in the Cartesian wind coordinate axes (see appendix~\ref{CoMm} for more details) that \color{black}satisfy the incompressible NSE and \color{black}help to conserve momentum as 
\begin{equation}
\label{Velpol}
\begin{array}{l}
\mathbf{V}=\nabla{}\tilde{\kappa}\\[10pt]
\nabla{}\left(\dfrac{\partial{}\tilde{\kappa}}{\partial{}t}+\dfrac{p}{\rho{}}+\dfrac{\left|\nabla \tilde{\kappa}\right|^2}{2}-\left[\dfrac{\lambda}{\rho}+2\nu{}\right]{\nabla{}}^2\tilde{\kappa}\right)=0\hspace{6pt} (t\ge 0)
\hspace{3pt}
\end{array}
\end{equation}
because the pressure field is scalar and invariant to coordinate transformations.

$\tilde{\kappa}$ is a solution of Laplace's equation, satisfying the viscous boundary conditions and incorporating unsteady characteristics of the cylinder wake flow\cite{Amoloye2024}. \color{black}Some of these characteristics including the Strouhal number and the energy cascade correlate agreeably with CFD and experiments in the Reynolds number and non-dimensional time ranges of $30 < Re < 10,000$ and $0.2 \le V_\infty t/R \le 77,047$ respectively \cite{Amoloye2024} (where $R$ is the radius of the cylinder). \color{black}

Amoloye \cite{Amoloye2024} provides details of the refinement of the classical potential flow theory \color{black}of a cylinder flow \color{black}and the derivation of $\tilde{\kappa}$. Here, a compressible form of $\tilde{\kappa}$ is defined as
\begin{equation}
	\label{CompKappa}
	\kappa=\rho\tilde{\kappa} \hspace{3pt}
\end{equation} 
to include compressibility effects. Attention is focused on the process of propagating sound waves to describe the density field. This process is a one-dimensional steady isentropic one that is governed by the combined continuity and Euler's equation
\begin{equation}
	\label{isentropicEuler}
	\begin{array}{l}
		\dfrac{d\rho{}}{\rho{}}=-\dfrac{d(V^2)}{2a^2}\\[10pt]
	\end{array}
\end{equation}
where $a$ is the local speed of sound. Substituting the corresponding energy equation for a polytropic gas
\begin{equation}
	\label{energy}
	\begin{array}{l}
		a^2=a_0^2-\dfrac{\gamma{}-1}{2}V^2\\
	\end{array}\hspace{3pt},
\end{equation}
gives a first order partial differential equation in terms of $\rho$ and $V^2$ as
\begin{equation}
	\label{isentropicEuler2}
	\begin{array}{l}
		\dfrac{d\rho{}}{\rho{}}=
		-\dfrac{1}{2\left(a_0^2-\dfrac{\gamma{}-1}{2}V^2\right)}d(V^2)\hspace{3pt}\\
	\end{array}
\end{equation}
where $a_0$ is the stagnation speed of sound, $\gamma$ is the ratio of specific heats, and $V$ is obtained from the incompressible flow i.e. $\tilde{\kappa}$. \color{black}Although, sound propagation is analytically modeled as a one-dimensional process, Eq.~\ref{isentropicEuler} applies independently in each dimension because sound propagates radially outwards from a source and Euler's equation applies throughout the flow field, not just on a streamline \cite{Anderson2011}. Equations~\ref{isentropicEuler} and ~\ref{isentropicEuler2} are also employed to obtain the full velocity potential equation in standard textbooks on aerodynamics \cite{Anderson2011,Anderson2003}. The difference in the approach here is that $V$ is a known quantity from $\tilde{\kappa}$, which makes Eq.~\ref{isentropicEuler2} analytically integrable. \color{black}Just before the impulsive start, $a_0$ equals $a_\infty$.
Therefore, the density field 
\begin{equation}
	\label{Rhobeta}
	\dfrac{\rho{}}{{\rho{}}_{\infty{}}}={\left(\dfrac{1-\dfrac{\gamma{}-1}{2}M_{\infty{}}^2}{1-\dfrac{\gamma{}-1}{2}M^2}\right)}^{\dfrac{1}{\gamma{}-1}}
\end{equation}
is analytically integrated from the differential equation. The local Mach number, $M$, is equal to $V/a_\infty$. 

The velocity components in the cylindrical coordinate system are derived as
\begin{equation}
	\label{compVelCyl}
	\left\{\begin{array}{
			ccc}
		V_r\\[10pt] V_\theta\\[10pt] V_{\tilde{z}} 
	\end{array} \right\}=\dfrac{1}{\rho}\nabla \times \overrightarrow{\kappa}=\left\{\begin{array}{
			ccc}
		\dfrac{1}{\rho r}\dfrac{\partial \kappa}{\partial \theta}\\[10pt]-\dfrac{1}{\rho}\dfrac{\partial \kappa}{\partial r}\\[10pt]0 
	\end{array} \right\}
	\hspace{3pt},
\end{equation}
 and in the Cartesian \color{black}wind \color{black} coordinate system as
\begin{equation}
	\label{compVel}
	\left\{\begin{array}{
			ccc}
		V_{\tilde{x}}\\[10pt] V_{\tilde{y}}\\[10pt] V_{\color{black}z\color{black}} 
	\end{array} \right\}=\dfrac{1}{\rho}\nabla \kappa=\left\{\begin{array}{
			ccc}
		\dfrac{1}{\rho}\dfrac{\partial \kappa}{\partial x}\\[10pt]\dfrac{1}{\rho}\dfrac{\partial \kappa}{\partial y}\\[10pt]\dfrac{1}{\rho}\dfrac{\partial \kappa}{\partial z}\ 
	\end{array} \right\}
	\hspace{3pt}.
\end{equation}
\color{black}Eqs.~\ref{compVelCyl} and~\ref{compVel} are not equivalent, but the magnitude of the physical velocity field that they describe is invariant to coordinate transformation. \color{black}

$\kappa$ satisfies the conservation of mass in
\begin{equation}
	\label{compcontphi}
	\begin{array}{l}
		\oint_\mathcal{V}\left[\dfrac{\partial{}\rho{}}{\partial{}t}+\underbrace{\nabla{}\cdot\left(\rho{}\mathbf{V}\right)}_{=0}\right]d\mathcal{V}
        \\[10pt]=\dfrac{\partial{}}{\partial{}t}\left[\oint_\mathcal{V}\rho{}d\mathcal{V}\right]\\[10pt]=0\hspace{10pt} (t\ge 0),
	\end{array}
\end{equation}
because the volume integral of the transient variation of the density is zero in the \color{black}infinite \color{black}domain. 

\color{black}
Using a vector identity \cite{Arfenetal2013}, the third term on the left-hand side of Eq.~\ref{NSEDiff} is expanded and simplified as
\begin{equation}
\label{ConvectTerm1}
\begin{array}{l}
\left[(\rho{}\mathbf{V})\cdot\nabla{}\mathbf{V}\right]\\[10pt]=\nabla{}\left(\rho\mathbf{V}\cdot\mathbf{V}\right)-\mathbf{V}\cdot\nabla\left(\rho\mathbf{V}\right)\\[10pt]-\mathbf{V}\times\left(\nabla\times\rho\mathbf{V}\right)-\rho\mathbf{V}\times\left(\nabla\times\mathbf{V}\right)\\[10pt]
=\nabla{}\left(\rho\mathbf{V}\cdot\mathbf{V}\right)-\mathbf{V}\cdot\nabla\left(\rho\mathbf{V}\right)\\[10pt]-\mathbf{V}\times\left(\nabla\times\nabla\kappa\right)-\rho\mathbf{V}\times\omega.
\end{array}
\end{equation}
In the Cartesian wind coordinate axes, $\mathbf{V}\times\left(\nabla\times\nabla\kappa\right)$ is equal to zero because the curl of the gradient of a scalar field is a zero vector. $\rho\mathbf{V}\times\omega$ is also equal to zero because $\kappa$ is a \color{black} quasi-potential function in these axes\color{black}. Thus, Eq.~\ref{NSEDiff} reduces to
\begin{equation}
\label{NSEDiff3}
\begin{array}{l}
\dfrac{\partial{}\left(\rho{}\mathbf{V}\right)}{\partial{}t}+\nabla{}\left(\rho\mathbf{V}\cdot\mathbf{V}\right)=-\nabla{}p+\nabla{}\left(\left[\lambda + 2\mu{}\right]\nabla{}\cdot \mathbf{V}\right) \hspace{10pt} (t\ge 0).
\end{array}
\end{equation}
\color{black}Using Eq.~\ref{compVel}, \color{black}the conservation of momentum yields the pressure field as
\begin{equation}
	\label{ThermPressure}
	p=-\dfrac{\partial{}\kappa{}}{\partial{}t}-\color{black}\rho{}\color{black}\left|\dfrac{\nabla \kappa}{\rho}\right|^2+\color{black}\left[\lambda + 2\mu{}\right]\color{black}{\nabla{}}\cdot\left(\dfrac{\nabla{}\kappa{}}{\rho}\right)+F(t)
\end{equation}
where $F(t)$ is a function of time that is taken to be \color{black}the total pressure \color{black} of the free stream, \color{black}$p_{0}$. \color{black}

When $\mathbf{V}=0$ (the surface no-slip condition), the pressure field must satisfy the hydrostatic pressure condition in which the sum of the normal stresses on the principal axes is a tensor invariant \cite{White2006,Joseph2003,Joseph2006}. The mechanical pressure \cite{White2006} (or the mean pressure \cite{GadelHak1995}), $\tilde{p}$, is defined as a negative one-third of the sum of the normal stresses on a fluid element. Thus, rearranging Eq.~\ref{ThermPressure} by separation of the isotropic and deviatoric parts of the stress component and taking the mean gives 
\begin{widetext}
\begin{equation}
	\label{ThermPressure2}
	p=-\dfrac{\partial{}\kappa{}}{\partial{}t}-\rho{}\left|\dfrac{\nabla \kappa}{\rho}\right|^2+\dfrac{4}{3}\mu{\nabla{}}\cdot\left(\dfrac{\nabla{}\kappa{}}{\rho}\right)+\left(\lambda{}+\frac{2}{3}\mu{}\right){\nabla{}}\cdot\left(\dfrac{\nabla{}\kappa{}}{\rho}\right)+p_{0}
\end{equation}
so that\color{black}
\begin{equation}
	\label{MechPressure}
	\tilde{p}=\left[\dfrac{\partial{}\kappa{}}{\partial{}t}+\color{black}\rho{}\left|\dfrac{\nabla \kappa}{\rho}\right|^2-\dfrac{4}{3}\mu{\nabla{}}\cdot\left(\dfrac{\nabla{}\kappa{}}{\rho}\right)-p_{0}\right]-\left(\lambda{}+\frac{2}{3}\mu{}\right){\nabla{}}\cdot\left(\dfrac{\nabla{}\kappa{}}{\rho}\right)\hspace{3pt}.
\end{equation}
\end{widetext}
The expression in the square brackets is negative of the thermodynamic pressure field. \color{black} Equation~\ref{MechPressure} is dimensionally consistent (see Appendix~\ref{DimAnal} for details) .\color{black}

Stokes' hypothesis\cite{White2006} assumes that $\lambda{}$ is 
\begin{equation}
	\label{Stokes' hypothesis}
	\lambda{}=-\dfrac{2}{3}\mu{} \hspace{3pt}.
\end{equation}
However, \color{black}$\kappa$ is defined on \color{black}the \color{black}principal \color{black}axes \color{black}of the flow \color{black}where the vorticity vector is identically zero\color{black} \cite{Amoloye2024}. Thus, $\lambda$ and the viscous normal stresses are \color{black}significant\cite{Joseph2006,Joseph2003,GadelHak1995,Schlichting1979,White2006} \color{black}in \color{black}Eq.~\ref{MechPressure}. Quantitative analyses on $\lambda$ and the pressure field in RPT are the subject of further publications.
\color{black}
\section{Results and Discussion}\label{RnD}
\begin{table}
	\begin{center}
		\caption{Cylinder radius and flow conditions.}
		\label{Flow Conditions}
		\begin{tabular}{rrrr}
			\hline
			 Radius, $R$ ($m$) &  $\nu_{\infty}$ ($m^{2}s^{-1}$) & $\mu_g$ ($m^{3}s^{-2}$) &   \\\hline
			$0.047$&  $1.46069\times 10^{-5}$ &$6.82\times 10^{-11}$ &\\
			\hline
		\end{tabular}
	\end{center}
\end{table}
The conditions presented in table~\ref{Flow Conditions} were used for the evaluation of the results unless otherwise stated. The choices for $R$ and the wake analogy factor, $\mu_g$, are informed by a preceding study of the incompressible cylinder flow with RPT \cite{Amoloye2024}. $\nu_\infty$ is for air at a temperature of about $15^\circ c$. \color{black}In another work\cite{Amoloyeetal2026}, the sensitivity of RPT to $\nu$ is explored, showing that the appropriate selection of $\nu$ is critical to influence flow stability, symmetry and energy spectra. This is similar to tuning artificial viscosity in some CFD simulations \cite{Amoloyeetal2026}. The $\gamma$ values used for the test cases are typical values in aerospace ($\gamma=1.4$; $\gamma=1.67$)\cite{Anderson2006,SinclairandCui2017} and astrophysical ($\gamma=1.67$; $\gamma=3.00$) \cite{Shivamoggi2025,KeshetandNaor2016,Jelineketal2022,Schereretal2016,Livadiotis2015} applications. However, the specific combinations of the values of these two variables serve as parametric studies and may not correspond to an achievable physical scenario. \color{black}All results are for the $z/D=0$ plane, corresponding to a two-dimensional flow. 

\begin{figure}
	\centering
	\includegraphics[width=0.5\textwidth,trim={0cm 0cm 0cm 0cm},clip]{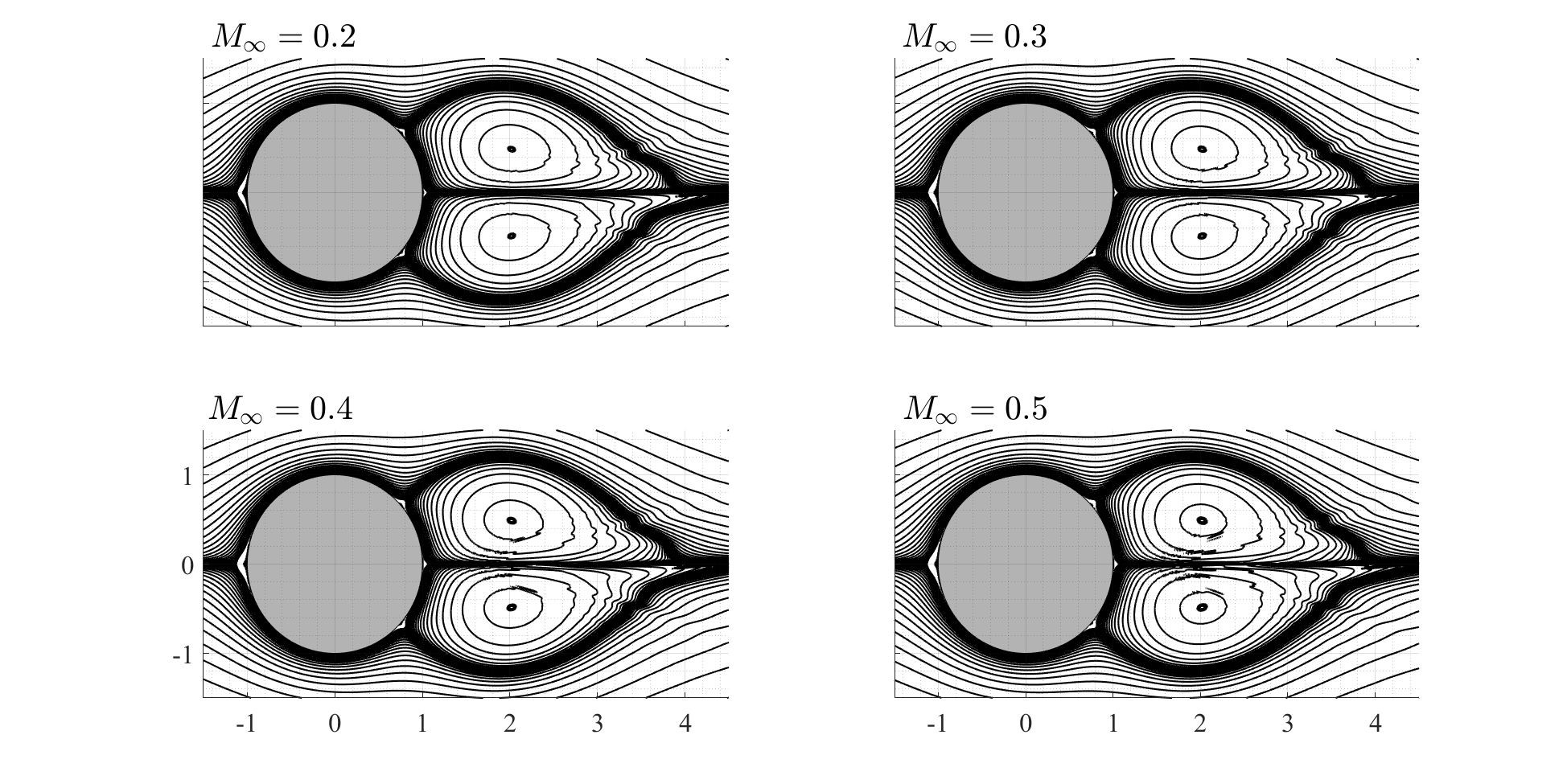}
	\caption{Instantaneous streamlines for varying $M_\infty$ at $Re = 34$ and $t=43R/V_\infty$.}\label{Fig2}
\end{figure}

The subtle effect of compressibility at low $Re$ is presented in figure~\ref{Fig2} that shows the instantaneous streamline contours (in the cylinder reference frame) of compressible flow in RPT at $Re=34$, $t = 43R/V_\infty$, and at varying $M_\infty$. As $M_\infty$ progresses, the strength of the wake bubbles increases evidently from the tightening of the streamlines around the eyes of the vortices. Consequently, the flow around the wake centerline becomes progressively distorted. This agrees with the observation in the literature that compressibility marginally increases vortex enstrophy \cite{Taoetal2025}.

\begin{figure}
	\centering
	\includegraphics[width=0.45\textwidth,trim={0cm 0cm 0cm 0cm},clip]{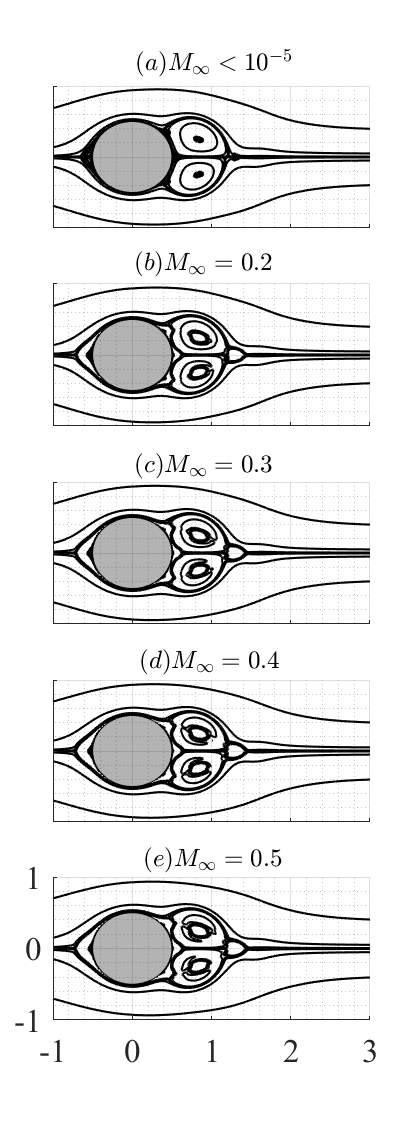}
	\caption{Instantaneous streamlines for varying $M_\infty$ at $Re = 3,900$ and $t=3R/V_\infty$.}\label{Fig3}
\end{figure}

Similar observations can be made from figure~\ref{Fig3} showing the RPT streamline contours of compressible flow at $Re=3,900$, $t = 3R/V_\infty$, $\gamma=1.4$ and varying $M_\infty$. A significantly lower viscosity value compared to Amoloye's analysis \cite{Amoloye2024} was used to delay the vortex shedding and allow a comparison with the CFD results of Xue et al. \cite{Xueetal2024}. Whereas Xue et al.\cite[figure~$17$, p.~$12$]{Xueetal2024} present a regression in the length of the recirculation zone with a progression in $M_\infty$ in the range $0.2\le M_\infty \le 0.5$, the RPT results show a nearly constant streamwise dimension of the recirculation zone, in agreement with Nagata et al.'s \cite{Nagataetal2020b} experimental results at $Re=2,000$. This suggests an insensitivity of the RPT wake flow stability to $M_\infty$ at $Re =3,900$ contrary to the CFD results  \cite{Xueetal2024}. \color{black} This insensitivity is due to the use of a constant viscosity in the model. Compressibility increases the local temperature and consequently $\mu$, as $M_\infty$ progresses. The more viscous the fluid, the shorter the recirculation zone. This process is treated in the CFD results with a coupling of the energy equation and Sutherland's law in the simulation\cite{Xueetal2024}, whereas the present RPT results do not feature this. \color{black}Nevertheless, RPT captures the existence and growth in intensity with $M_\infty$ of a secondary recirculation flow structure at the back of the cylinder similar to CFD \cite{Xueetal2024}.

\begin{figure}
	\centering
	\includegraphics[width=0.5\textwidth,trim={0cm 0cm 0cm 0cm},clip]{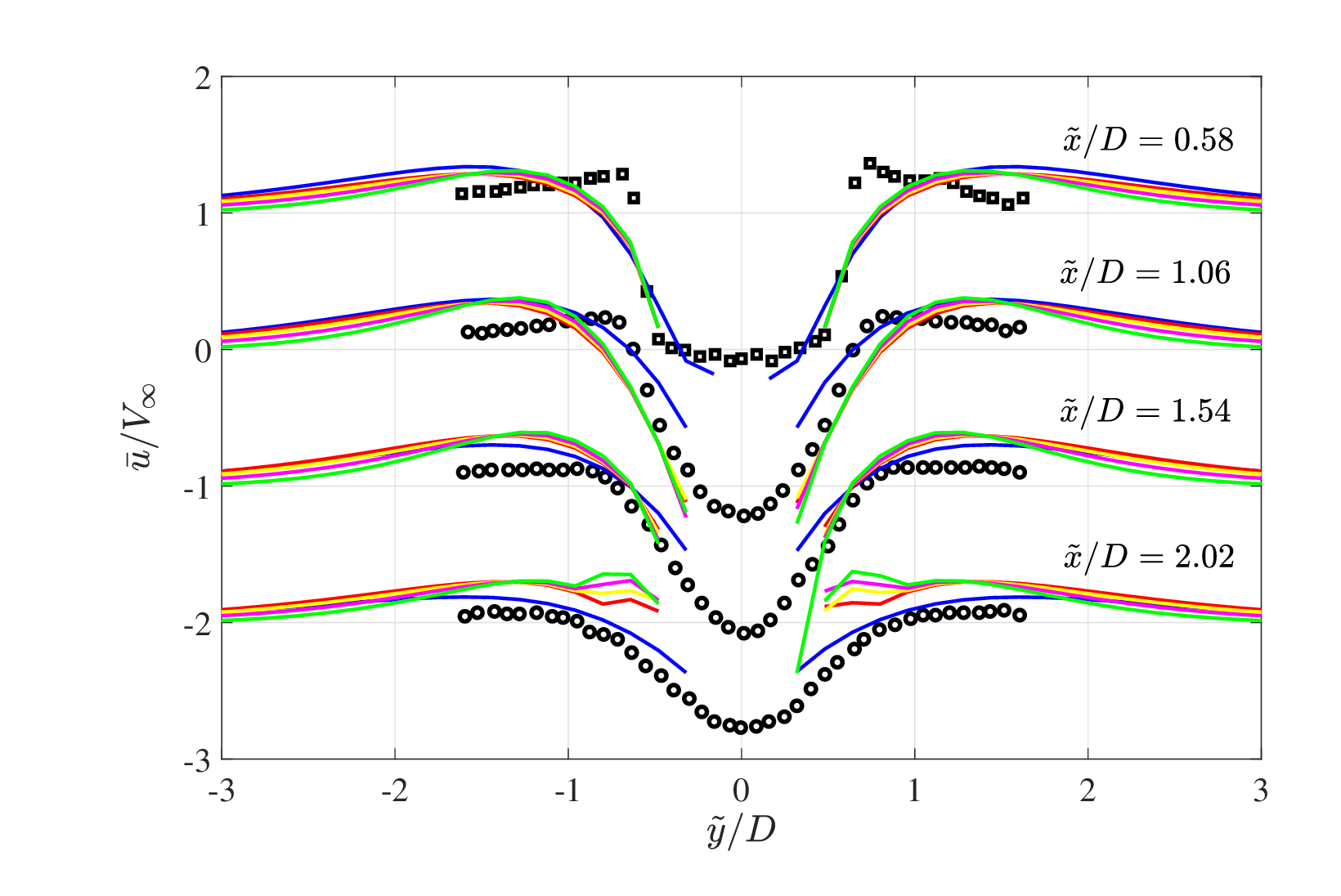}
	\caption{Time-averaged streamwise velocity in a cylinder wake for varying $M_\infty$ at $Re = 3,900$. $M_\infty<10^{-5}~(\color{blue}-\color{black})$$;~M_\infty=0.2~(\color{black}-\color{black})$$;~M_\infty=0.3~(\color{yellow}-\color{black})$$;~M_\infty=0.4~(\color{magenta}-\color{black})$$;~M_\infty=0.5~(\color{green}-\color{black})$}\label{Fig4}
\end{figure}

\begin{figure}
	\centering
	\includegraphics[width=0.5\textwidth,trim={0cm 0cm 0cm 0cm},clip]{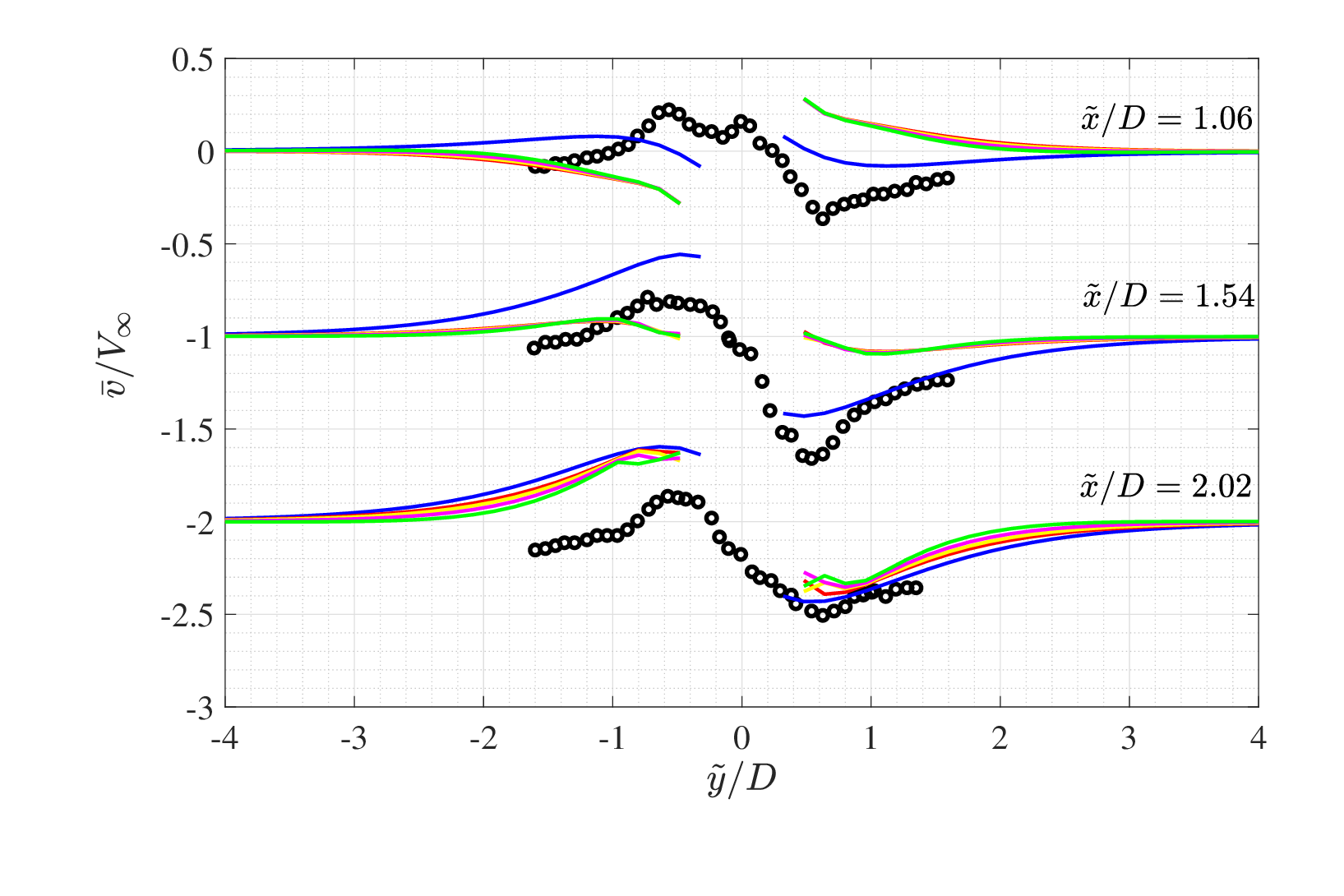}
	\caption{Time-averaged crossflow velocity in a cylinder wake for varying $M_\infty$ at $Re = 3,900$. $M_\infty<10^{-5}~(\color{blue}-\color{black})$$;~M_\infty=0.2~(\color{black}-\color{black})$$;~M_\infty=0.3~(\color{yellow}-\color{black});$$~M_\infty=0.4~(\color{magenta}-\color{black})$$;~M_\infty=0.5~(\color{green}-\color{black})$}\label{Fig5}
\end{figure}

The streamwise extent of the primary recirculation region captured at $Re =3,900$, which is nearly constant throughout the sampled $M_\infty$, is generally approximately smaller by a diameter magnitude for the RPT compared to the CFD results of Xue et al.\cite{Xueetal2024}. There is a match at $M_\infty =0.5$, where the length of the recirculation zone is the shortest for the CFD results. The length of the recirculation region has a bearing on the shape of the time-averaged streamwise velocity profiles at different stations within the wake, as presented in figure~\ref{Fig4} for varying $M_\infty$ at $Re = 3,900$, $\nu=1.46069 \times 10^{-8} m^2/s$, and $\gamma =1.4$. Figure~\ref{Fig5} also shows the time-averaged crossflow velocity in a cylinder wake under the same conditions. In these figures, Kravchenko and Moin's\cite{KravchenkoandMoin2000} experimental results of the incompressible flow at the same $Re$ have been included for a complete comparison. From the first downstream station on the figures, every successive profile has been shifted by $-1$ from the preceding one. The RPT averages were numerically obtained from instantaneous samples in the non-dimensional time range $1\le V_\infty t/R \le 77,047$ using the default settings of an in-built integrator in \textit{Matlab\texttrademark}\cite{Matlab}. $\tilde{x}$ and $\tilde{y}$ in these figures are equivalent to $x$ and $y$ respectively in the experimental data \cite{KravchenkoandMoin2000,Amoloye2024}. Outlier points \cite{Amoloye2024} at the base of the trough were left out to aid in a general assessment of the shape of the profiles. The streamwise velocity profiles after $\tilde{x}/D =0.58$ are generally V-like for the sampled $M_\infty$, characterizing the short recirculation zones in figure~\ref{Fig3}. In particular, the shape of the $\tilde{x}/D =1.06$ profile is consistent up to $M_\infty\color{black}=0.5$\color{black}, while its outline is sensitive to variations in $M_\infty$ in the CFD results \cite{Xueetal2024}. Since the length of the recirculation bubble and the consequent shape of the velocity profile are determined by the shear layer transition \cite{Xueetal2024}, the consistency of the V-shaped streamwise velocity profile in the near wake seems suggestive of a marginal effect of compressibility on the shear layer transition in RPT at $Re=3,900$ for $M_\infty \le 0.5$. Additionally, figure~\ref{Fig5} shows insignificant changes in the crossflow velocity profile for $0.2\le M_\infty \le 0.5$. However, at $\tilde{x}/D =2.02$, it is evident that compressibility reduces the velocity deficit in the RPT wake. This agrees with CFD result at similar conditions at the same downstream location \cite[figure~$16$, p.~$11$]{Xueetal2024}.   

\begin{figure*}
	\centering
	\includegraphics[width=\textwidth,trim={0cm 0cm 0cm 0cm},clip]{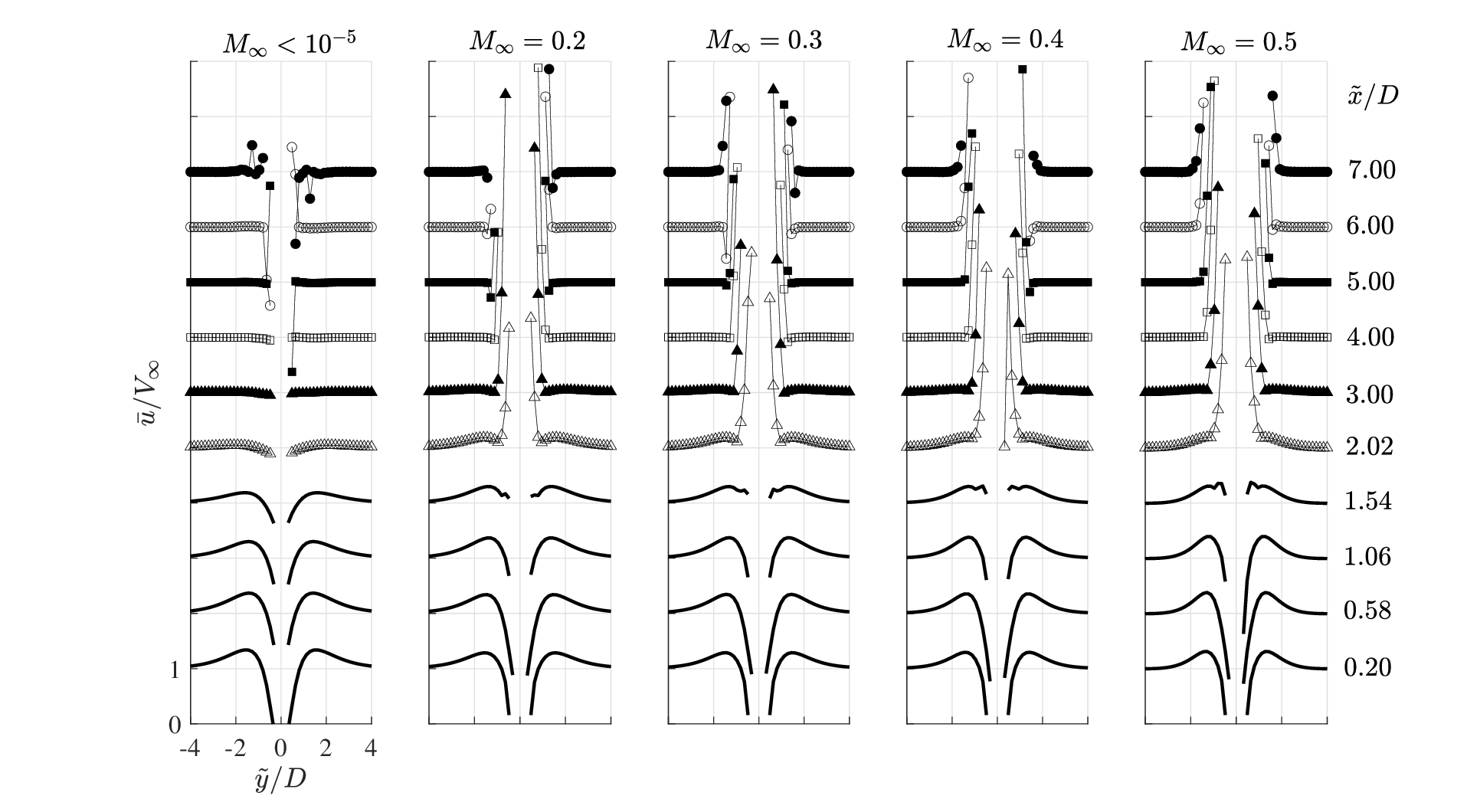}
	\caption{Time-averaged streamwise velocity in a cylinder wake for varying $M_\infty$ at $Re = 3,900$.}\label{Fig6}
\end{figure*}

\begin{figure*}
	\centering
	\includegraphics[width=\textwidth,trim={0cm 0cm 0cm 0cm},clip]{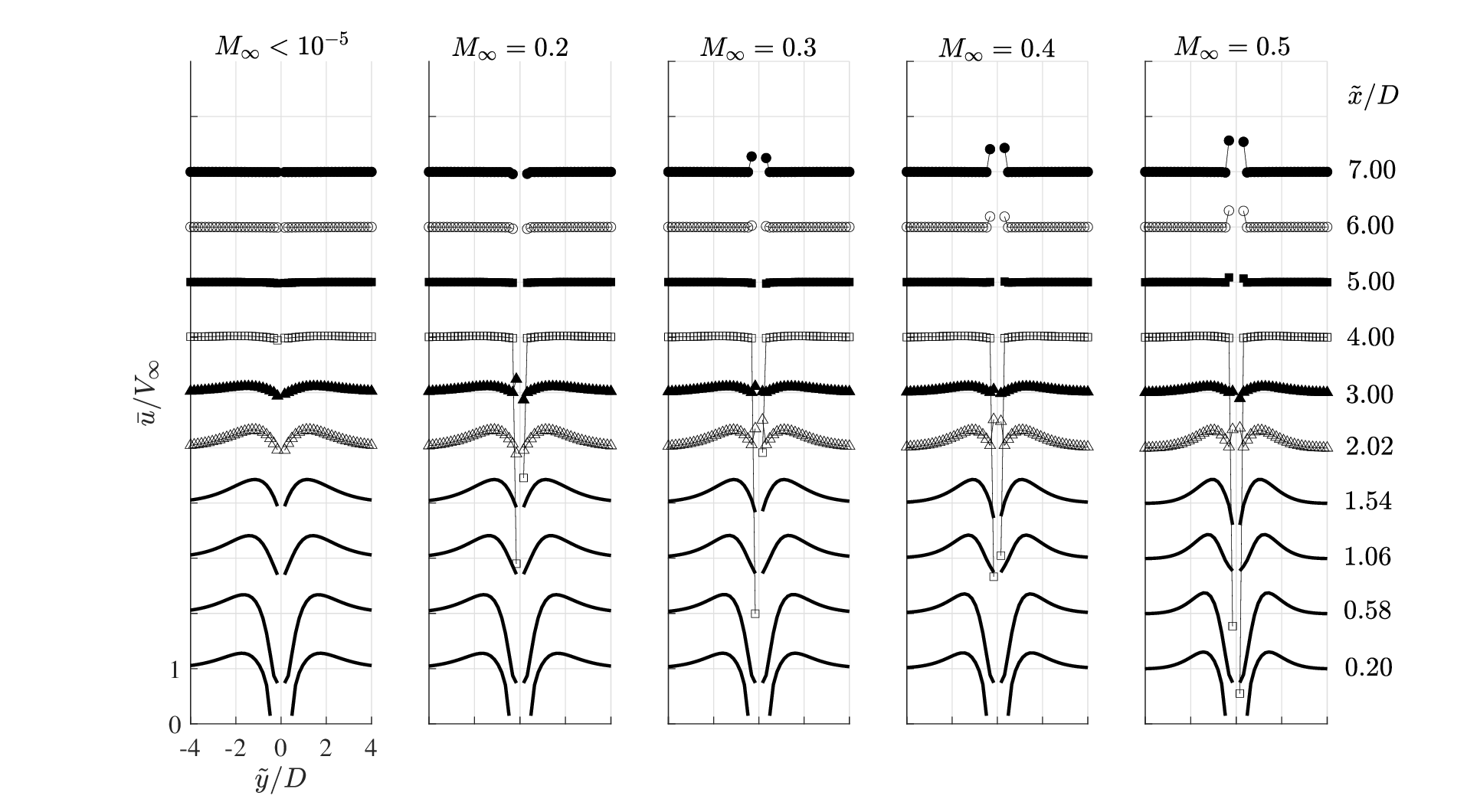}
	\caption{Time-averaged streamwise velocity in a cylinder wake for varying $M_\infty$ at $Re = 56$.}\label{Fig7}
\end{figure*}

\begin{figure*}
	\centering
	\includegraphics[width=\textwidth,trim={0cm 0cm 0cm 0cm},clip]{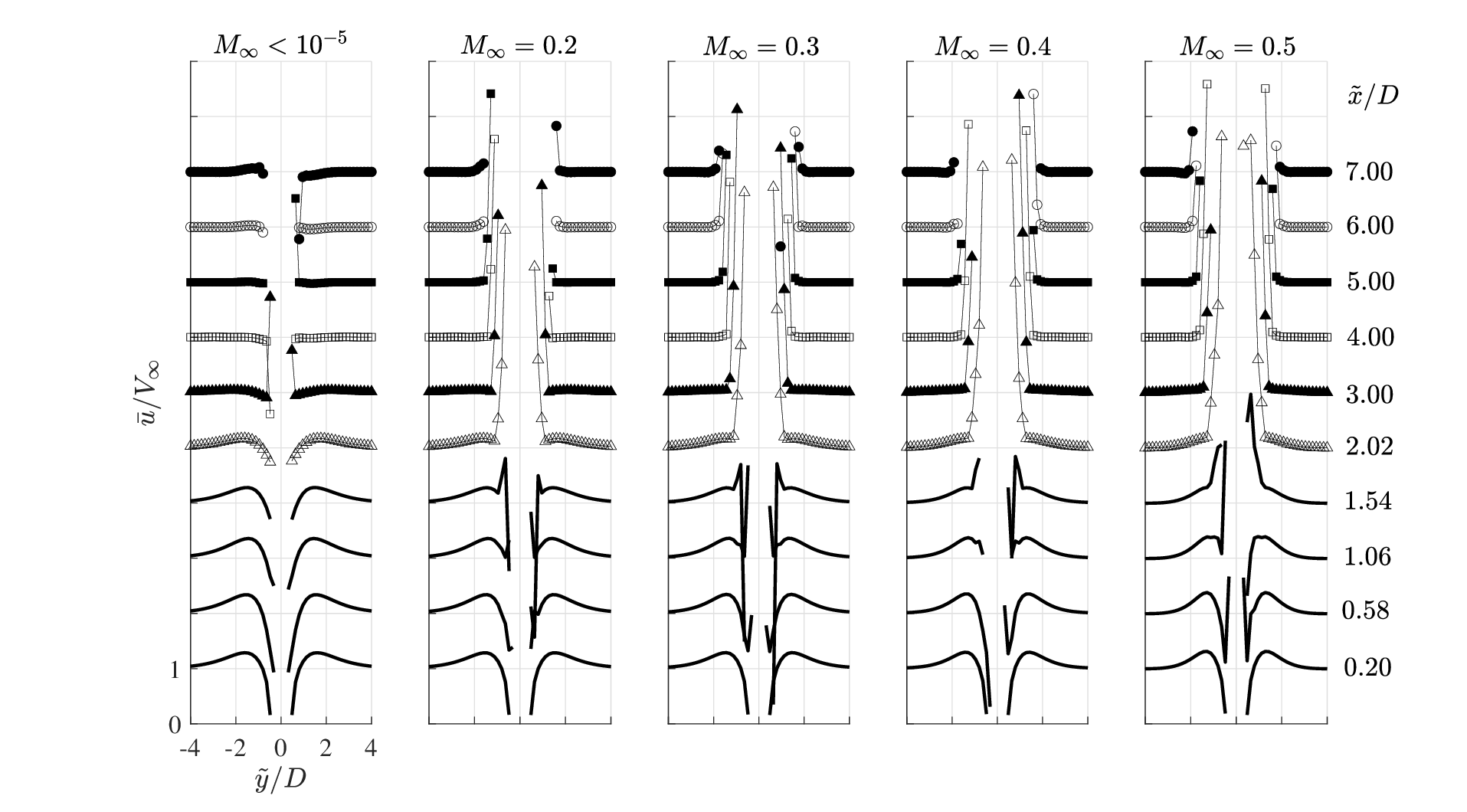}
	\caption{Time-averaged streamwise velocity in a cylinder wake for varying $M_\infty$ at $Re = 9,500$.}\label{Fig8}
\end{figure*}

Further in the wake, compressibility increases vortex enstrophy in agreement with CFD observations \cite{Taoetal2025}, exacerbating the outlier points \cite{Amoloye2024}. This is revealed when going from left to right in figure~\ref{Fig6} that shows time-averaged streamwise velocity in a cylinder wake for varying $M_\infty$ at $Re = 3,900$, $\nu=1.46069 \times 10^{-8} m^2/s$, and $\gamma =1.4$. The profiles are presented for several downstream locations in the wake. From $\tilde{x}/D=0.58$, each successive profile is shifted upwards by $-1$ for clarity. The averaging period is the same as in figure~\ref{Fig4}. Figures~\ref{Fig7} and \ref{Fig8} present similar plots at the same conditions for $Re=56$ and $Re=9,500$, respectively. An irregular pattern of velocity deficit and acceleration is experienced for $\tilde{x}/D \ge 3.00$, with the irregularity least severe in the laminar regime ($Re=56$) and most severe in the turbulent regime ($Re=9,500$).

\begin{table}
	\begin{center}
		\caption{Comparison of predicted Strouhal number, $St$, for varying $M_\infty$.}
		\label{Strouhal}
        \begin{tabular}{rrrr}
			\hline
			 $M_\infty$ &  RPT & CFD \cite{Xueetal2024}  &   \\\hline
			$0.01$&  $0.207$& $0.212$& \\
            $0.1$&   $0.174$& & \\
            $0.2$&   $0.239$& $0.208$ & \\
            $0.3$&   $0.228$& $0.206$ & \\
            $0.4$&   $0.228$& $0.203$ & \\
            $0.5$&   $0.228$& $0.198$ & \\
			\hline
		\end{tabular}
	\end{center}
\end{table}

\begin{figure*}
	\centering
	\subfloat[\color{black}$(\tilde{x}/D,\tilde{y}/D)=(3.00,0.56)$: $M_\infty<10^{-5}~(\color{blue}-\color{black})$$;~M_\infty=0.2~(\color{black}-\color{black})$$;~M_\infty=0.3~(\color{yellow}-\color{black})$$;~M_\infty=0.4~(\color{magenta}-\color{black})$$;~M_\infty=0.5~(\color{green}-\color{black})$\color{black}]{\includegraphics[width=0.5\textwidth,trim={0cm 0cm 0cm 0cm},clip]{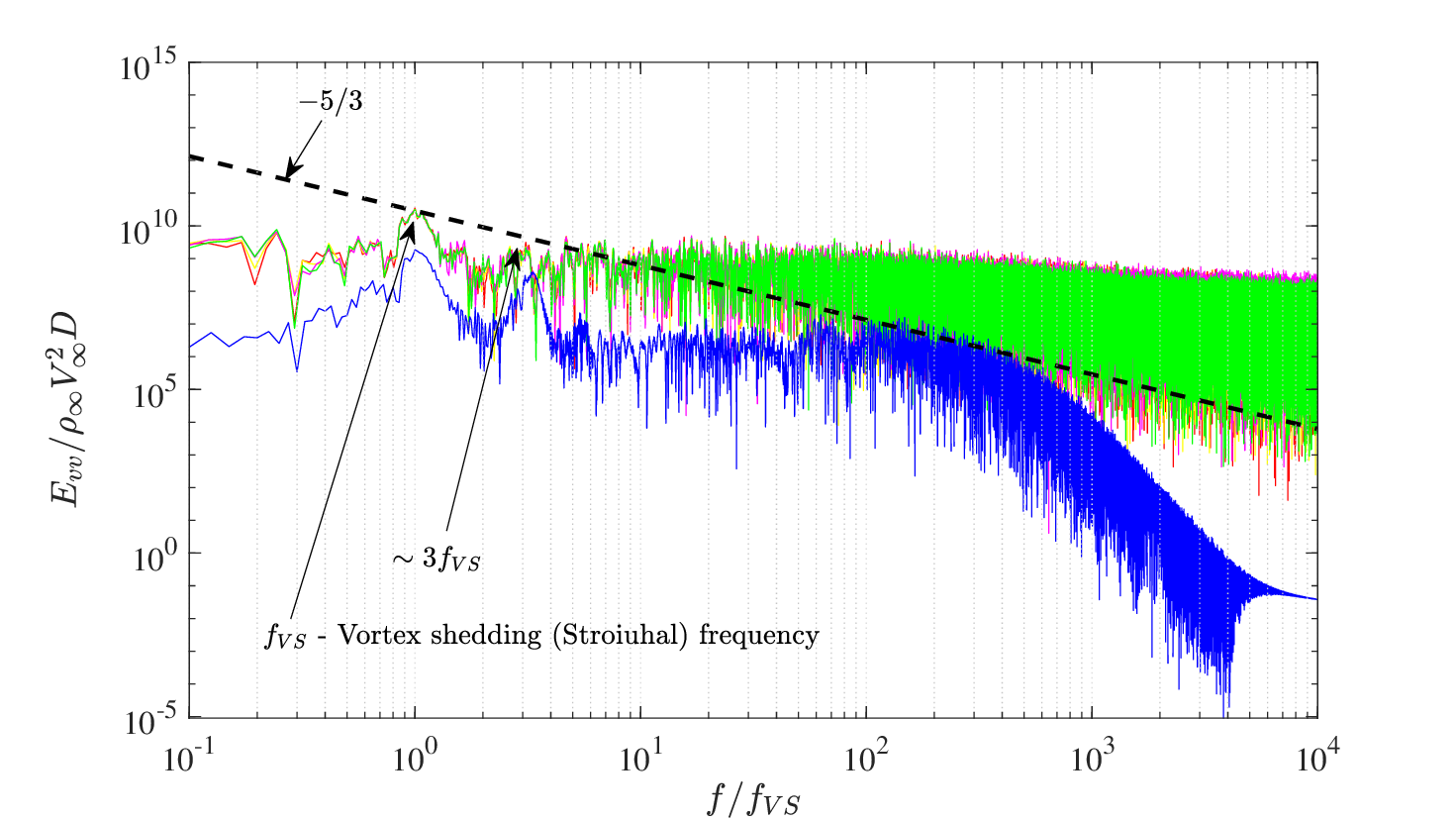}\label{Fig9a}}
    \subfloat[\color{black}$(x/D,y/D)=(3.00,0)$ \cite{Xueetal2024}. Reproduced from Kuiju Xue, Qinling Li, and Liangyu Zhao, Physics of Fluids, $36$, $085165$ ($2024$), with the permission of AIP Publishing.\color{black}]{\includegraphics[width=0.5\textwidth,trim={0cm 0.2cm 0cm 0cm},clip]{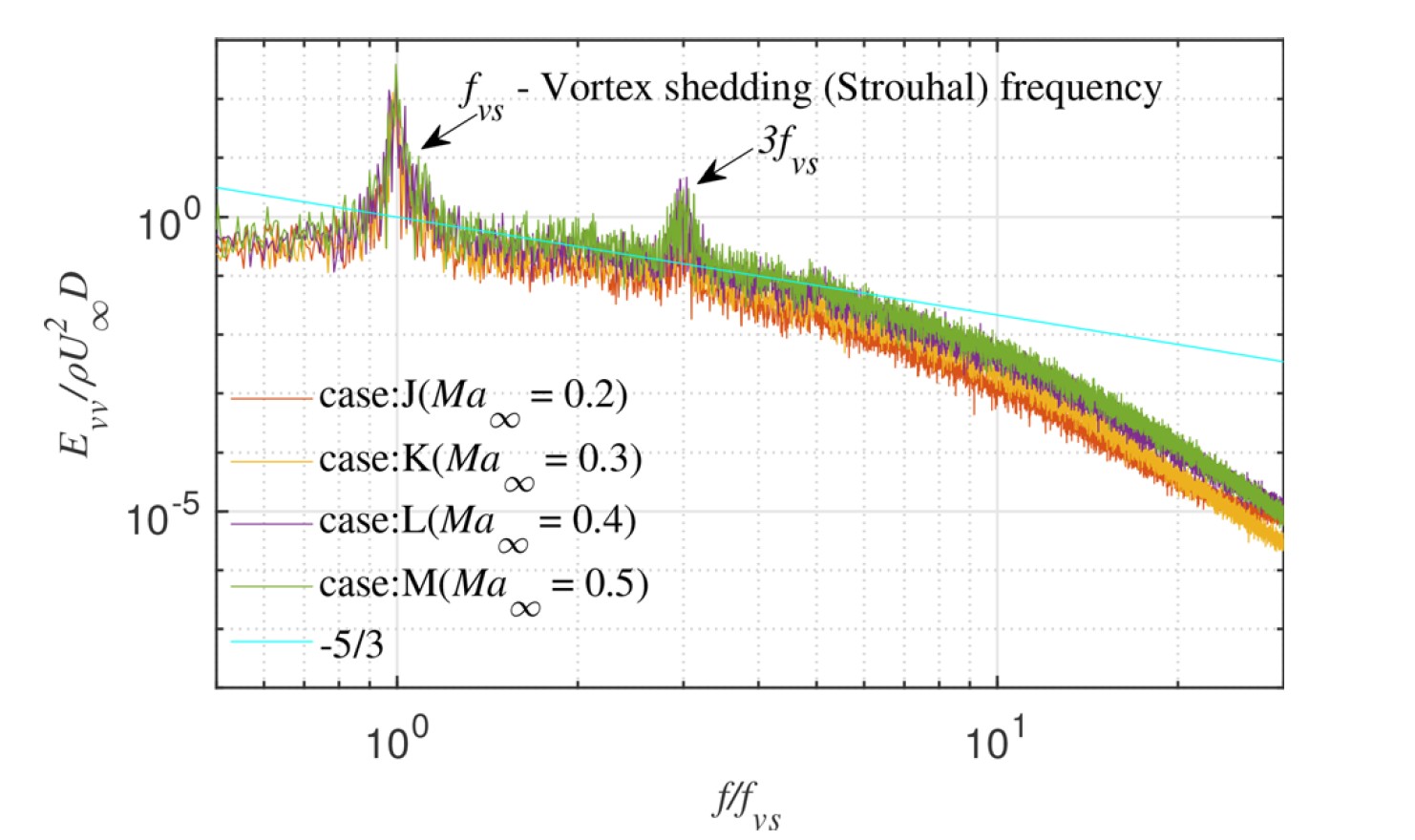}\label{Fig9b}}
	\caption{Energy spectra of the crossflow velocity in a cylinder wake for varying $M_\infty$ at $Re = 3,900$ \color{black}for RPT $(a)$ and CFD \cite{Xueetal2024} $(b)$\color{black}.}
\end{figure*}

\begin{figure}
	\centering
    \subfloat[]{\includegraphics[width=0.5\textwidth,trim={0cm 0cm 0cm 0cm},clip]{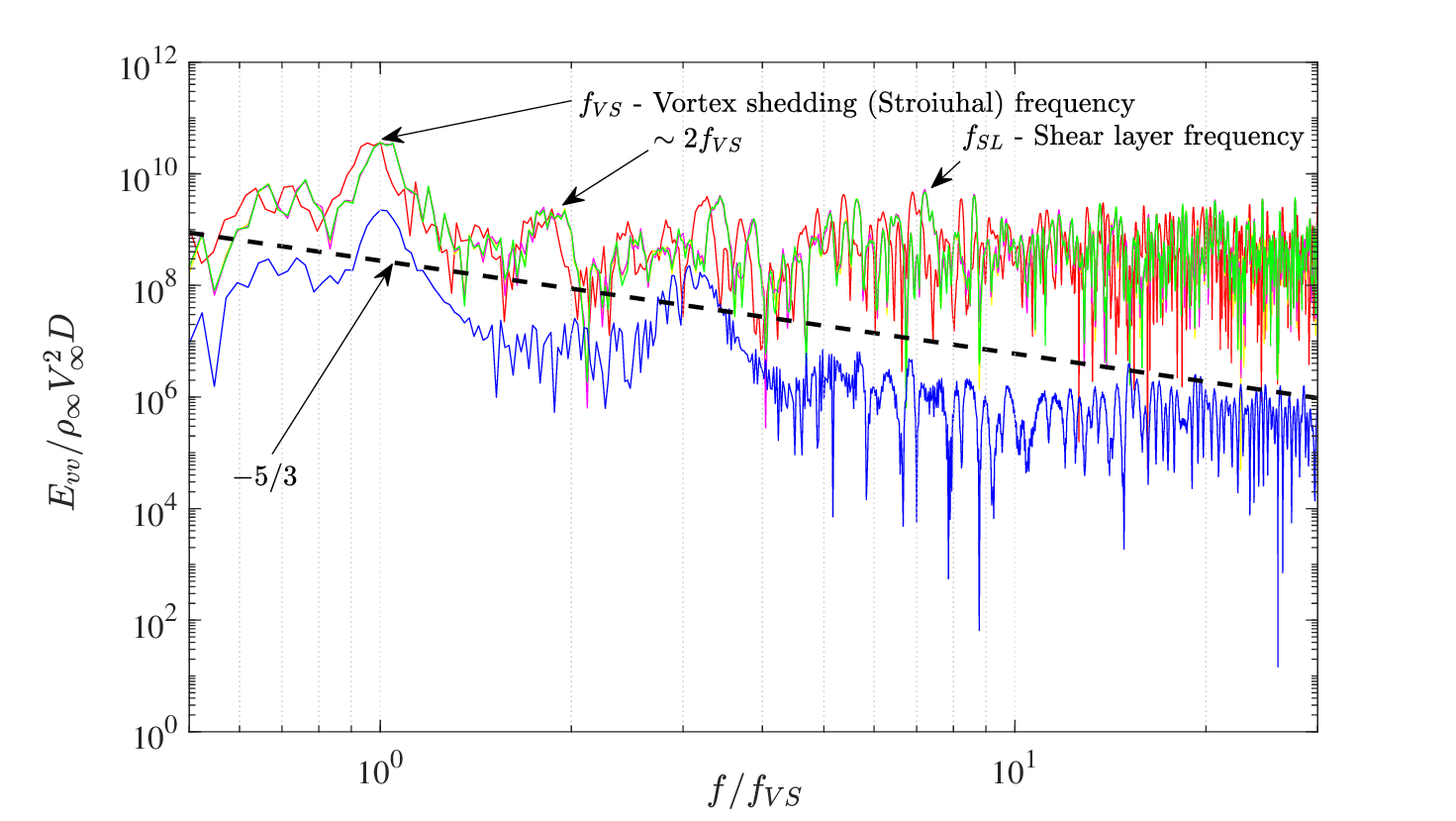}\label{Fig10a}}\\
	\subfloat[]{\includegraphics[width=0.5\textwidth,trim={0cm 0cm 0cm 0cm},clip]{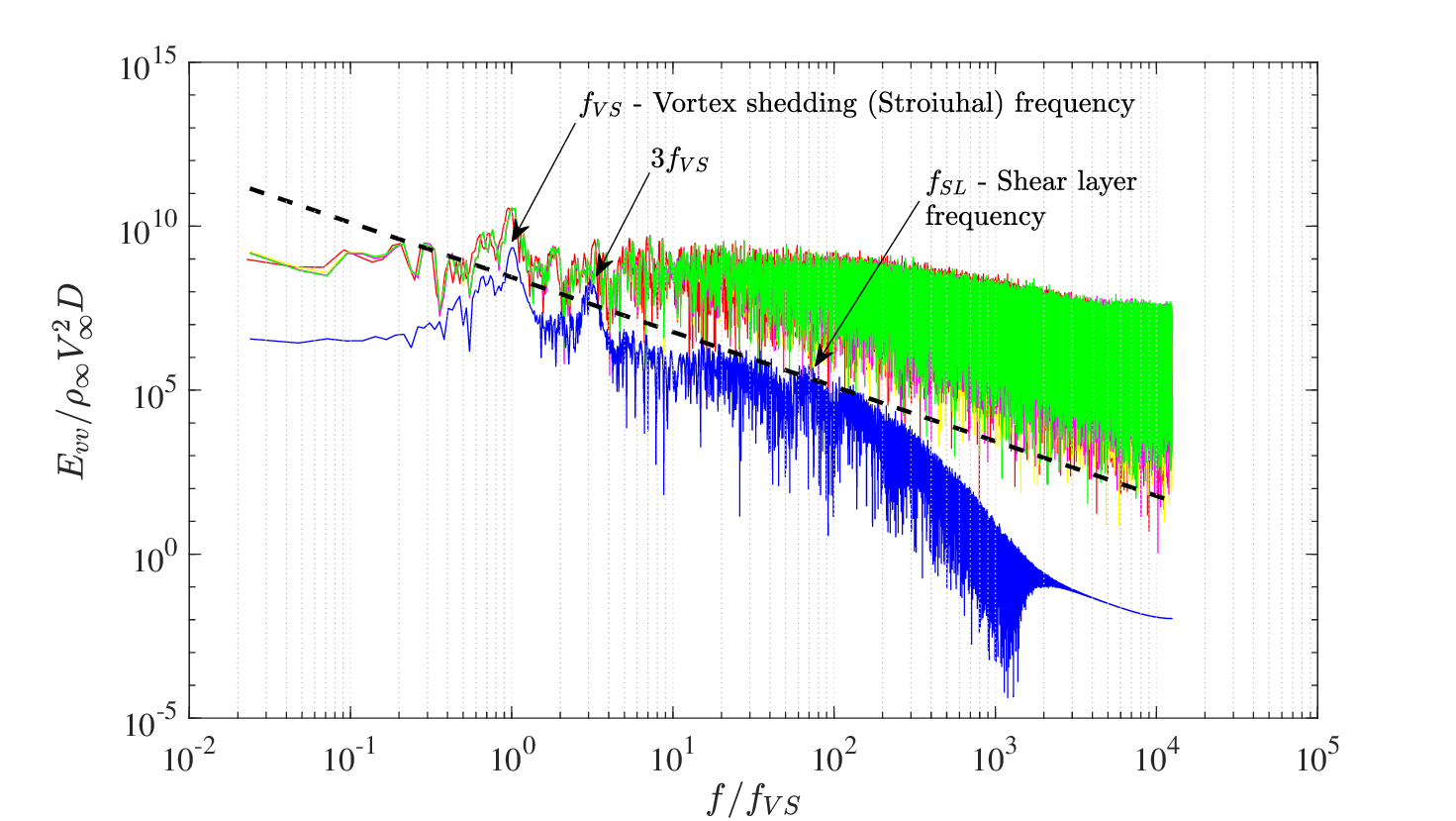}\label{Fig10b}}
	\caption{Energy spectra of the crossflow velocity in a cylinder wake for varying $M_\infty$ at $Re = 3,900$ and $(\tilde{x}/D,\tilde{y}/D)=(0.69,0.69)$.$M_\infty<10^{-5}~(\color{blue}-\color{black})$$;~M_\infty=0.2~(\color{black}-\color{black})$$;~M_\infty=0.3~(\color{yellow}-\color{black})$$;~M_\infty=0.4~(\color{magenta}-\color{black})$$;~M_\infty=0.5~(\color{green}-\color{black})$,}\label{Fig10}
\end{figure}

Figure~\ref{Fig9a} shows the one-dimensional energy spectra of the crossflow velocity in a cylinder wake for varying $M_\infty$, $Re = 3,900$, and $\gamma =1.4$ at $(\tilde{x}/D,\tilde{y}/D)=(3.00,0.56)$. These are wavenumber spectra \cite{Amoloye2024} produced from $829, 250$ samples that were obtained over a duration of $11.28$s while advancing downstream at $V_\infty$ from each sampled location where the theoretical flow was frozen at $t_0 = 75,500R/V_\infty$. The Fast-Fourier Transform (FFT) was used to compute the spectra. The frequency is non-dimensionalized by the vortex shedding frequency, $f_{VS}$. This Strouhal frequency and its first harmonic (at $\sim3f_{VS}$) are distinctly visible in the $M_\infty < 10^{-5}$ spectrum in which there are decades of wavenumber where the slope of the spectrum is $-5/3$, signifying the inertial range of turbulence in excellent agreement with experimental and CFD results in the incompressible regime \cite[figure~$4$,~p.$7$]{Xueetal2024}. The inertial range of the turbulence is disrupted by compressibility effects as $M_\infty$ increases, introducing some background noise into the spectra that slightly conceals the first harmonic at $\sim3f_{VS}$. This is in contrast to the CFD energy spectra under similar $Re$ and $M_\infty$ conditions but a different vertical wake location on the center line \color{black}(figure~\ref{Fig9b}\cite{Xueetal2024}) \color{black}in which both the inertial range and first harmonic at $3f_{VS}$ are distinctively discernible across the sampled Mach number range. \color{black} As $M_\infty$ increases, a corresponding change in the fluid viscosity that is appropriately computed via a coupling of the governing equations, including the energy equation, and Sutherland's law maintains the energy cascade in the CFD as opposed to the RPT model. \color{black}Figure~\ref{Fig9a} is not evaluated on the wake center line because the finite velocity jump on the rear axis of the cylinder in RPT present some difficulty for spectra analysis at that location \cite{Amoloye2024}. However, the spectra analyses in figure~\ref{Fig10}, with flow conditions similar to that of figure~\ref{Fig9a}, \color{black}are at \color{black} equivalent $(\tilde{x}/D,\tilde{y}/D)=(0.69,0.69)$ to CFD results \cite[figure~$22$,~p.$14$]{Xueetal2024}, giving a more direct comparison and the opportunity to assess the behavior of the shear layer in RPT under compressibility effects. The streamwise velocity spectra (not shown here) are much noisier and present greater difficulty in reading the Strouhal frequency off the spectra as $M_\infty$ increases. Thus, in Table~\ref{Strouhal}, only the RPT Strouhal numbers for $M_\infty\le 0.01$ were read off the spectra of the streamwise velocity. The crossflow velocity oscillates at a dominant frequency that is higher than the CFD and Nagata et al.'s\cite[figure~$13$, p.$15$]{Nagataetal2020b} experimental results at the corresponding $M_\infty$. The relevant part of Nagata et al.'s\cite[figure~$13$, p.$15$]{Nagataetal2020b} experimental results to the current discussion are at a close $Re=4,000$, contrary to the RPT and CFD results that are at $Re=3,900$. However, one of the general trends in the RPT predictions that agrees with both CFD and the experimental results is that compressibility has only a marginal effect on the Strouhal number for $0.2\le M_\infty \le 0.5$. The shape of the spectra remains the same in this Mach range and markedly different for $M_\infty < 10^{-5}$. From $M_\infty < 10^{-5}$, the energy increases with $M_\infty$, but remains unchanged in the range $0.2\le M_\infty \le 0.5$, contrary to CFD \cite{Xueetal2024}. \color{black}However, \color{black}the second harmonic shifts from $\sim3f_{VS}$ to $\sim2f_{VS}$ as the Mach number increases from $M_\infty < 10^{-5}$ to the $0.2\le M_\infty \le 0.5$ range in an excellent agreement with CFD\cite{Xueetal2024}. The harmonic at $\sim3f_{VS}$ is reduced in energy but becomes part of a system of pseudo-peaks that includes the shear layer frequency, $f_{SL}$, at $\sim7f_{VS}$ (in figure~\ref{Fig10a}). This value of shear layer frequency is about a tenth of its value for $M_\infty <10^{-5}$ (in figure~\ref{Fig10b}). This suggests that compressibility in RPT suppresses shear layer instability, favoring lower frequency oscillations as observed in both experimental \cite{Nagataetal2020b} and CFD \cite{Xueetal2024} investigations. 

\begin{figure}
	\centering
	\includegraphics[width=0.5\textwidth,trim={0cm 0cm 0cm 0cm},clip]{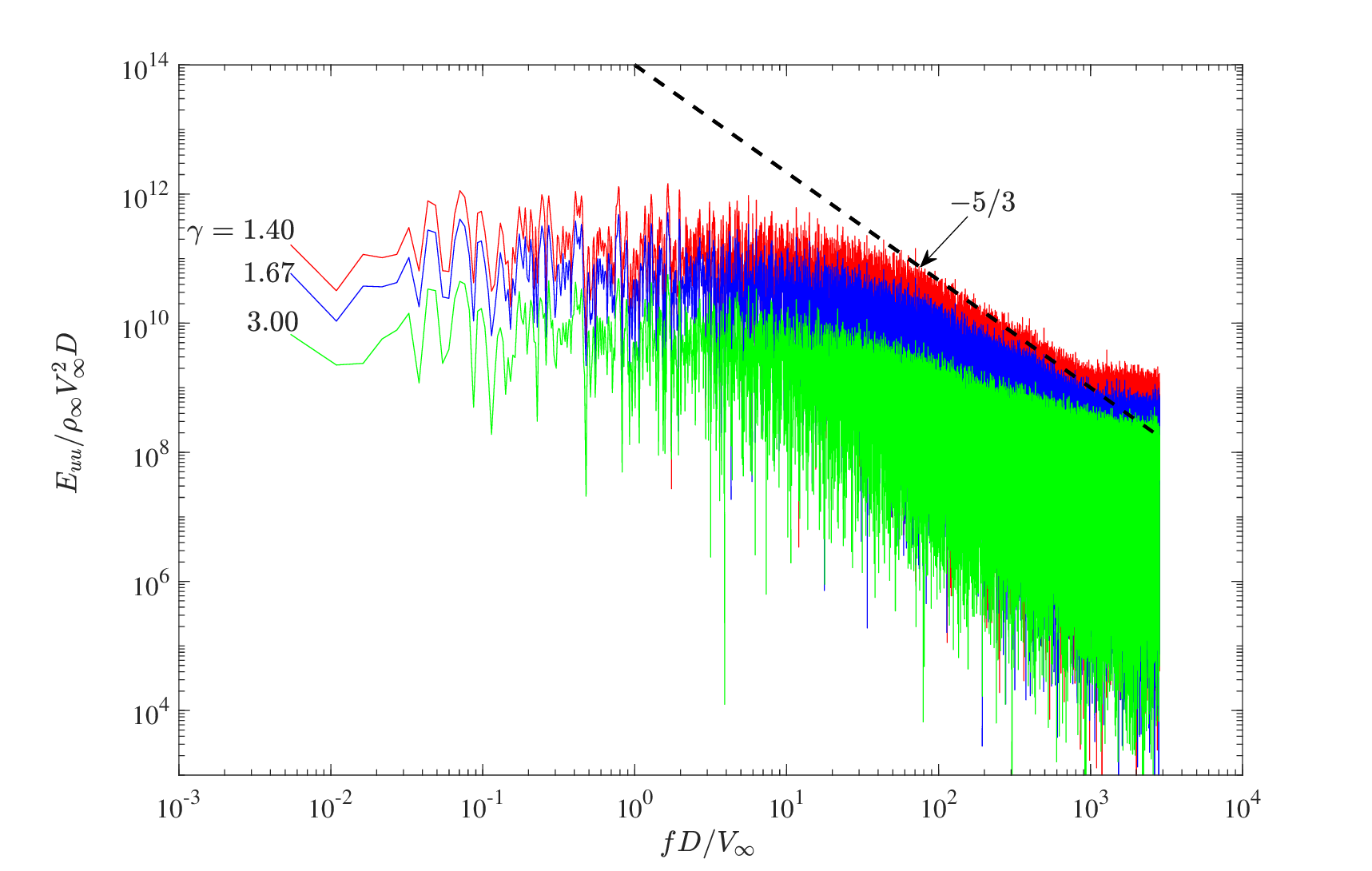}
	\caption{Energy spectra of the streamwise velocity in a cylinder wake for varying $\gamma$ at $Re = 3,900$, $M_\infty=0.5$ and $(\tilde{x}/D,\tilde{y}/D)=(0.69,0.69)$.}\label{Fig11}
\end{figure}

The effect of the change in $\gamma$ on the compressible flow at this location is also explored and presented in figure~\ref{Fig11}. The energy spectra of the streamwise velocity for three values of $\gamma =1.4$ (diatomic gas), $1.67$ (monoatomic gas), and $3.00$ (magnetized plasma with a one degree of freedom \cite{Schereretal2016,Livadiotis2015}) are presented for $Re=3,900$ and $M_\infty=0.5$. The energy of the spectra reduces as the degree of freedom reduces and $\gamma$ rises. The shape of the spectra also changes as the range of wavenumbers for which the slope is $-5/3$ shrinks in the process, gradually energizing higher frequencies. This suggests that RPT captures the exchange of energy between kinetic and internal energy modes. 

\begin{figure}
	\centering
	\includegraphics[width=0.5\textwidth,trim={0cm 0cm 0cm 0cm},clip]{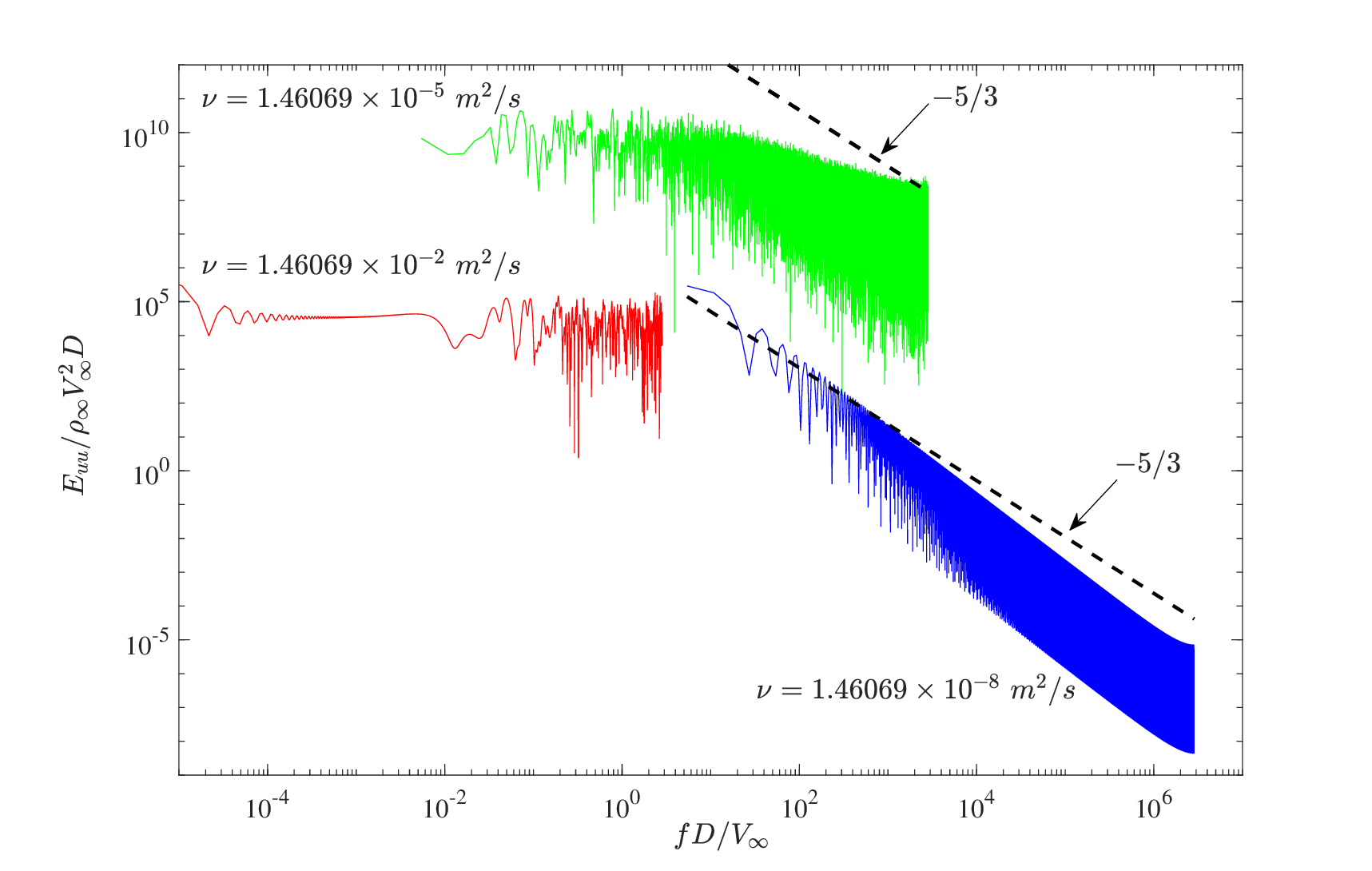}
	\caption{Energy spectra of the streamwise velocity in a cylinder wake for varying $\nu$ at $Re = 3,900$, $M_\infty=0.5$, $\gamma=3.0$ and $(\tilde{x}/D,\tilde{y}/D)=(0.69,0.69)$.}\label{Fig12}
\end{figure}

When the conditions in figure~\ref{Fig11} are maintained at $\gamma=3.0$ but the viscosity is varied, more drastic changes to the spectra are observed in figure~\ref{Fig12}. From $\nu=1.46069\times 10^{-5} m^2/s$, increasing the viscosity by three orders of magnitude depletes the energy on all scales and moves the spectrum to the low-frequency range with no frequency fall-off characteristics. However, the spectrum moves into the very high frequency dissipation range with a much steeper fall-off slope than the inertial range while at sufficiently lower energies on all scales when the viscosity is decreased by three orders of magnitude to $\nu=1.46069\times 10^{-8} m^2/s$. These arbitrarily chosen values of viscosity demonstrate how RPT wake characteristics are affected by the freestream temperature dependent properties.     

\begin{figure*}
	\centering
	\includegraphics[width=\textwidth,trim={0cm 0cm 0cm 0cm},clip]{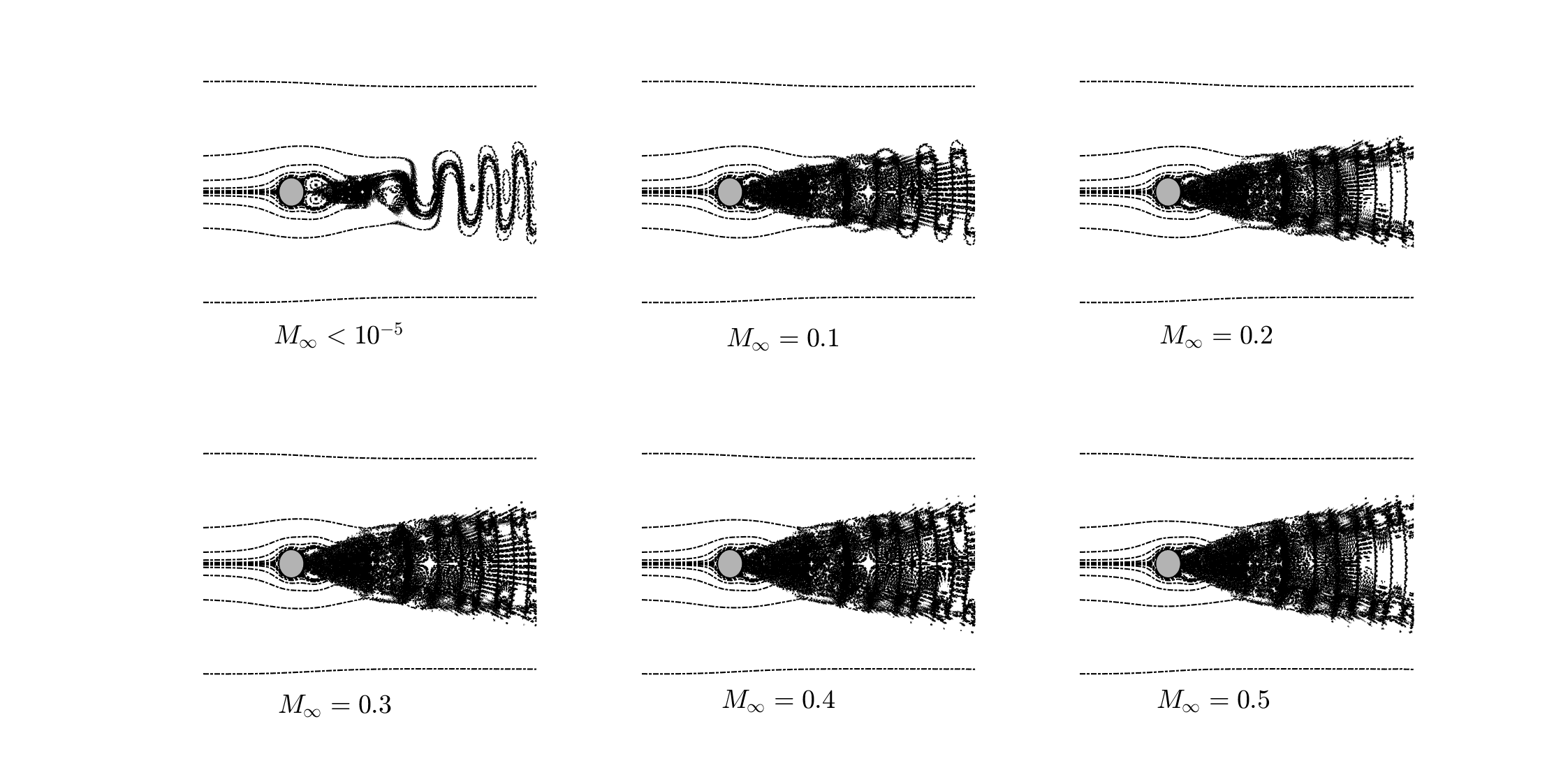}
	\caption{Streamline for $Re = 3,900$ and $t=5,000R/V_\infty$,$\nu=1.46069 \times 10^{-8} m^2/s$, and $\gamma =3.0$.}\label{Fig13}
\end{figure*}

\begin{figure*}
	\centering
	\includegraphics[width=\textwidth,trim={0cm 0cm 0cm 0cm},clip]{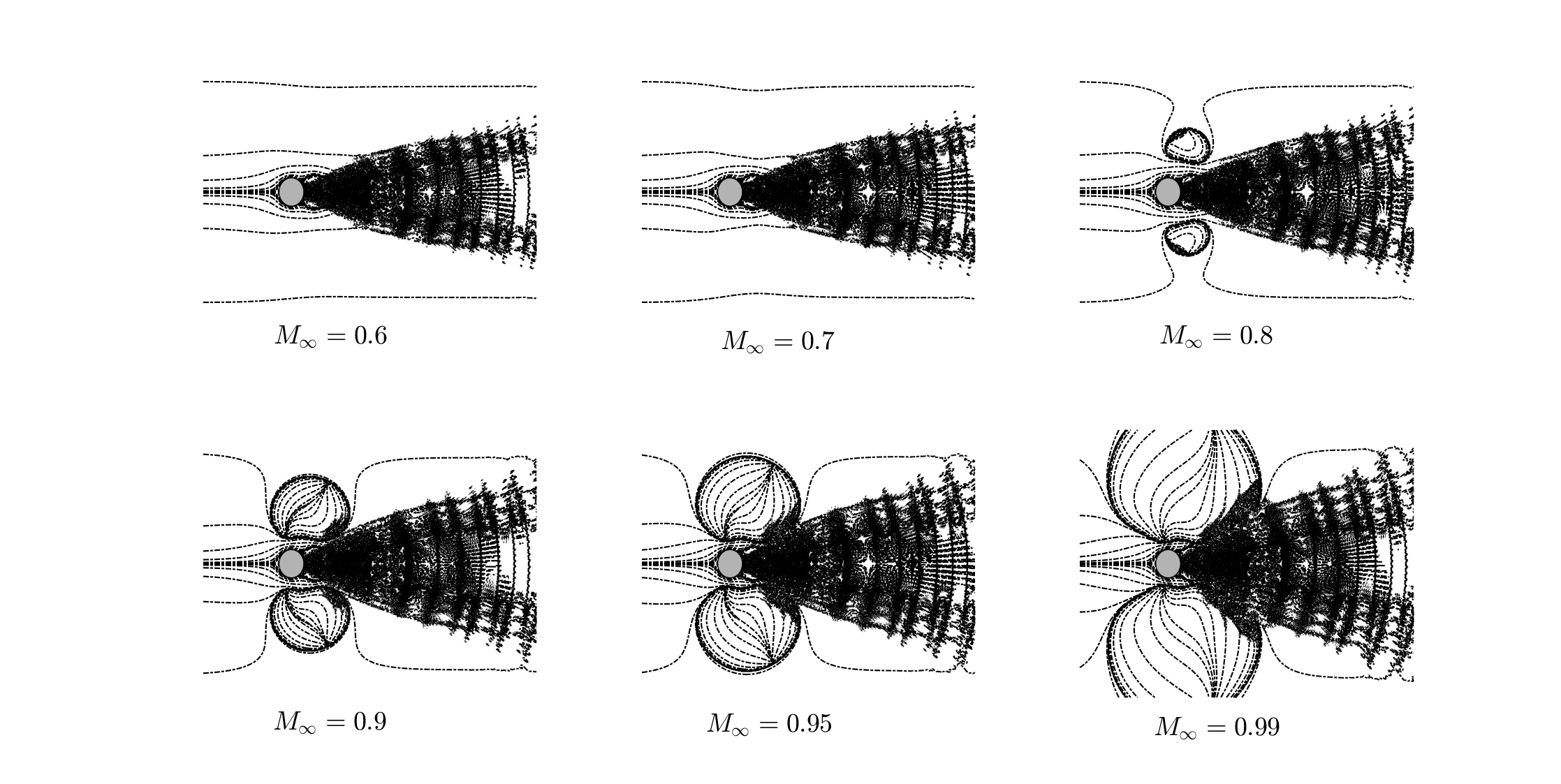}
	\caption{Streamline for $Re = 3,900$ and $t=5,000R/V_\infty$,$\nu=1.46069 \times 10^{-8} m^2/s$, and $\gamma =3.0$.}\label{Fig14}
\end{figure*}

\begin{figure}
	\centering
	\includegraphics[width=0.45\textwidth,trim={0cm 0cm 0cm 0cm},clip]{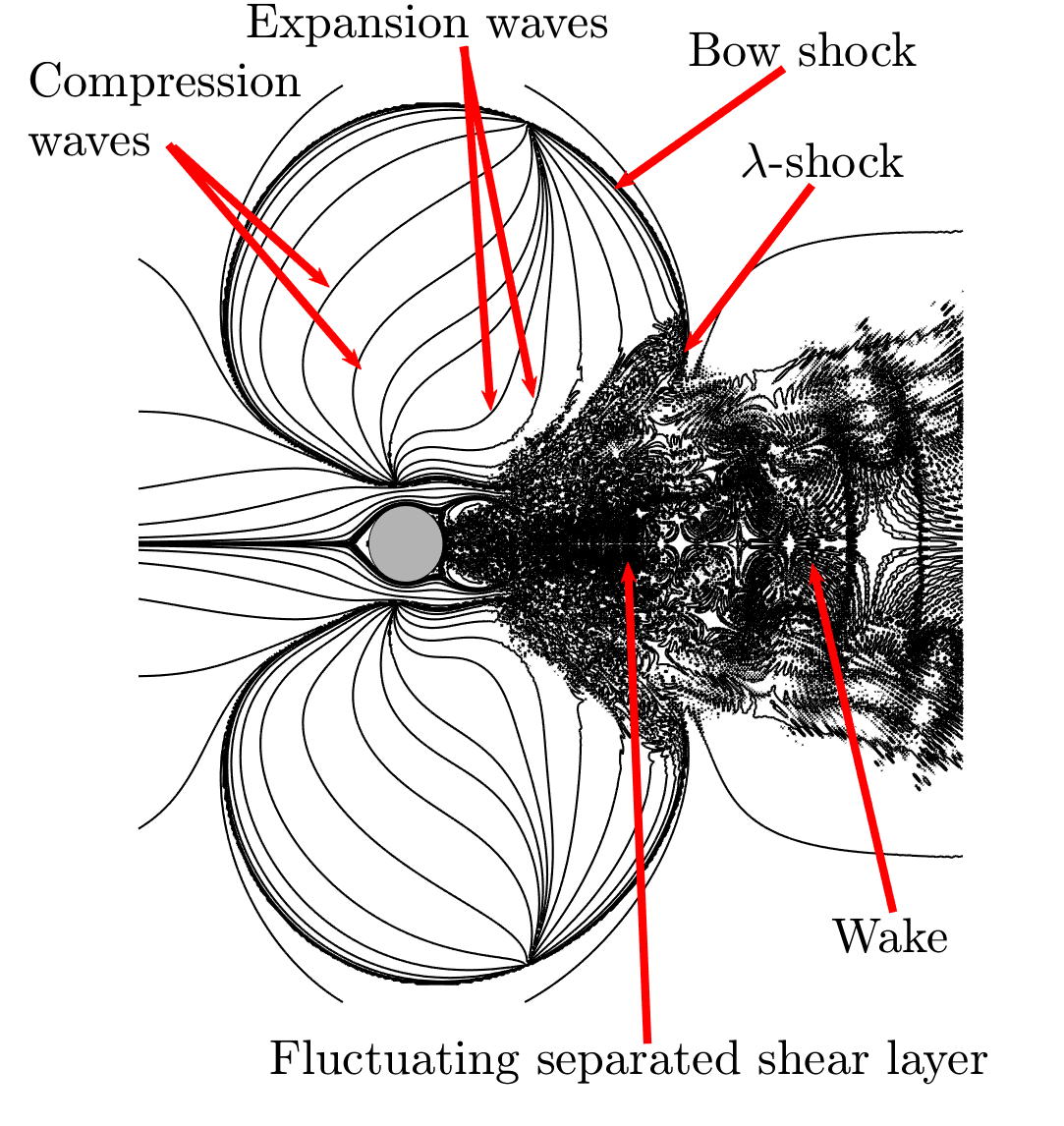}
	\caption{Streamline for $Re = 3,900$ and $t=5,000R/V_\infty$, $\gamma =3.0$,$\nu=1.46069 \times 10^{-8} m^2/s$, $M_\infty=0.99$.}\label{Fig15}
\end{figure}

The instantaneous flow fields presented in figures~\ref{Fig13} and \ref{Fig14} show the RPT streamline contours at $Re=3,900$, $t = 5,000R/V_\infty$, $\nu=1.46069 \times 10^{-8} m^2/s$, and $\gamma =3.0$ for $10^{-5}>M_\infty\le 0.99$ in the cylinder reference frame \cite{Amoloye2024}. At $M_\infty < 10^{-5}$, the large scale staggered von K\'{a}rm\'{a}n vortices are distinctively discernible  and break off into the wake at some distance from the cylinder after the recirculation zone. This is why the spectral peaks at the Strouhal frequency and its harmonics are sharp in the energy spectra plot. As the Mach number rises to $M_\infty =0.5$, compressibility effects introduce energetic smaller scale structures that almost overwhelm the periodicity of the wake, bringing the pinch-off location \cite{Nagataetal2020b} of the von K\'{a}rm\'{a}n vortices closer to the cylinder. Migration of the pinch-off location towards the cylinder with compressibility effects has also been experimentally observed at $Re=4,000$ \cite{Nagataetal2020b}. By $M_\infty =0.5$, the width of the wake begins to grow with the distance from the cylinder as the wake fans out. The curvature of the cylinder accelerates the flow over the crests so that the streamlines begin to depress towards the cylinder at $M_\infty =0.6$. As $M_\infty$ \color{black} progressively increases\color{black}, the acceleration effect becomes stronger until pockets of locally supersonic flow bounded by Mach waves appear above the recirculation region by $M_\infty =0.8$. Initially, the bases of these pockets are almost flat and inclined away from the freestream direction. This inclination further expands the local supersonic flow in the interior. The expanded flow recompresses when it hits the upward slope of the wake, forming a rear shock. As $M_\infty$ gradually approaches one ($1$), the pockets grow in extent. The gaps between their bases and the cylinder crests close. The bases become wavy with a trough-crest-trough outline. Starting from the upstream trough, the upward slope of the base's crest turns the supersonic flow into itself, forming a series of compression waves. These waves coalesce into internal bow shocks at $M_\infty =0.9$. The shocks lose their strength towards $M_\infty=0.99$. However, the downward slope of the base's crest turns the flow away from itself. Consequently, the flow is expanded through a series of expansion waves towards the rear bow shocks that are formed as the upward slope of the wake recompresses the flow. The rear bow shocks interact with the recirculation region in the wake, coming together to form a complex system of lamda ($\lambda$) shocks \cite{Li2025} by $M_\infty =0.99$ as shown in figure~\ref{Fig15}. \color{black}This figure \color{black} has striking \color{black}similarities \color{black} to a CFD result under similar conditions ($Re=3,900$ and $M_\infty=0.9$) \cite[figure~$3$, p.~$7$]{Li2025} and an experimental flow visualization at an unknown $Re$ but $M_\infty =0.98$ \cite[figure~$222$, p.~$131$]{VanDyke1982}. However, there are notable differences including the displacement of the supersonic region from the cylinder and the shortened separated shear layers compared to experiments and CFD. 

\begin{figure*}
	\centering
	\includegraphics[width=\textwidth,trim={0cm 0cm 0cm 0cm},clip]{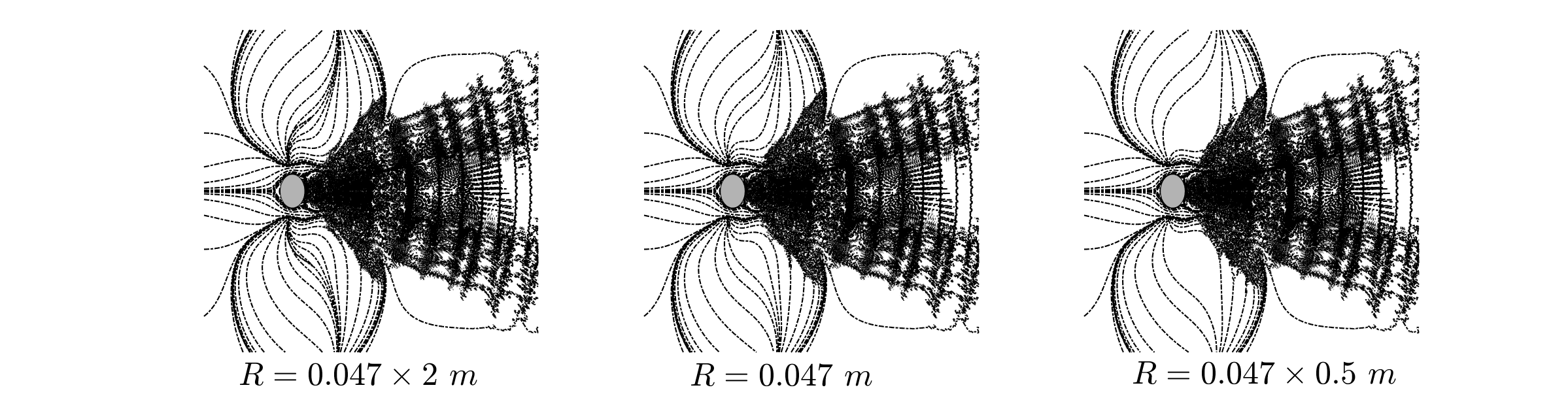}
	\caption{Streamline for $Re = 3,900$ and $t=5,000R/V_\infty$,$M_\infty =0.99$, $\nu=1.46069 \times 10^{-8} m^2/s$, $\gamma =3.0$.}\label{Fig16}
\end{figure*}

\begin{figure*}
	\centering
	\includegraphics[width=\textwidth,trim={0cm 0cm 0cm 0cm},clip]{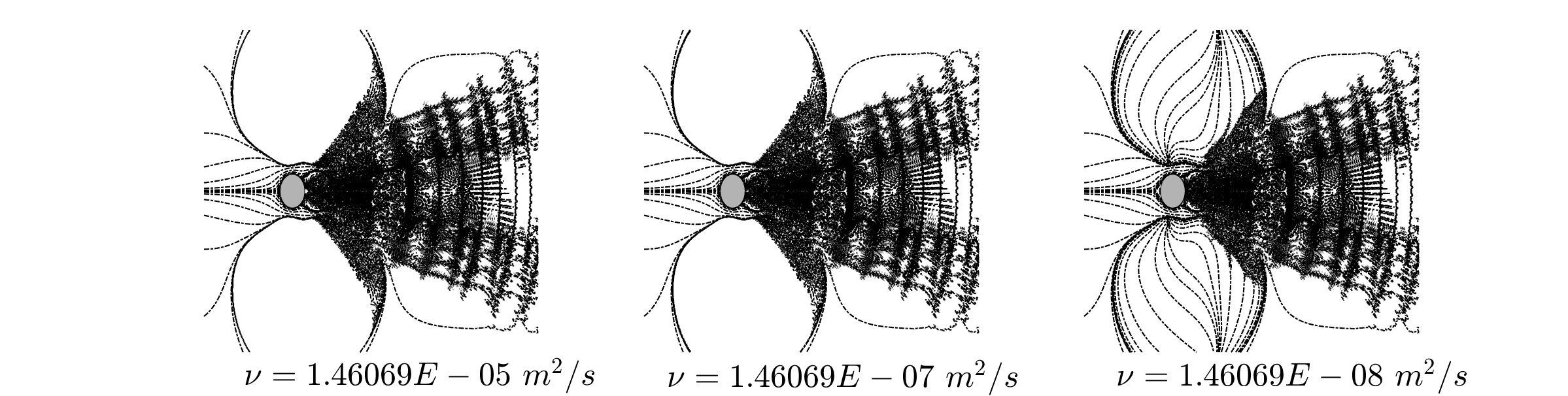}
	\caption{Streamline for $Re = 3,900$ and $t=5,000R/V_\infty$,$M_\infty =0.99$, $\gamma =3.0$.}\label{Fig17}
\end{figure*}

\begin{figure*}
	\centering
	\includegraphics[width=\textwidth,trim={0cm 0cm 0cm 0cm},clip]{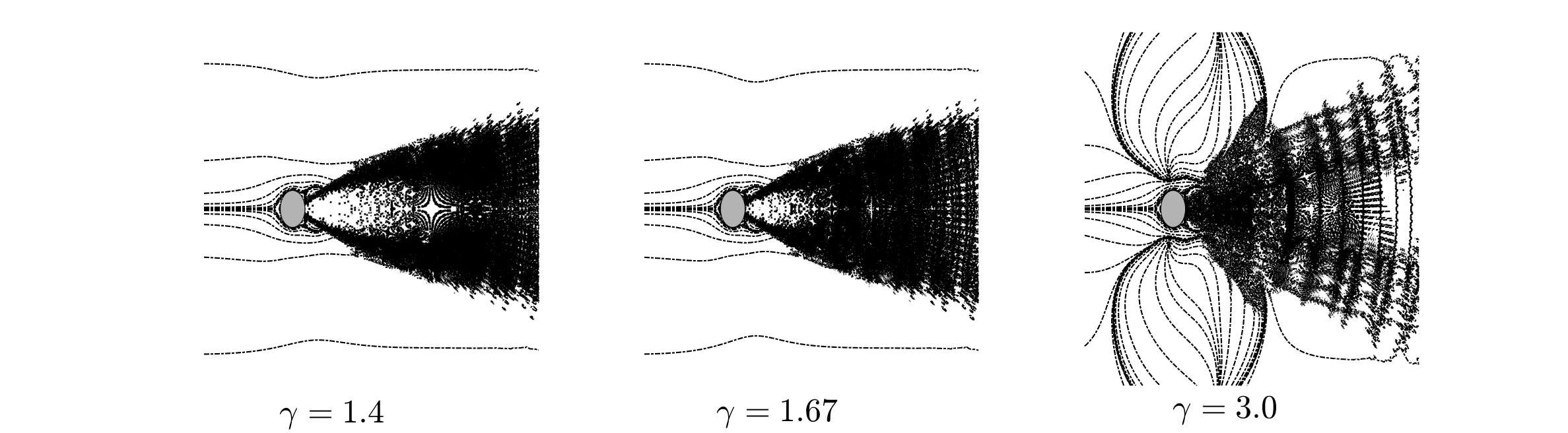}
	\caption{Streamline for $Re = 3,900$ and $t=5,000R/V_\infty$,$\nu=1.46069 \times 10^{-8} m^2/s$, $M_\infty=0.99$.}\label{Fig18}
\end{figure*}

\begin{figure*}
	\centering
	\includegraphics[width=\textwidth,trim={0cm 0cm 0cm 0cm},clip]{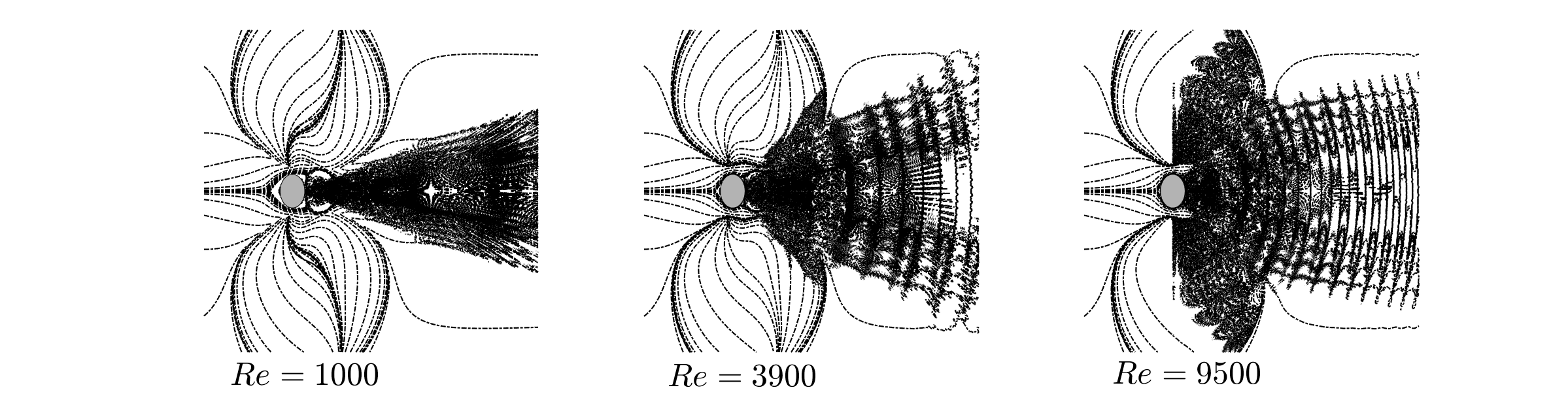}
	\caption{Streamline for varying $Re$ and $t=5,000R/V_\infty$,$M_\infty =0.99$, $\nu=1.46069 \times 10^{-8} m^2/s$, $\gamma =3.0$.}\label{Fig19}
\end{figure*}

\begin{figure*}
	\centering
	\includegraphics[width=\textwidth,trim={0cm 0cm 0cm 0cm},clip]{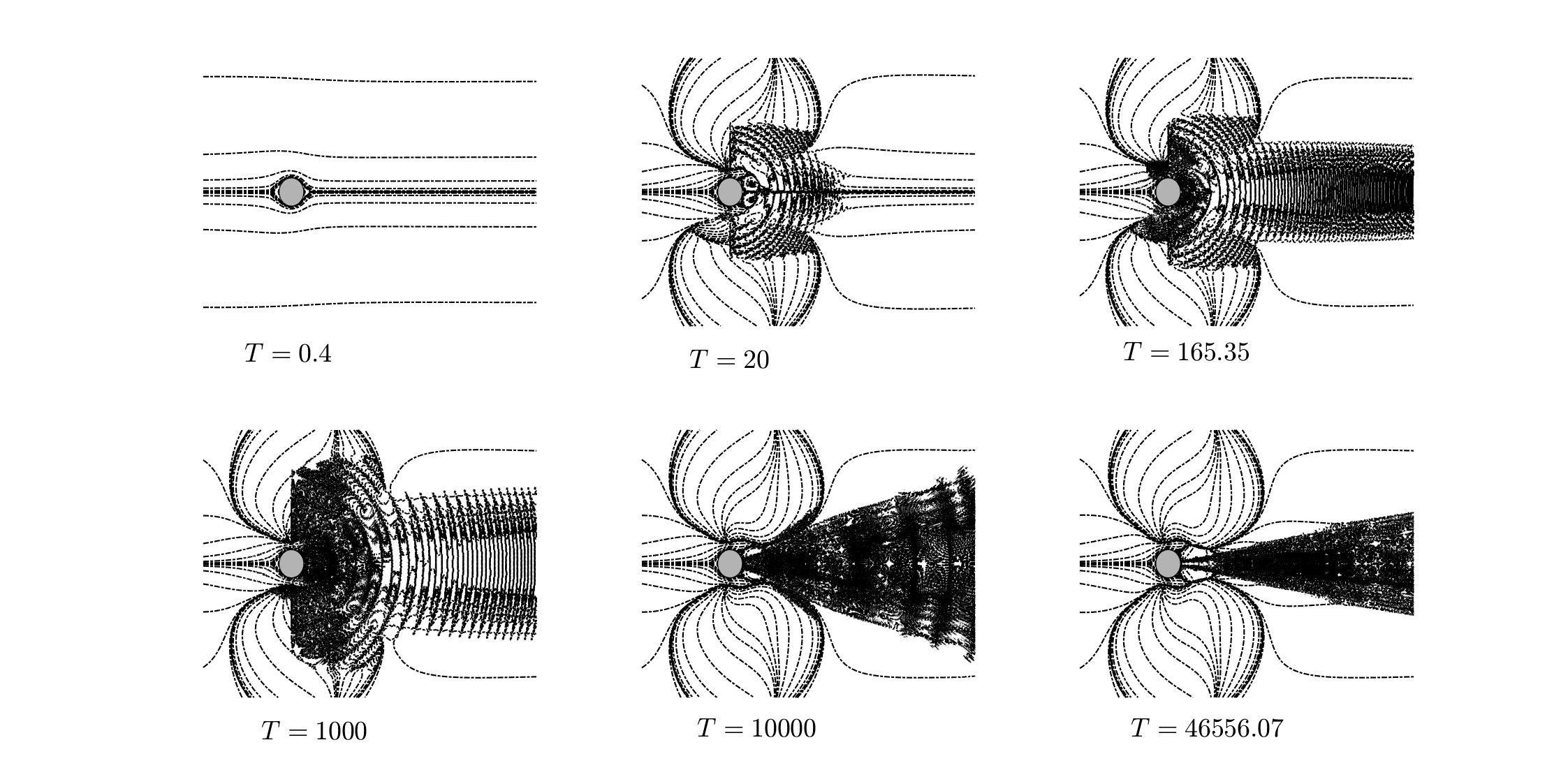}
	\caption{Streamline for $Re = 3,900$ and $M_\infty =0.99$, $\nu=1.46069 \times 10^{-8} m^2/s$, $\gamma =3.0$. $T=V_\infty t/R$}\label{Fig20}
\end{figure*}

The physics of the supersonic pocket can be modified by varying some parameters. The strength of its internal shock is proportional to the radius of the cylinder as illustrated in figure~\ref{Fig16}. The internal shock can be smoothed out by increasing the viscosity as shown in figure~\ref{Fig17}. Figure~\ref{Fig18} shows that the pocket can be delayed with a decrease in $\gamma$. Reducing $Re$ to $1,000$ strengthens the internal shock and positions the base of the pocket on the cylinder. However, increasing it to $9,500$ strengthens the internal shock to a normal shock wave without dislocating the base of the pocket downward to the cylinder. In addition, this introduces a much more complex interaction downstream of the internal shock, as depicted in figure~\ref{Fig19}. Some of these changes, including the elongation of the separated shear layers, can also be achieved by varying just the non-dimensional time as figure~\ref{Fig20} presents, featuring transient disruptions of flow symmetry in the unsteady flow development. 


\begin{figure*}
	\centering
	\subfloat[]{\includegraphics[width=0.33\textwidth,trim={4cm 4cm 4cm 4cm},clip]{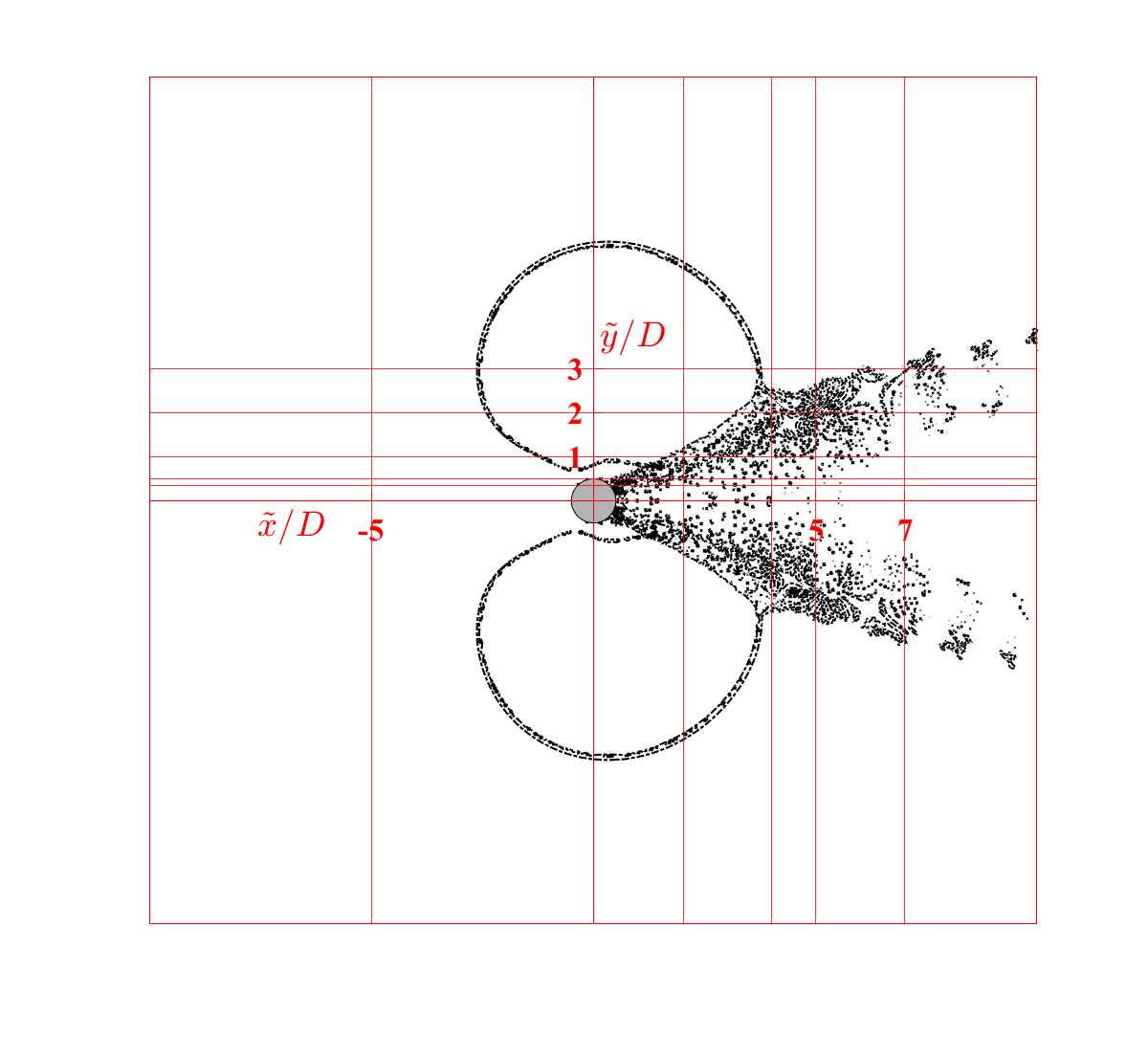}\label{Fig21a}}
    \subfloat[]{\includegraphics[width=0.33\textwidth,trim={4cm 4cm 4cm 4cm},clip]{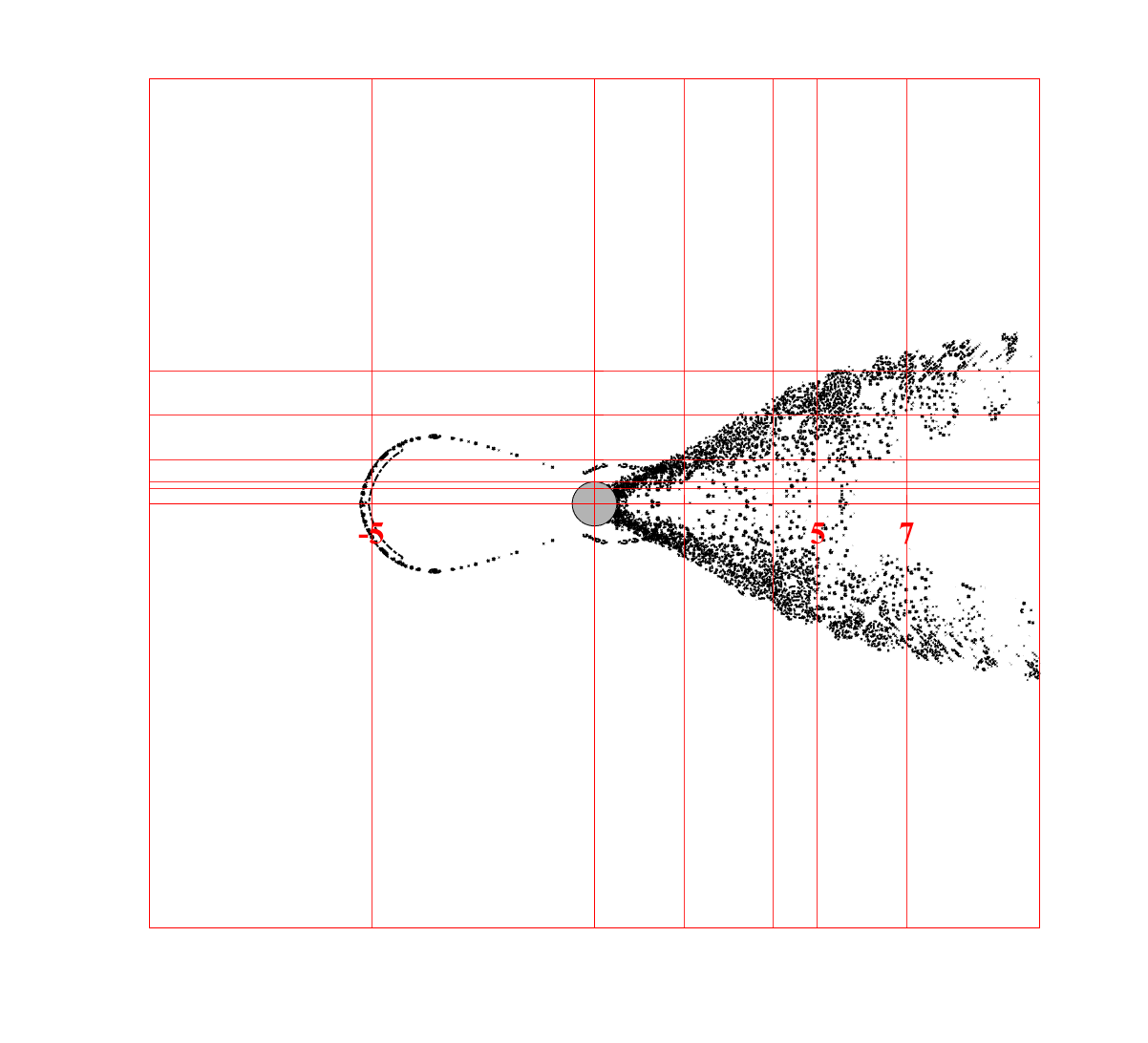}\label{Fig21b}}
    \subfloat[]{\includegraphics[width=0.33\textwidth,trim={4cm 4cm 4cm 4cm},clip]{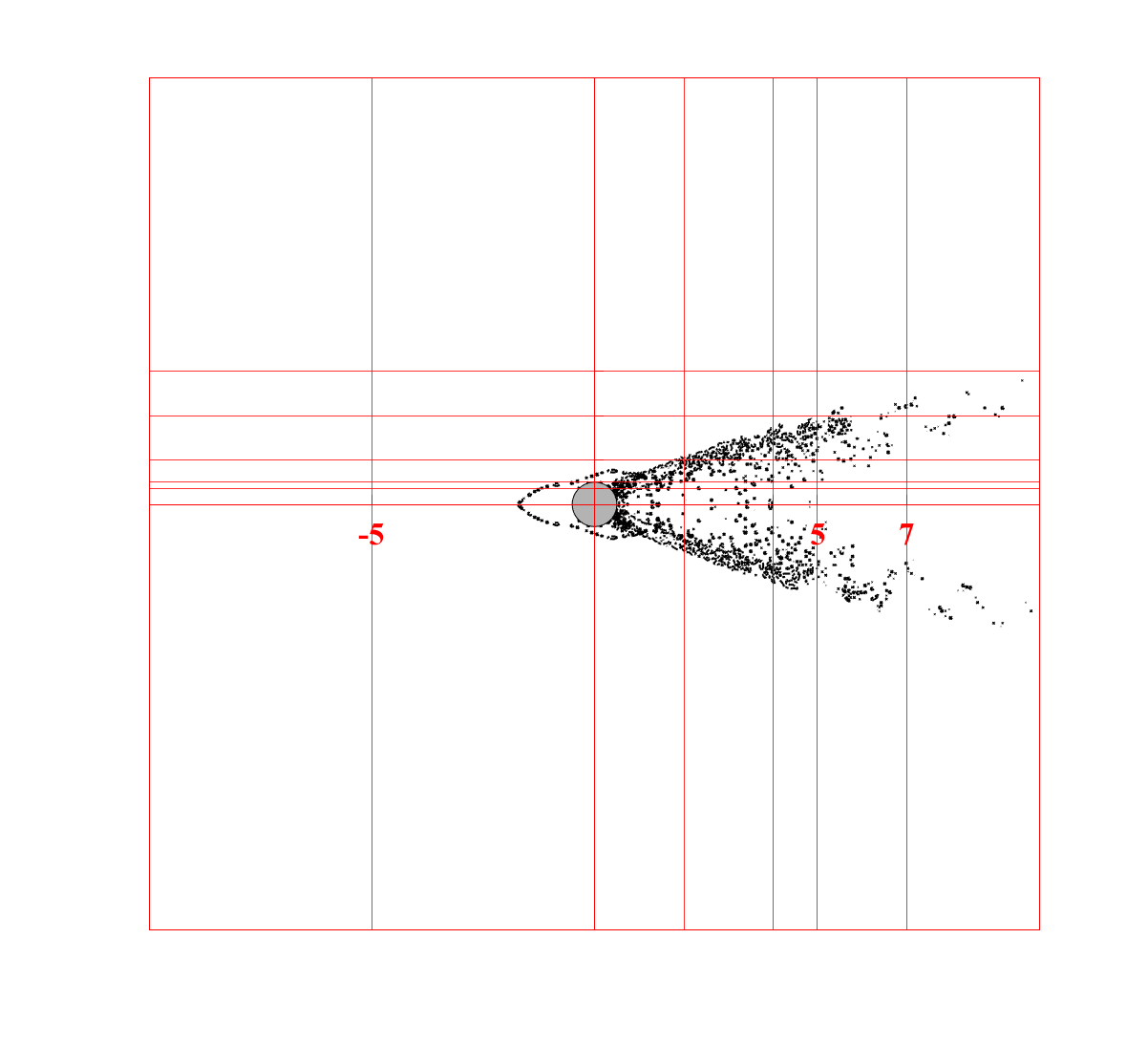}\label{Fig21c}}
	\caption{Density gradient field for $Re = 3,900$, $t=5,000R/V_\infty$, $\gamma =3.0$,$\nu=1.46069 \times 10^{-8} m^2/s$, and (a)$M_\infty=0.99$ (b)$M_\infty=1.01$ (c)$M_\infty=1.35$.}\label{Fig21}
\end{figure*}

Figure~\ref{Fig21} employs a numerical schlieren technique \color{black}(using the function $S(x,y)$\cite{HadjadjandKudryavtsev2005,Lysenkoetal2012} defined in \color{black} Eq.~\ref{Schlieren}) to visualize the instantaneous density gradient field of RPT at $t= 5,000R/V_\infty$, $\nu=1.46069 \times 10^{-8} m^2/s$, $\gamma =3.0$ and $Re=3,900$ for $M_\infty = 0.99,~1.01$ and $1.35$. The density field is evaluated from Eq.~\ref{Rhobeta}. Equation~\ref{Schlieren} contains two adjustable parameters, $\beta$ and K, whose values were set at $0.95$ and $15$, respectively. $\beta$ controls the zero gradient and K regulates the amplification of small gradients \cite{Lysenkoetal2012}. $|\nabla\rho|_{max}$ is the maximum absolute density gradient in the entire flow field. The contour levels are spaced $1/140$ between zero ($0$) and one ($1$). 
\begin{equation}
	\label{Schlieren}
	S(x,y)=\beta exp\left(-\dfrac{K|\nabla\rho|}{|\nabla\rho|_{max}}\right)
\end{equation}   
As the free stream reaches a slightly supersonic $M_\infty=1.01$, the $\lambda$-shock system in figure~\ref{Fig21a} disappears and is replaced by an upstream bow shock at a distance from the cylinder, as shown in figure~\ref{Fig21b}. The shock detachment distance closes and the bow shock curvature intensifies when $M_\infty =1.35$ in figure~\ref{Fig21c}. These observations are in good qualitative agreement with CFD and experimental investigations of the transonic flow field of a cylinder. 

\begin{figure}
	\centering
	\includegraphics[width=0.5\textwidth,trim={0cm 0cm 0cm 0cm},clip]{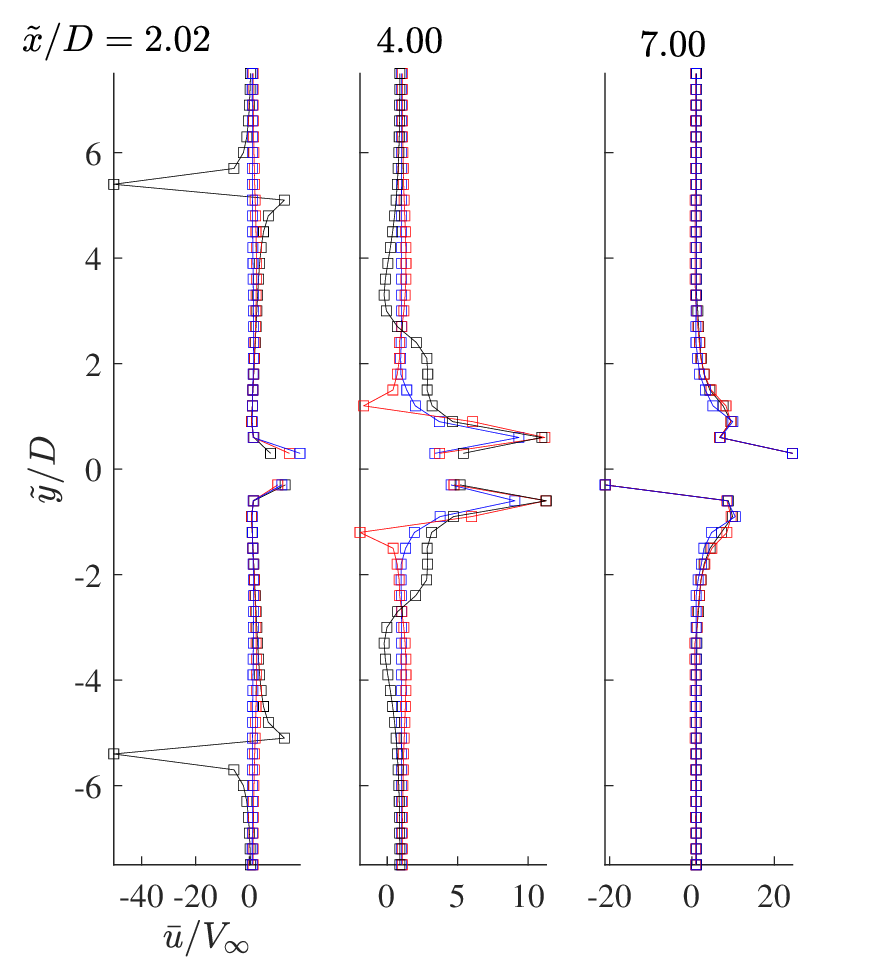}
	\caption{Mean streamwise velocity profile at three downstream locations for $Re = 3,900$, $\nu=1.46069 \times 10^{-8} m^2/s$, and $\gamma =3.0$. $M_\infty=0.99~(\color{black}-\color{black})$$;~M_\infty=1.01~(\color{black}-\color{black})$$;~M_\infty=1.35~(\color{blue}-\color{black})$}\label{Fig22}
\end{figure}

\begin{figure}
	\centering
	\includegraphics[width=0.5\textwidth,trim={0cm 0cm 0cm 0cm},clip]{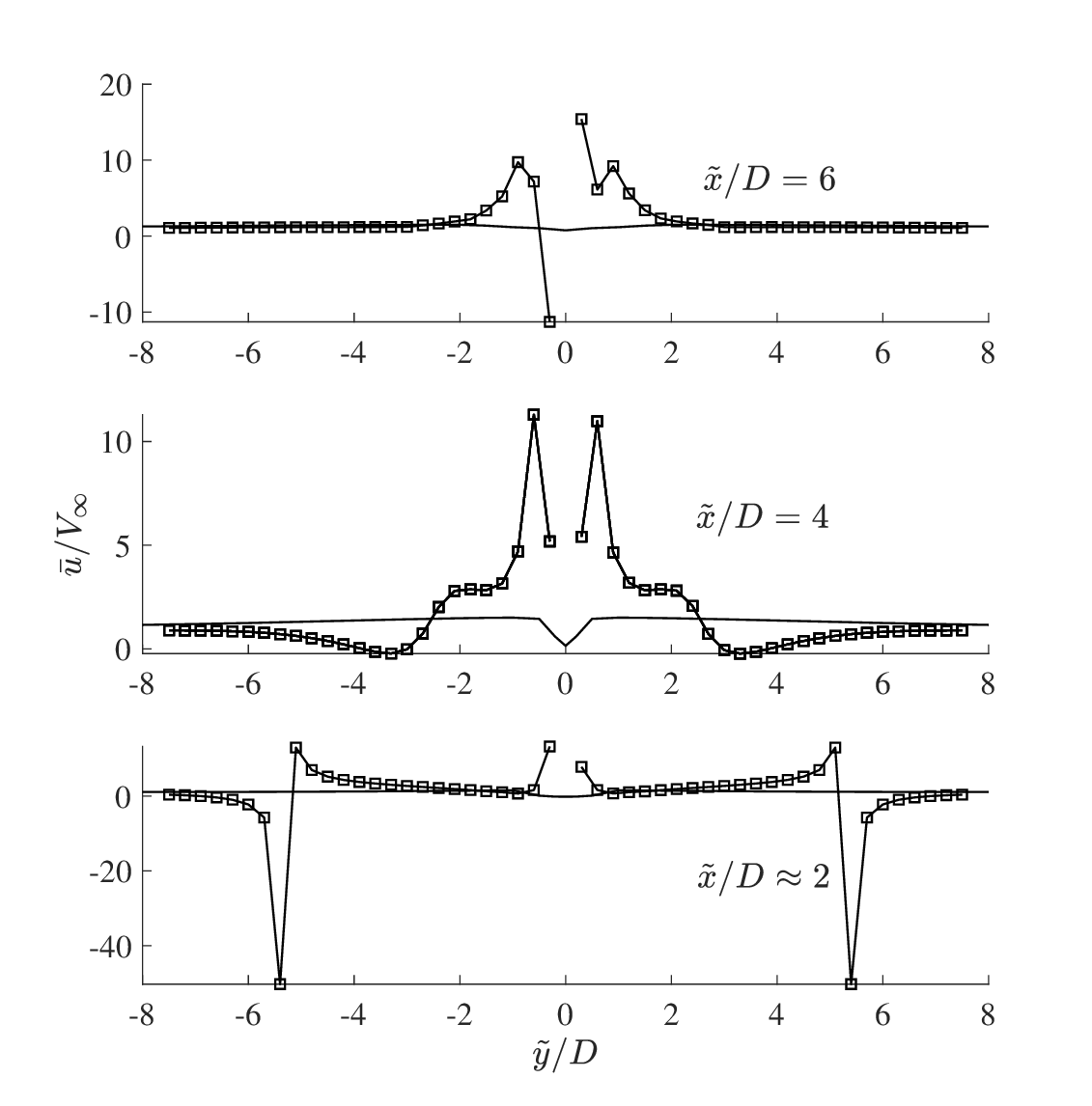}
	\caption{Comparison of RPT mean streamwise velocity profile at three downstream locations for $Re = 3,900$, $\nu=1.46069 \times 10^{-8} m^2/s$, and $\gamma =3.0$ $(M_\infty=0.99:~\color{black}-\square-\color{black})$ with digitized CFD data\cite{Li2025}$(M_\infty=0.9:~\color{black}-\color{black})$. Shuai Li, Phys. Rev. Fluids $8$, $034603$,$2025$; licensed under a Creative Commons Attribution $4.0$ International (CC BY $4.0$) license.}\label{Fig23}
\end{figure}

\begin{figure*}
	\centering
     \subfloat[$\tilde{y}/D=0.35$]{\includegraphics[width=0.5\textwidth,trim={0cm 0cm 0cm 0cm},clip]{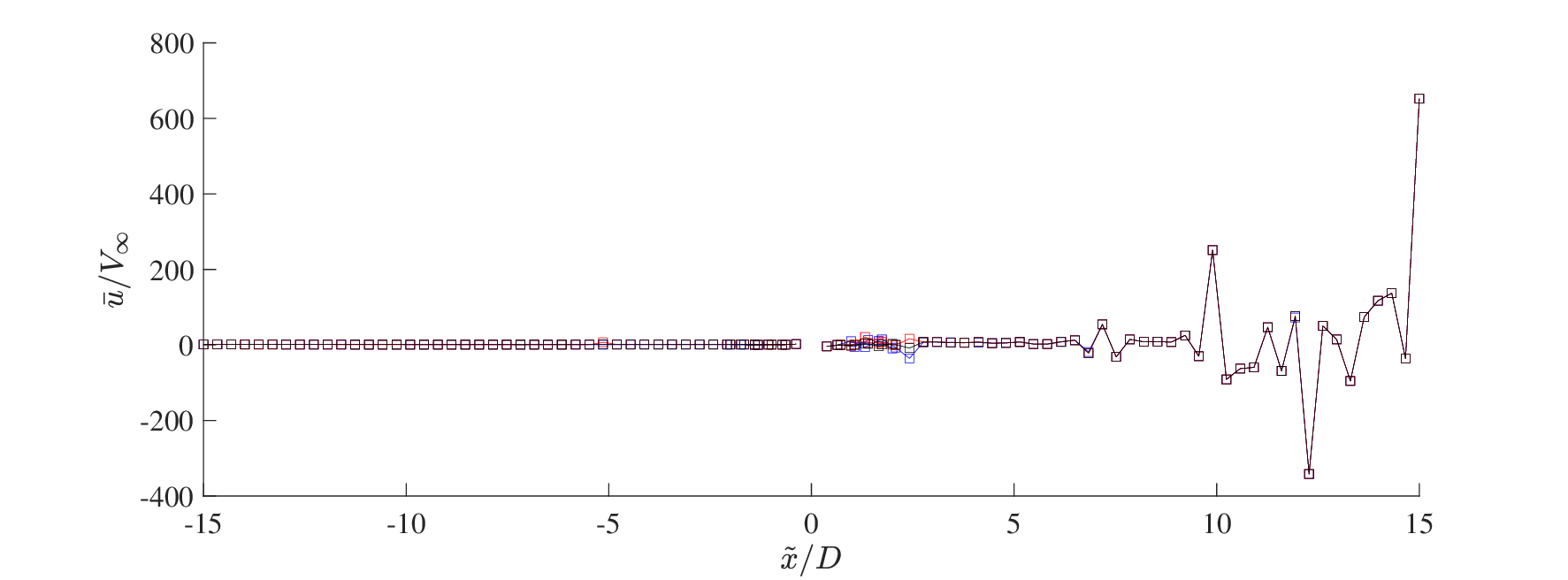}\label{Fig24a}}
     \subfloat[$\tilde{y}/D=0.50$]{\includegraphics[width=0.5\textwidth,trim={0cm 0cm 0cm 0cm},clip]{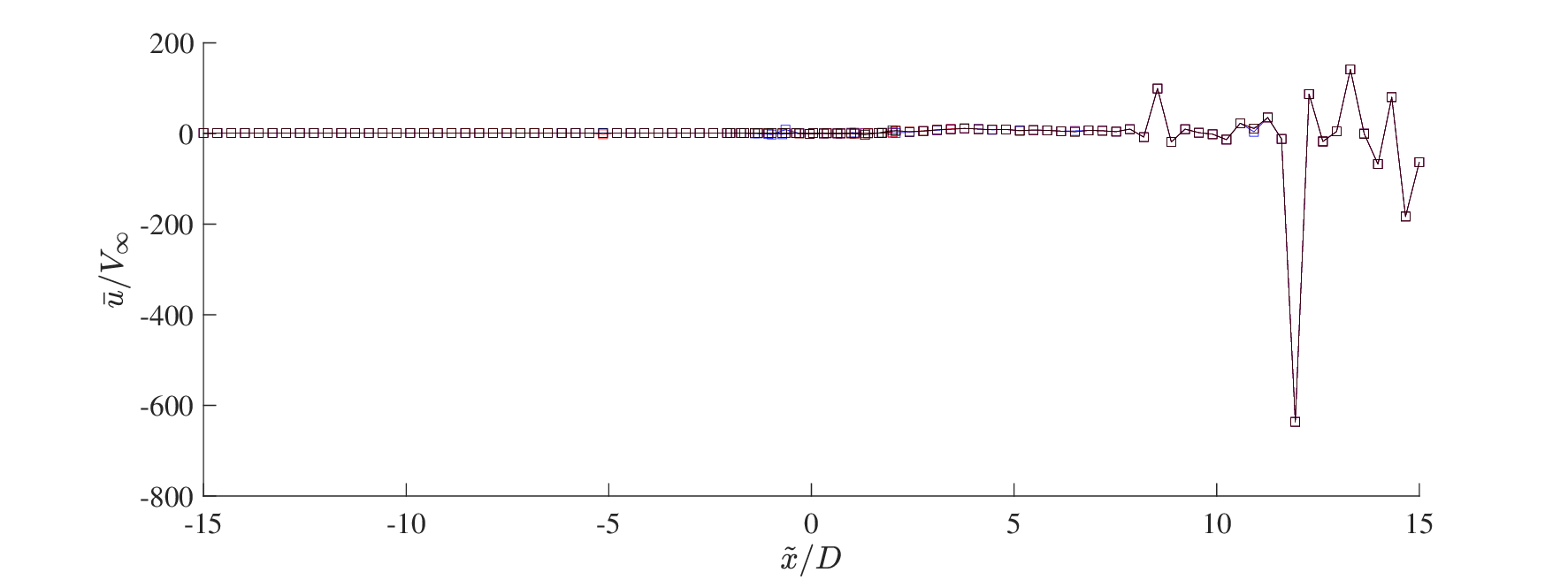}\label{Fig24b}}
	\caption{Mean streamwise velocity in a cylinder flow at vertical locations for $Re = 3,900$, $\gamma =3.0$,$\nu=1.46069 \times 10^{-5} m^2/s$, and varying $M_\infty$. $M_\infty=0.99~(\color{black}-\color{black})$$,~M_\infty=1.01~(\color{black}-\color{black})$$,~M_\infty=1.35~(\color{blue}-\color{black})$.}\label{Fig24}
\end{figure*}

\begin{figure}
	\centering
    \subfloat[$\tilde{y}/D=3.00$]{\includegraphics[width=0.5\textwidth,trim={0cm 0cm 0cm 0cm},clip]{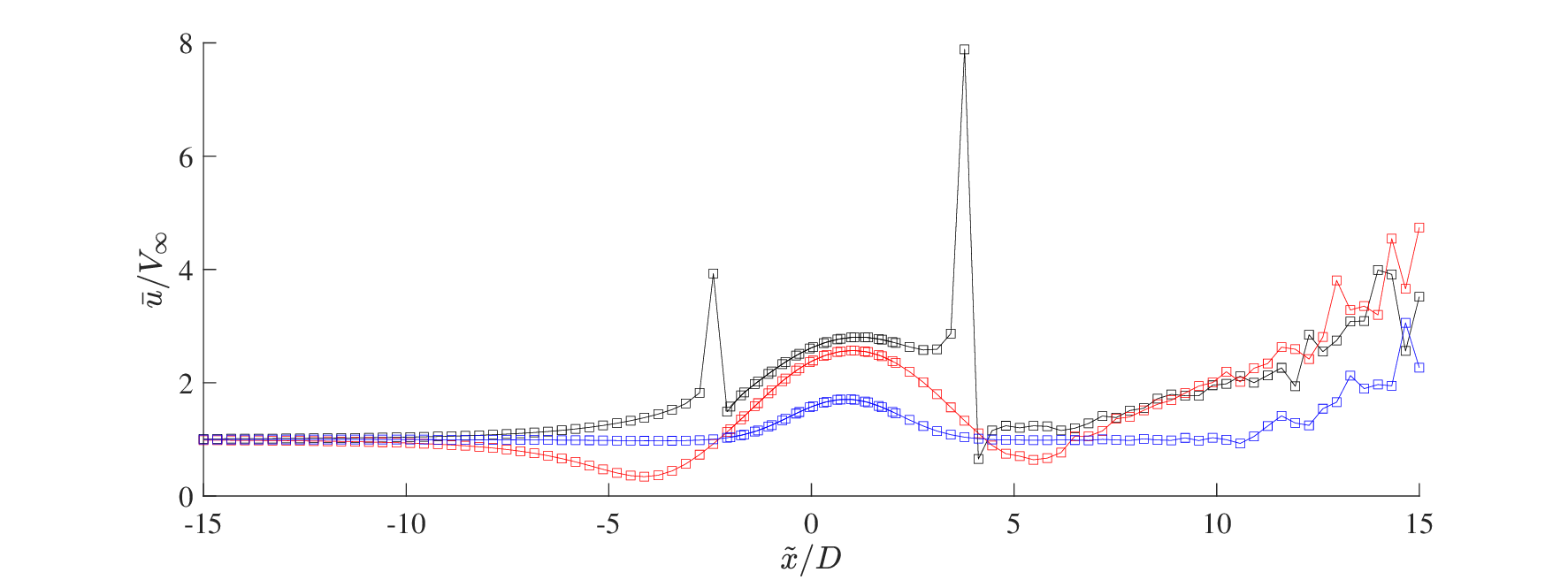}\label{Fig25a}}\\
    \subfloat[$\tilde{y}/D=2.00$]{\includegraphics[width=0.5\textwidth,trim={0cm 0cm 0cm 0cm},clip]{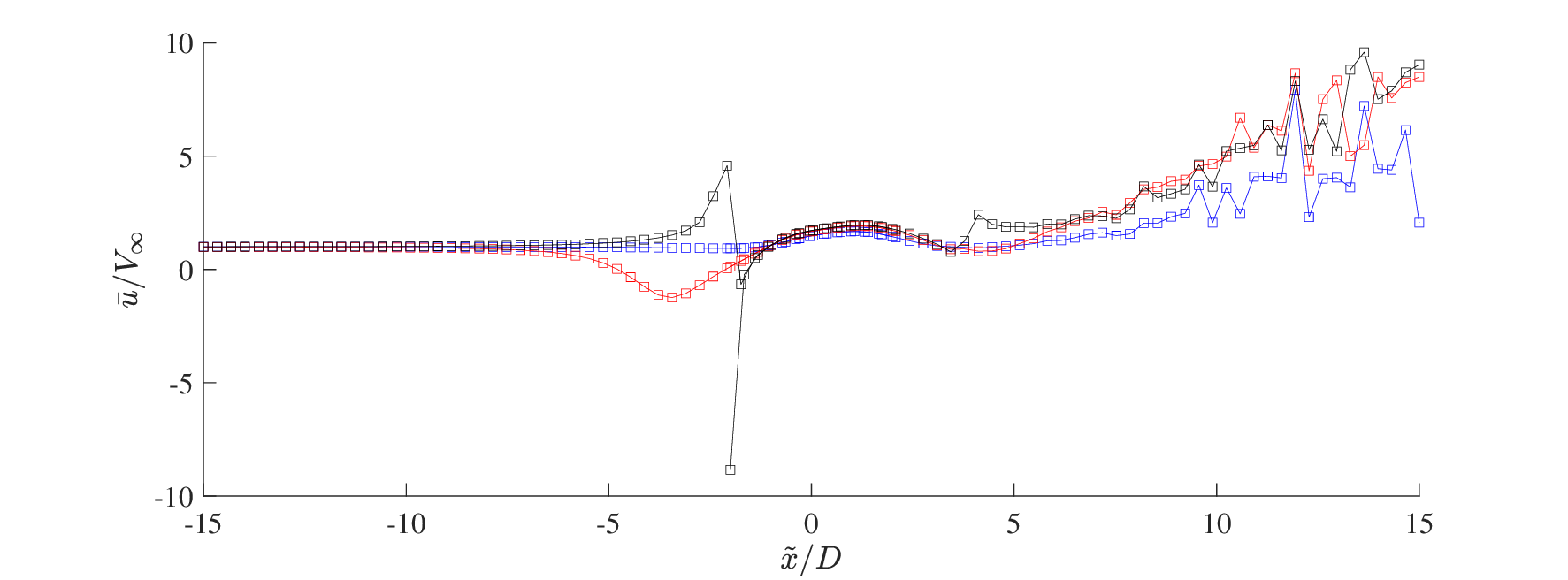}\label{Fig25b}}\\
     \subfloat[$\tilde{y}/D=1.00$]{\includegraphics[width=0.5\textwidth,trim={0cm 0cm 0cm 0cm},clip]{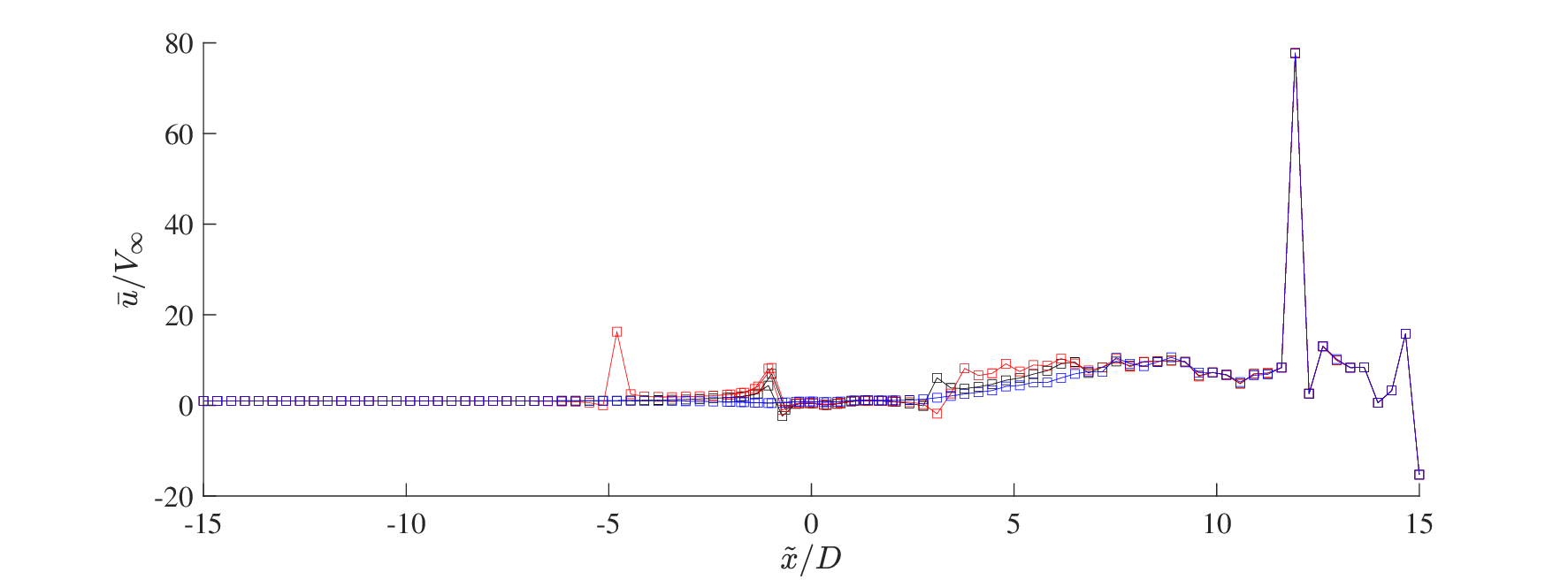}\label{Fig25c}}
	\caption{Mean streamwise velocity in a cylinder flow at vertical locations for $Re = 3,900$, $\gamma =3.0$,$\nu=1.46069 \times 10^{-5} m^2/s$, and varying $M_\infty$. $M_\infty=0.99~(\color{black}-\color{black})$$,~M_\infty=1.01~(\color{black}-\color{black})$$,~M_\infty=1.35~(\color{blue}-\color{black})$.}\label{Fig25}
\end{figure}

To aid in a quantitative assessment, the mean streamwise velocity profiles are evaluated along some of the overlaying grid lines in figure~\ref{Fig21} and presented in figures~\ref{Fig22} to \ref{Fig25} for the three $M_\infty$. These mean values are the average quantities numerically integrated over the same time period as in figure~\ref{Fig4}. The wake streamwise velocity profiles at three vertical downstream locations are shown in figure~\ref{Fig22}. Outlier points still exist on the rear axis and are omitted. At $M_\infty=0.99$, station $\tilde{x}/D=2.02$ is upstream of the $\lambda$-shock system but cuts through the rear bow shocks. Consequently, there are discontinuities in the velocity profile at $\tilde{y}/D\sim \pm 5$ corresponding to the shock locations where the magnitude of the streamwise velocity is greater than $40V_\infty$. Compared with Lin's\cite[figure~$10a$, p.$13$]{Li2025} CFD result at a similar $M_\infty=0.9$, the rear bow shocks do not feature in the profile at $\tilde{x}/D=2.00$ \color{black}(in figure~\ref{Fig23}) \color{black}but further downstream at $\tilde{x}/D=8.00$ (due to the elongated free shear layers) where the general outline of the profile (between the shocks, including the protrusion in the separated shear layer region) and the lateral locations of the discontinuity of the profile at $\tilde{y}/D\sim \pm 5$ are similar to RPT. Whereas station $\tilde{x}/D=4.00$ is directly downstream of the rear bow shock in the present RPT result, it is an upstream station to the entire shock system in the CFD result. \color{black}This explains the difference in the RPT and CFD profiles at this station in figure~\ref{Fig23}. \color{black}Both stations $\tilde{x}/D=7.00$ for the RPT and $\tilde{x}/D=10.00$ for the CFD are sufficiently downstream of the respective shock-systems and have very similar profiles. The RPT velocity profiles at $M_\infty=1.01$ and $1.35$ are similar for all three downstream locations. They are also similar to the profile at $\tilde{x}/D=7.00$ for $M_\infty=0.99$. 

The streamwise velocity profiles along horizontal locations in the range $0.35\le \tilde{y}/D\le 3$ are shown for the three $M_\infty$ in figures~\ref{Fig24} and \ref{Fig25}. These locations were chosen for comparison with Li's \cite[figure~$9$, p.$13$]{Li2025} CFD results at $M_\infty=0.9$. Although the CFD results do not include the \color{black}windward \color{black}freestream, the present profiles cover the upstream far field and the wake. This is to facilitate a comparison of the complete transonic flow field phenomena as $M_\infty$ is varied. The upstream centerline also has a singular behavior similar to the cylinder rear axis and is therefore left out. In the freestream, the profiles coincide for the three $M_\infty$. As the flow approaches the cylinder, it encounters a weak shock wave at $M_\infty =1.01$. This is identifiable as discontinuities in the profiles around $\tilde{x}/D=-5$ for $0.35\le\tilde{y}/D\le1.00$. This discontinuity is more visible at $\tilde{y}/D=1.00$, while it appears as a kink in the profiles at  $\tilde{y}/D=0.35$ and $0.5$ due to the relatively large excursions in the wake velocity variations in the respective figures. The predicted shock detachment distance, $\delta/D$, at $M_\infty=1.01$ is $\sim 4.5$, which is consistent with published experimental and computational trends \cite[figure~$7$, p.$7$]{SinclairandCui2017}. There is no upstream shock at this location for $M_\infty=0.99$ and $M_\infty=1.35$. Moving closer to the cylinder ($-1.74\ge\tilde{x}/D\ge -4.80$), the profiles remain overlapped for $0.35\le\tilde{y}/D\le 0.5$ because the incident weak shock when $M_\infty=1.01$ reduces the flow velocity by a negligible amount and the curvature of the cylinder has a minimal effect. However, the situations are different for $1.00\le\tilde{y}/D\le 3.00$. At $\tilde{y}/D= 1.00$, the post shock $M_\infty=1.01$ flow and the $M_\infty =0.99$ freestream flow are both subsonic and accelerate to higher velocities by the cylinder curvature. The post shock $M_\infty=1.01$ flow hits the sonic line, and the $M_\infty=0.99$ freestream encounters the leading shock of the supersonic pocket around $\tilde{x}/D\approx-0.98$.  At $\tilde{y}/D= 2.00$ and $3.00$, the $M_\infty=0.99$ freestream continues to expand, crossing the leading shock of the supersonic pocket at $\tilde{x}/D\approx -2.08$ and $\tilde{x}/D\approx -2.42$, respectively. At $\tilde{y}/D= 2.00$ and $3.00$, the curvature effect of the cylinder gradually turns the $M_\infty =1.01$ freestream into itself, compressing and slowing it down to subsonic conditions. This compression happens without a shock. When the subsonic condition is reached, the flow velocity begins to increase as the flow expands around the cylinder.

The $M_\infty =1.35$ freestream encounters a shock around $\tilde{x}/D\approx -1.7$ and $\delta/D\sim 1.2$. This value of $\delta/D$ is lower than the computational and experimental results for a gas with $\gamma=1.4$ \cite{SinclairandCui2017}. However, $\delta/D$ has a proportional relationship with $\gamma$ \cite{Yildizetal2023}. So,it is expected that the present results with $\gamma=3.0$ have a higher value of $\delta/D$. \color{black}This is due to the breakdown of the one-way coupling of the density field on an incompressible base flow, offering insufficient feedback to the modeled compressible flow field at this supersonic regime. \color{black} Between $-0.5\ge\tilde{x}/D\ge -1.7$, the velocity profiles again coincide for $0.35\le\tilde{y}/D\le 0.5$ because the flows are in similar locally subsonic conditions. The $M_\infty=0.99$ profiles dip for $1.00\le\tilde{y}/D\le 3.00$ due to the re-compression of the local flow by the upward slope of the base of the supersonic pocket. However, the $M_\infty=1.01$ and $1.35$ profiles are increasing under the effect of the cylinder curvature. For $\tilde{x}/D> -0.5$, the three profiles overlap at almost every point except in the very near wake for $\tilde{y}/D= 0.35$ and $0.50$. Between $1.00\le\tilde{y}/D\le 3.00$, flows expand at different rates. The $M_\infty=1.35$ recompresses in the wake without any shocks but compression waves. Re-compression of the $M_\infty=1.01$ flow occurs through a trailing shock at $\tilde{y}/D=1.00$, but gradually through compression waves at higher locations. The $M_\infty=0.99$ flow recompresses in the wake through the $\lambda$-shock system for $1.00\le\tilde{y}/D\le 3.00$. At a similar $M_\infty=0.9$ in Li's\cite[figure $9$, p.$13$]{Li2025} CFD results, the flow also recompresses through shocks for $1.00\le\tilde{y}/D\le 3.00$. However, the RPT flow does not recompress to the freestream in contrast to Li's result. This is likely because the time period is not sufficient to average out the large velocity fluctuations in the wake eddies.

\begin{figure*}
	\centering
	\subfloat[]{\includegraphics[width=0.33\textwidth,trim={0cm 0cm 0cm 0cm},clip]{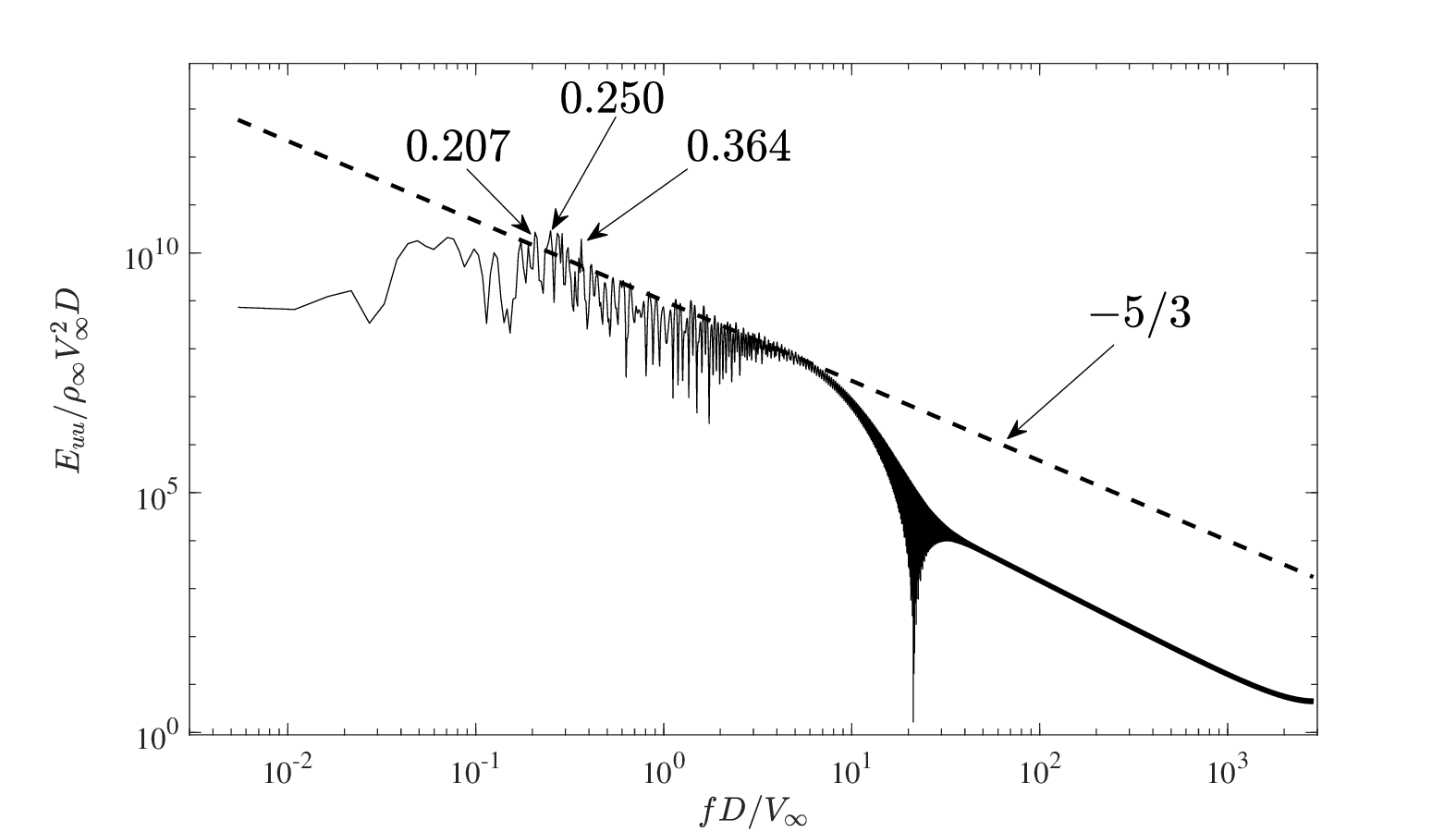}\label{Fig26a}}
    \subfloat[]{\includegraphics[width=0.33\textwidth,trim={0cm 0cm 0cm 0cm},clip]{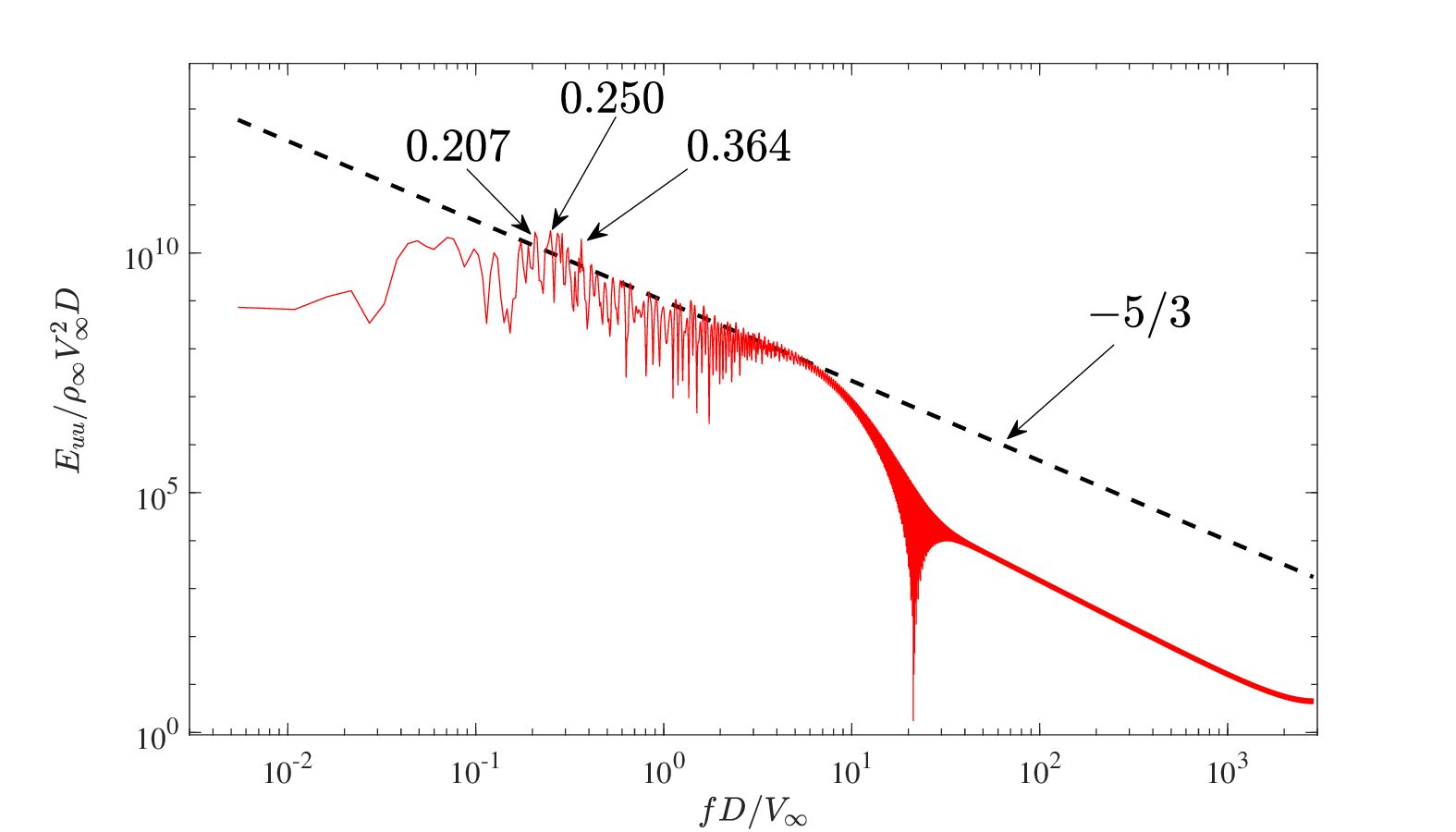}\label{Fig26b}}
    \subfloat[]{\includegraphics[width=0.33\textwidth,trim={0cm 0cm 0cm 0cm},clip]{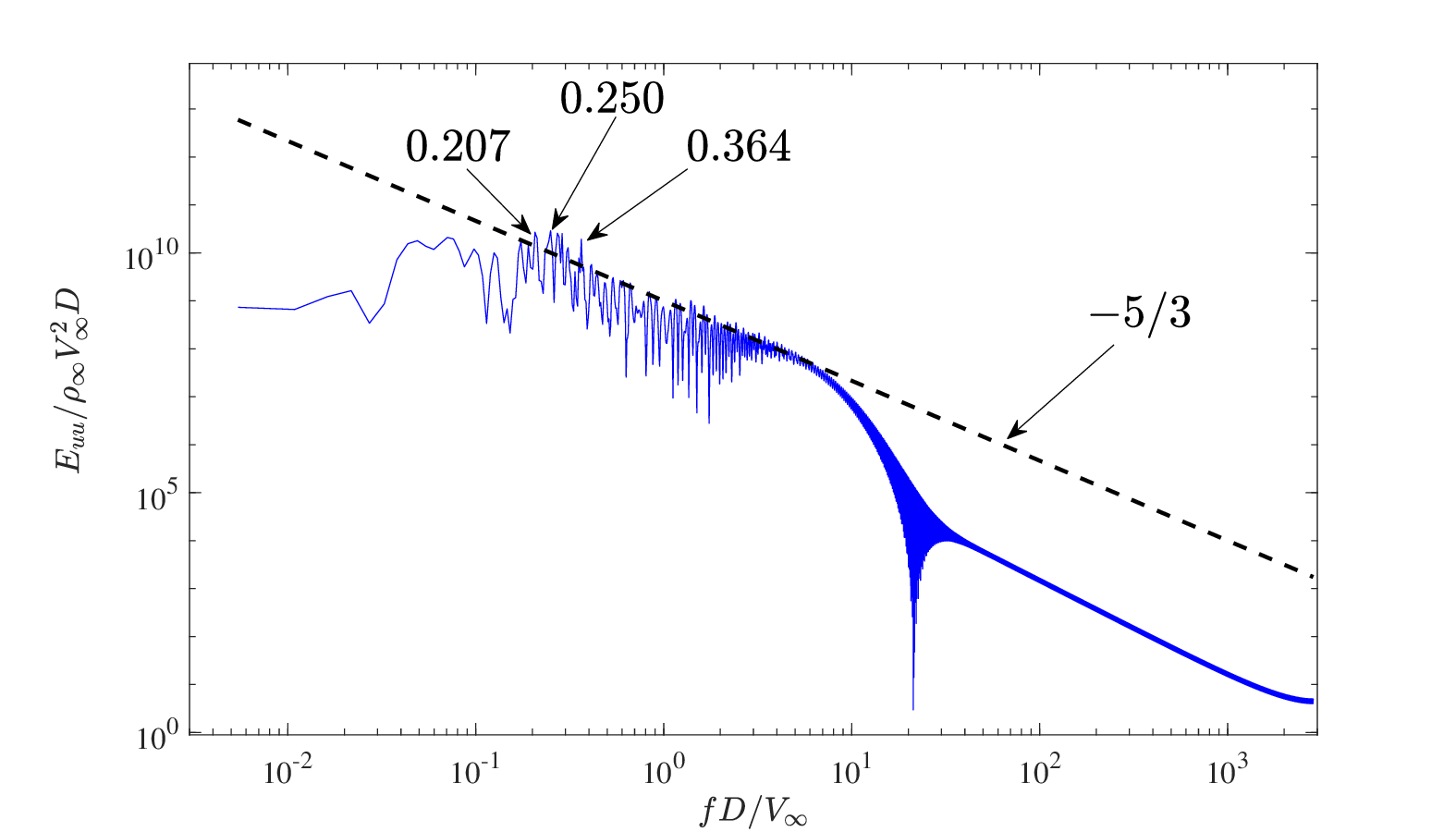}\label{Fig26c}}
	\caption{Energy spectra of the streamwise velocity in a cylinder wake at $(\tilde{x}/D,\tilde{y}/D)=(32,0.69)$ for $Re = 3,900$, $\gamma =3.0$,$\nu=1.46069 \times 10^{-5} m^2/s$, and (a)$M_\infty=0.99$ (b)$M_\infty=1.01$ (c)$M_\infty=1.35$.}\label{Fig26}
\end{figure*}

The one-dimensional energy spectra of the streamwise velocity at a far field location in the wake are presented for the three $M_\infty$ in figure~\ref{Fig26}. These are obtained in a similar way to figure~\ref{Fig9a}. The spectra are identical across $M_\infty$. The turbulence decays at a rate that is consistent with Kolmogorov's Five-Thirds law. The Strouhal frequency of the incompressible flow, $0.207$, is no more dominant. Energy is transferred to a range of frequencies close to $0.3$ starting with $0.25$. There are peaks at $0.364$. These are significant predictions because they agree with experimental \cite{Awasthietal2025} and computational \cite{Li2025} trends. In particular, at a similar $M_\infty=0.9$ and in the far field, the separated shear layers fluctuate at a Strouhal frequency $0.4$ in Li's\cite{Li2025} direct numerical simulation (DNS) results. This is very close to the observed peak at $0.364$ in the present result at $M_\infty=0.99$. The Strouhal frequency at $0.2$ is also deprived of dominant energy in DNS results. Furthermore, Awasthi et al.\cite{Awasthietal2025} experimentally observed that the energy at this hydrodynamic frequency is weaker than the energy at an aeroacoustic shedding frequency of $0.42$ in a supersonic cylinder wake. Additionally, the universal Strouhal number is approximately $0.3$. These are in agreement with the present observations at a supersonic $M_\infty=1.35$ where there are dominant shedding frequencies around $0.3$ and the spectral peak at $0.364$ is within a $13\%$ range of the measured aeroacoustic frequency.
\color{black}

However, the compressible extension of RPT is not a fully predictive tool in strongly transonic/supersonic and highly turbulent conditions but an approximate model with delimited validity. Where the model departs significantly from CFD and experimental data, it is partly because it rests on an isentropic density closure built on an incompressible base flow, employs a constant viscosity, and is without an explicit coupling with the energy equation. In the literature, the one-way coupling of the density field is valid at $M_\infty<0.3$ where the modest density variations cause minimal or negligible volume dilatation \cite{Economon2020}. At higher $M_\infty$, the volume dilation due to these changes is not negligible, and there is a mutual dependence between velocity, density, and pressure\cite{Economon2020}. However, in the present analysis it produces qualitatively reliable results in $30<Re<10,000$ for $M_\infty \le 1.01$ possibly because the velocity field is kept divergence-free.

\section{Conclusion}\label{C}
This article explores the compressible flow field around an impulsively started circular cylinder using RPT. The purpose of the study is \color{black}an attempt \color{black}to address \color{black}some of \color{black}the limitations of experimental and CFD methods by providing an analytical model to complement these approaches. The research focuses on subsonic compressible and transonic \color{black}($M_\infty \le 1.01$) \color{black}flow physics in the subcritical regime \color{black}($30<Re<10,000$)\color{black}, with potential applications in aerospace engineering, space exploration, and astrophysics.
\subsection{Key Findings}
\begin{enumerate}
\item Compressibility Effects:
Compressibility marginally increases vortex enstrophy at low $Re$, as evidenced by tighter streamlines around vortex centers. At higher $Re$ (e.g., $3,900$), compressibility effects are less pronounced in the wake flow stability, with nearly constant recirculation zone dimensions across varying Mach numbers ($M_\infty \le 0.5$).

\item Wake Characteristics:
Compressibility reduces the velocity deficit in the wake and suppresses shear layer instability, favoring lower frequency oscillations. The recirculation zone length and wake stability are less sensitive to compressibility at $Re = 3,900$ for $M_\infty \le 0.5$, contrary to CFD results.
\item Strouhal Number:
The Strouhal number predictions from RPT show marginal changes for $0.2 \le M_\infty \le 0.5$, aligning with experimental and CFD trends.
Compressibility introduces energetic smaller-scale structures, disrupting the inertial range of turbulence and altering the energy spectra.

\item Local Supersonic Flow Features:
At higher Mach numbers ($M_\infty \ge 0.6$), local supersonic pockets and shock waves appear in the wake, with complex interactions such as $\lambda$-shocks and bow shocks forming at $M_\infty = 0.99$ and above. The supersonic pocket's characteristics, such as internal shock strength and location, are influenced by parameters like cylinder radius, viscosity, and specific heat ratio ($\gamma$).

\item Energy Spectra:
Compressibility effects introduce background noise into energy spectra, reducing the dominance of Strouhal frequency and favoring lower frequency oscillations. At supersonic Mach numbers, energy is transferred to higher frequencies, consistent with experimental and computational observations.
\end{enumerate}

\subsection{Conclusions}
RPT provides a valuable analytical tool for modeling compressible flow over circular cylinders, capturing key flow features such as wake dynamics, vortex shedding, and compressibility effects. Although RPT aligns with the experimental and CFD results in some aspects, there are discrepancies, particularly in the wake flow stability and the dimensions of the recirculation zone at higher $Re$ and $M_\infty$.

The study highlights the potential of RPT to complement experimental and numerical methods, offering insights into compressible flow physics for applications in aerospace and astrophysics. Future work could focus on addressing the limitations of RPT, such as its inability to fully replicate the elongated free shear layers and wake recompression observed in experiments and CFD. \color{black}These limitations stem from the use of an isentropic density closure on an incompressible base flow, the application of constant viscosity, and the lack of an explicit feedback from the energy equation. \color{black} \color{black} In addition, quantitative analysis of the pressure field and surface quantities, including base pressure and drag, is the subject of further publications. Analysis of the pressure and vorticity distribution in the flow field will aid in assessing the entropy production and the physical credibility of detailed shock predictions. More in-depth quantitative comparisons of key quantities, including recirculation length, Strouhal number, and shock standoff distance, are also in the direction of future analysis.
\color{black}

This research contributes to understanding compressible flows over bluff bodies and provides a foundation for further exploration of refined analytical models in fluid dynamics.

\begin{acknowledgments}
This work is based on Ph.D. research supervised by Professor Marilyn J. Smith and funded by Kwara State University, Malete, Nigeria. The premise of the research was first presented at $2018$ AIAA SciTech Forum (AIAA 2018-1288). 
\end{acknowledgments}
\section*{Data Availability Statement}
Data that support the findings of this study are available within the article.
\section*{Conflict of Interest}
The authors have no conflicts to disclose.
\color{black}
\appendix
\section{Satisfying NSE}\label{CoMm}
Equation~\ref{NSEDiffinc} highlights the role of $\omega$ in incompressible NSE. \color{black}The velocity field $\mathbf{V} = \nabla \phi$ associated with a classical velocity potential, $\phi$, is irrotational \color{black} and satisfies Eq.~\ref{NSEDiffinc} identically \cite{Anderson2011}. However, \color{black}the velocity field $\mathbf{V} = \nabla \times ( \psi_{vis} \hat{\mathbf{e}}_z )$ \color{black}is rotational and cannot identically satisfy Eq.~\ref{NSEDiffinc}. $\tilde{\kappa}$ is a \color{black} quasi-potential \color{black} stream function that satisfies Eq.~\ref{NSEDiffinc} \cite{Amoloye2024,Amoloye2021}. $\tilde{\kappa}$ is derived by resolving $\psi_{vis}$ in the Cartesian wind coordinate system as follow. 

\begin{figure}
	\centering
	\includegraphics[width=0.35\textwidth,trim={0cm 0cm 0cm 0cm},clip]{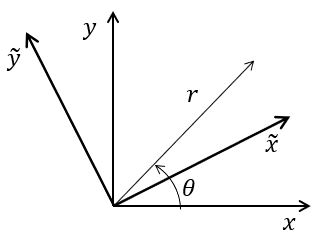}
	\caption{\color{black}The relationship between cylindrical polar and Cartesian coordinate systems in $\psi_{vis}$\cite{Amoloye2021}. Reprinted from "A Closed-Form Analytical Solution to the Complete Three-Dimensional Unsteady Compressible Navier-Stokes Equations", by Taofiq Omoniyi Amoloye, 2022, Research Square (DOI:\url{https://doi.org/10.21203/rs.3.rs-1157529/v1}). CC-BY $4.0.$\color{black}}\label{Fig27}
\end{figure}

The cylindrical polar coordinate variables are related to the Cartesian $x$-$y$ geometric axes as
\[r=\sqrt{x^2+y^2}~~~~~~\theta=\arctan{(y,x)}\hspace{3pt}.\]
This is illustrated in figure~\ref{Fig27}. There are at least $360$ other orientations of the $\tilde{x}$-$\tilde{y}$ (wind) axes that correlate with the same cylindrical polar coordinates in figure~\ref{Fig27}. These correspond to successive units of $\theta$. For a principal axis of the flow, the wind axes in $\psi_{vis}$ are rotated such that
\begin{equation}
\label{kwasutheta}
\theta{}=\arctan{(x,y)}\hspace{3pt} ,
\end{equation} 
and the negative $\tilde{x}$-axis is coincident with the positive $y$-axis.
Thus, $\psi_{vis}$ becomes $\tilde{\kappa}$ that is a \color{black} quasi-potential \color{black} function in the principal axis. $\tilde{\kappa}$ is summarized in table~\ref{Function Comparison}. 
\begin{figure}
	\centering
	\subfloat[$\psi=r\sin{\theta}\left(1-\dfrac{R^2}{r^2}\right)$]{\includegraphics[width=0.25\textwidth,trim={0cm 0cm 0cm 0cm},clip]{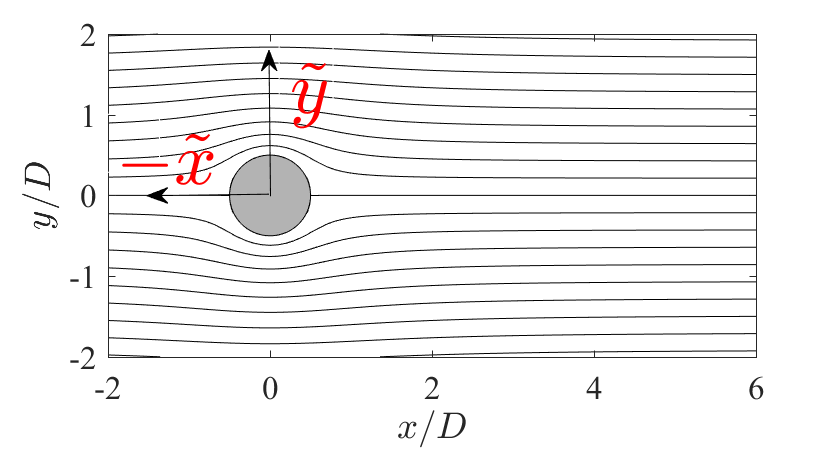}\label{Fig28a}}
	\subfloat[$\phi=r\cos{\theta}\left(1+\dfrac{R^2}{r^2}\right)$]{\includegraphics[width=0.25\textwidth,trim={0cm 0cm 0cm 0cm},clip]{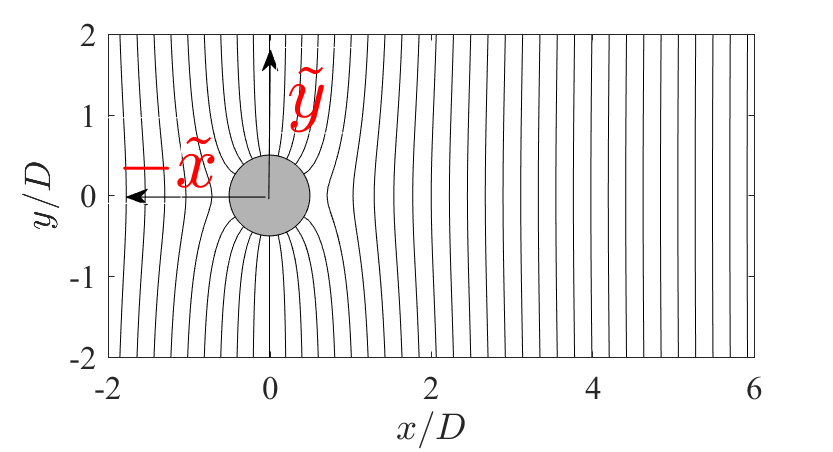}\label{Fig28b}}\\
	\subfloat[$\tilde{\kappa}=r\sin{\left(\arctan{(r\cos{\theta},r\sin{\theta})}\right)}\left(1-\dfrac{R^2}{r^2}\right)$]{\includegraphics[width=0.25\textwidth,trim={0cm 0cm 0cm 0cm},clip]{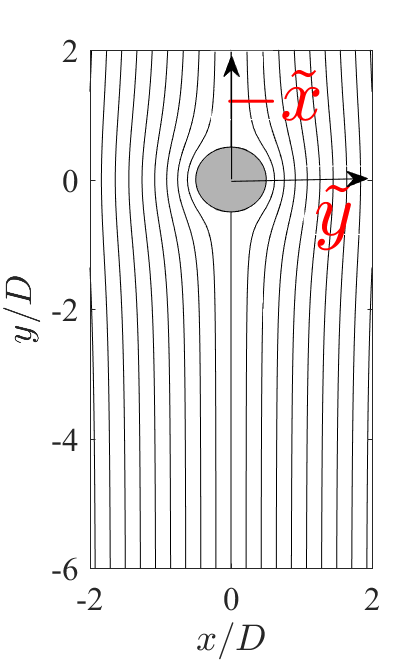}\label{Fig28c}}
	\caption{\color{black}A visual comparison of $\psi$ $(a)$, $\phi$ $(b)$ and $\tilde{\kappa}$ $(c)$ for a cylinder non-lifting flow \cite{Amoloye2021}. Reprinted from "A Closed-Form Analytical Solution to the Complete Three-Dimensional Unsteady Compressible Navier-Stokes Equations", by Taofiq Omoniyi Amoloye, 2022, Research Square (DOI:\url{https://doi.org/10.21203/rs.3.rs-1157529/v1}). CC-BY $4.0.$\color{black}}
	\label{Fig28}
\end{figure}
Table~\ref{Function Comparison} also compares $\tilde{\kappa}$ with $\phi$ and a generic stream function, $\psi$. This is illustrated in figures~\ref{Fig28} -~\ref{Fig31} that present plots of the level sets of the three functions and their velocity fields for the inviscid non-lifting cylinder flow \cite{Anderson2011,Amoloye2021}.
\begin{table*}
\begin{centering}
		\caption{Properties of the $\tilde{\kappa}$ in comparison to $\phi$ and $\psi$.}
		\label{Function Comparison}
		\begin{tabular}{rrrr}
			\hline
			\hspace{5pt}&$\phi{}$\hspace{20pt} & $\tilde{\kappa}$ &\hspace{20pt}$\psi{}$  \\\hline
			\hspace{5pt}$V_{r}$&$\dfrac{\partial\phi{}}{\partial r}$\hspace{20pt} & $\dfrac{1}{r}\dfrac{\partial \tilde{\kappa{}}}{\partial \theta}$ &\hspace{20pt}$\dfrac{1}{r}\dfrac{\partial \psi{}}{\partial \theta}$  \\
			\\
			\hspace{5pt}$V_{\theta}$&$\dfrac{1}{r}\dfrac{\partial \phi{}}{\partial \theta}$\hspace{20pt} & $-\dfrac{\partial\tilde{\kappa{}}}{\partial r}$ &\hspace{20pt}$-\dfrac{\partial\psi{}}{\partial r}$  \\
			\\
			\hspace{5pt}$V_{z}$&$\dfrac{\partial \phi{}}{\partial z}$\hspace{20pt} & $\dfrac{\partial\tilde{\kappa{}}}{\partial z}$ &\hspace{20pt}$-$  \\
			\\
			\hspace{5pt}$V_x$&$\dfrac{\partial \phi{}}{\partial x}$\hspace{20pt} & $\dfrac{\partial\tilde{\kappa{}}}{\partial x}$ &\hspace{20pt}$\dfrac{\partial\psi{}}{\partial y}$  \\
			\\
			\hspace{5pt}$V_y$&$\dfrac{\partial \phi{}}{\partial y}$\hspace{20pt} & $\dfrac{\partial\tilde{\kappa{}}}{\partial y}$ &\hspace{20pt}$-\dfrac{\partial\psi{}}{\partial x}$  \\
			\\
			\hspace{5pt}$V_{r}$&$V_x\cos{\theta}+V_y\sin{\theta}$\hspace{20pt} & $-V_x\sin{\theta}+V_y\cos{\theta}$ &\hspace{20pt}$V_x\cos{\theta}+V_y\sin{\theta}$  \\
			\\
			\hspace{5pt}$V_{\theta}$&$-V_x\sin{\theta}+V_y\cos{\theta}$\hspace{20pt} & $-\left(V_x\cos{\theta}+V_y\sin{\theta}\right)$ &\hspace{20pt}$-V_x\sin{\theta}+V_y\cos{\theta}$  \\
			\\
			\hspace{5pt}$V_{z}$&$-(V_x\sin{\theta}-V_y\cos{\theta})\cos{\varphi}$\hspace{20pt} & $-\left(V_x\cos{\theta}+V_y\sin{\theta}\right)\cos{\varphi}$ &\hspace{20pt}$-$  \\
			\hline
		\end{tabular}
\end{centering}
\end{table*}

In figures~\ref{Fig28a} and \ref{Fig28b}, the negative $x$-axis is coincident with the negative $\tilde{x}$-axis, and the freestream flow is parallel to them. However, the freestream flow approaches from the positive $y$-axis in figure~\ref{Fig28c}. The streamlines and the potential lines in figures~\ref{Fig28a} and \ref{Fig28b} respectively are orthogonal. However, they describe the same velocity field. Whereas $\psi$ is differentiated in a perpendicular direction to a flow to obtain the velocity components in figure~\ref{Fig29}, $\phi$ is differentiated in the flow direction to obtain the same velocity components in figure~\ref{Fig30}. Thus, $\phi$ results in an irrotational flow, and $\psi$ does not \cite{Anderson2011}.
\begin{figure}
	\centering
	\subfloat[]{\includegraphics[width=0.25\textwidth,trim={0cm 0cm 0cm 0cm},clip]{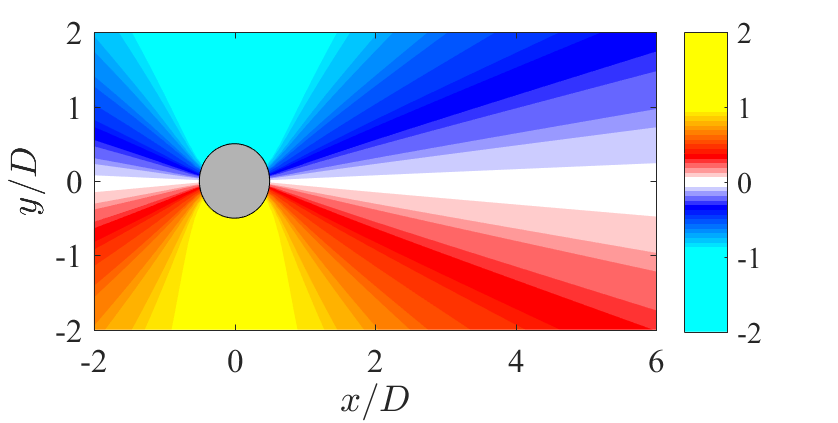}\label{Fig29a}}
	\subfloat[]{\includegraphics[width=0.25\textwidth,trim={0cm 0cm 0cm 0cm},clip]{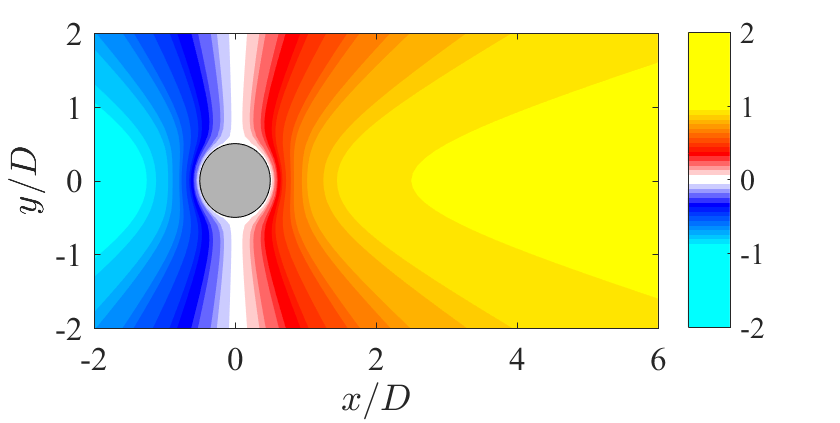}\label{Fig29b}}\\
	\subfloat[]{\includegraphics[width=0.25\textwidth,trim={0cm 0cm 0cm 0cm},clip]{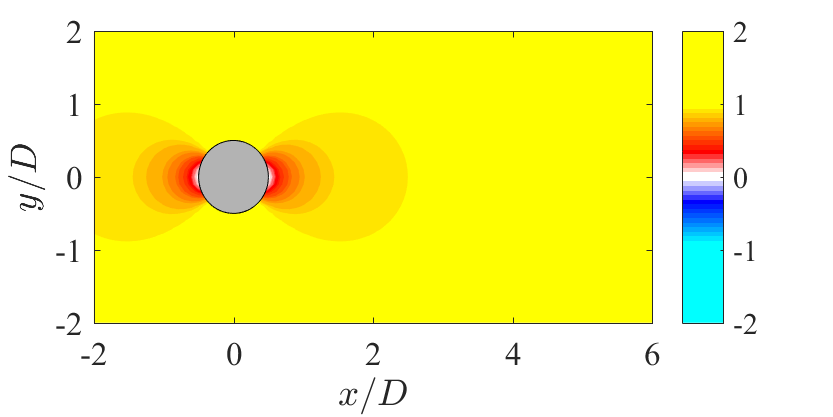}\label{Fig29c}}
	\subfloat[]{\includegraphics[width=0.25\textwidth,trim={0cm 0cm 0cm 0cm},clip]{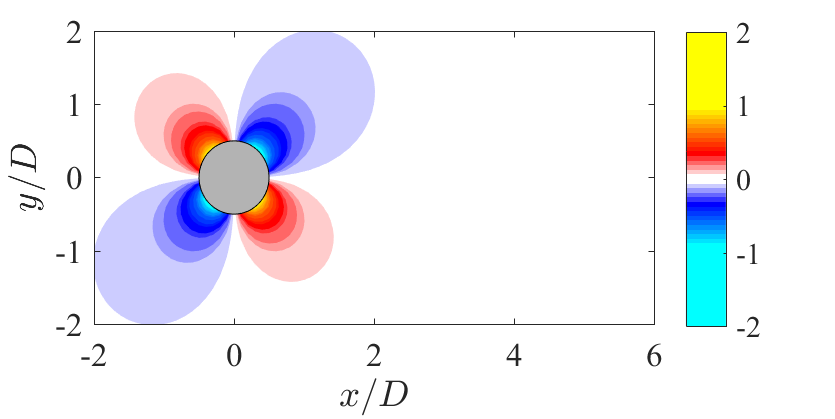}\label{Fig29d}}
	\caption{\color{black}An illustration of the non-dimensional planar velocity field of $\psi$ in the inviscid non-lifting cylinder flow. $(a)$ $V_{\theta}=-{\partial \psi}/{\partial r}$, $(b)$ $V_{r}={\partial \psi}/{r\partial \theta}$, $(c)$ $V_{x}={\partial \psi}/{\partial y}$ and $(d)$ $V_{y}=-{\partial \psi}/{\partial x}$ \cite{Amoloye2021}. Reprinted from "A Closed-Form Analytical Solution to the Complete Three-Dimensional Unsteady Compressible Navier-Stokes Equations", by Taofiq Omoniyi Amoloye, 2022, Research Square (DOI:\url{https://doi.org/10.21203/rs.3.rs-1157529/v1}). CC-BY $4.0.$\color{black}}
	\label{Fig29}
\end{figure}
\begin{figure}
	\centering
	\subfloat[]{\includegraphics[width=0.25\textwidth,trim={0cm 0cm 0cm 0cm},clip]{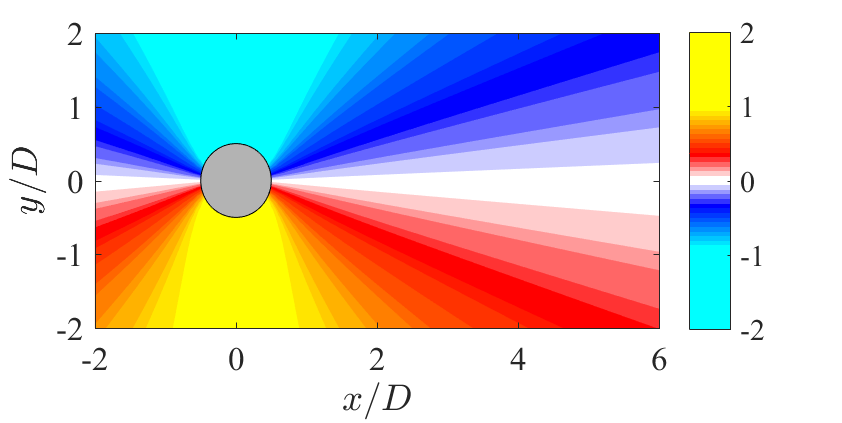}\label{Fig30a}}
	\subfloat[]{\includegraphics[width=0.25\textwidth,trim={0cm 0cm 0cm 0cm},clip]{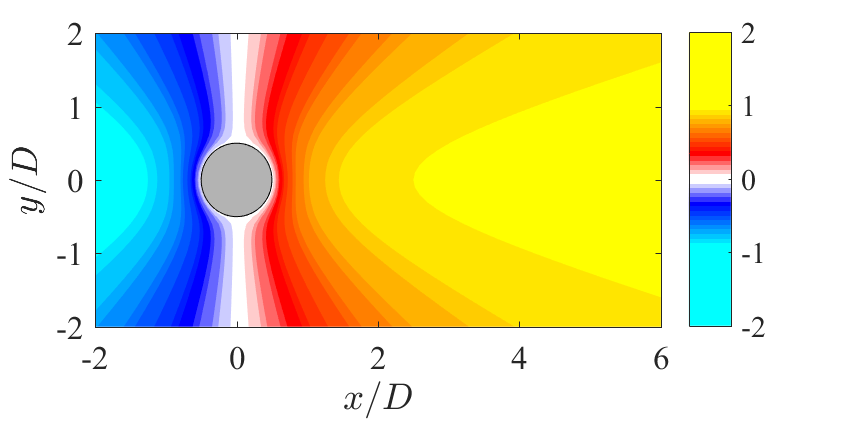}\label{Fig30b}}\\
	\subfloat[]{\includegraphics[width=0.25\textwidth,trim={0cm 0cm 0cm 0cm},clip]{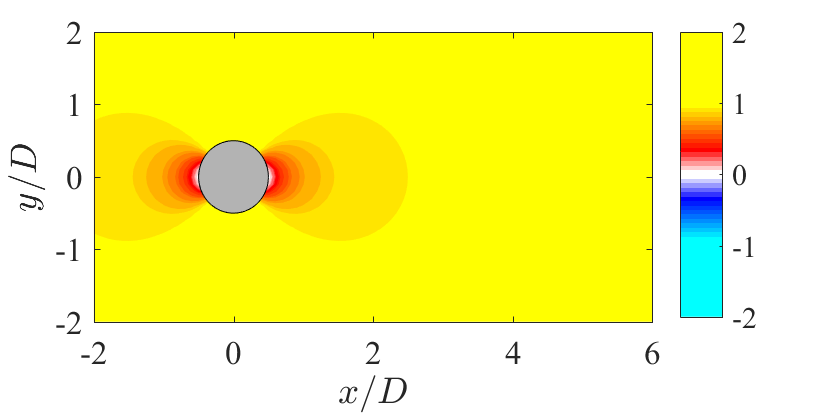}\label{Fig30c}}
	\subfloat[]{\includegraphics[width=0.25\textwidth,trim={0cm 0cm 0cm 0cm},clip]{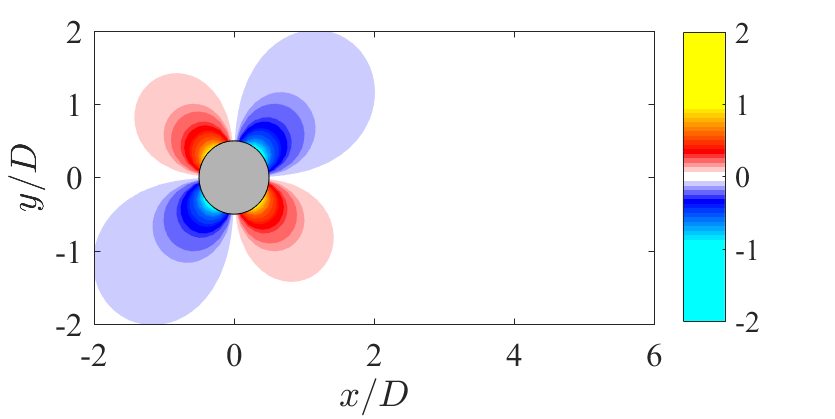}\label{Fig30d}}
	\caption{\color{black}An illustration of the non-dimensional planar velocity field of $\phi$ in the inviscid non-lifting cylinder flow. $(a)$ $V_{\theta}={\partial \phi}/{r\partial \theta}$, $(b)$ $V_{r}={\partial \phi}/{\partial r}$, $(c)$ $V_{x}={\partial \phi}/{\partial x}$ and $(d)$ $V_{y}={\partial \phi}/{\partial y}$ \cite{Amoloye2021}. Reprinted from "A Closed-Form Analytical Solution to the Complete Three-Dimensional Unsteady Compressible Navier-Stokes Equations", by Taofiq Omoniyi Amoloye, 2022, Research Square (DOI:\url{https://doi.org/10.21203/rs.3.rs-1157529/v1}). CC-BY $4.0.$\color{black}}
	\label{Fig30}
\end{figure}

In the cylindrical polar coordinate system, the velocity components are obtained from $\tilde{\kappa}$ in a similar way to $\psi$ as depicted in figures\ref{Fig31a} and \ref{Fig31b}. However, in the Cartesian wind coordinate system, $\tilde{\kappa}$ is defined on the principal axes for a viscous rotational flow about which $\mathbf{\omega}$ is identically zero \cite[pp.~57-58]{Schlichting1979}. These axes coincide with the wind axes, and the negative principal $x$-axis points into the incident flow. $\tilde{\kappa}$ is a potential function in these axes, and its velocity components are obtained as

\begin{equation}
	\label{Vkappa}
	\begin{array}{l}
		x=r\cos{\theta} \hspace{10pt} y=r\sin{\theta} \hspace{10pt}z=r\cos{\varphi}
		\\[10pt] 
		V_{\tilde{x}}=\dfrac{\partial{}\tilde{\kappa{}}}{\partial{}x}=\dfrac{\partial{}\tilde{\kappa{}}}{\partial{}r}\dfrac{\partial{}r}{\partial{}x}+\dfrac{\partial{}\tilde{\kappa{}}}{r\partial{}\theta{}}\dfrac{r\partial{}\theta{}}{\partial{}x}=-V_{\theta{}}\cos{\theta{}}-V_r\sin{\theta{}}
		\\[10pt]
		V_{\tilde{y}}=\dfrac{\partial{}\tilde{\kappa{}}}{\partial{}y}=\dfrac{\partial{}\tilde{\kappa{}}}{\partial{}r}\dfrac{\partial{}r}{\partial{}y}+\dfrac{\partial{}\tilde{\kappa{}}}{r\partial{}\theta{}}\dfrac{r\partial{}\theta{}}{\partial{}y}=-V_{\theta{}}\sin{\theta{}}+V_r\cos{\theta{}}
		\\[10pt]
		V_{z}=\dfrac{\partial{}\tilde{\kappa{}}}{\partial{}z}=\dfrac{\partial{}\tilde{\kappa{}}}{\partial{}r}\dfrac{\partial{}r}{\partial{}z}=-V_{\theta{}}\cos{\varphi{}}\\
	\end{array}
\end{equation}
\begin{figure*}
	\centering
	\subfloat[]{\includegraphics[width=0.25\textwidth,trim={0cm 0cm 0cm 0cm},clip]{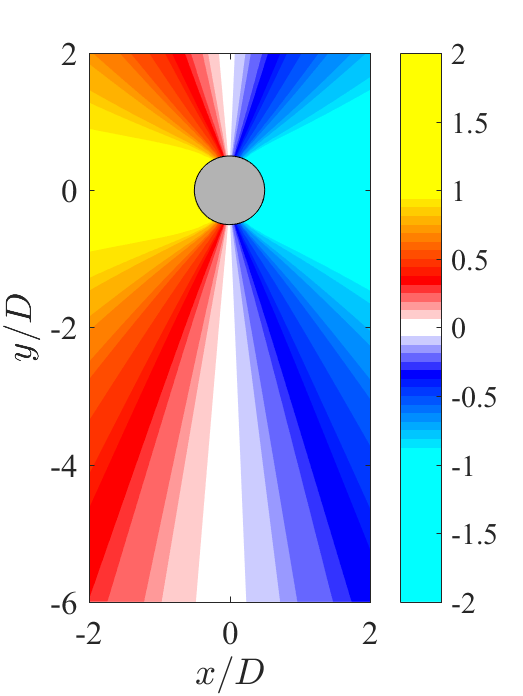}\label{Fig31a}}
	\subfloat[]{\includegraphics[width=0.25\textwidth,trim={0cm 0cm 0cm 0cm},clip]{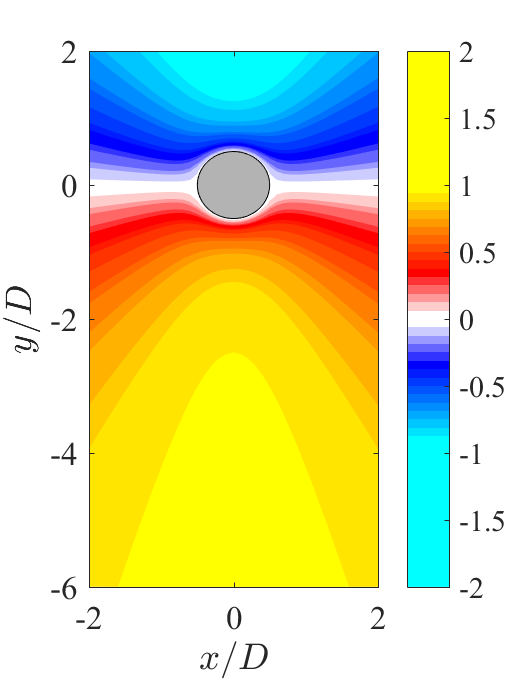}\label{Fig31b}}\\
	\subfloat[]{\includegraphics[width=0.25\textwidth,trim={0cm 0cm 0cm 0cm},clip]{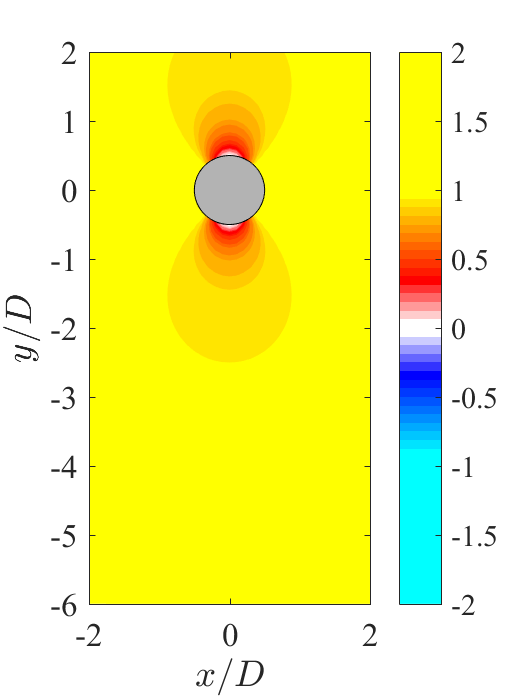}\label{Fig31c}}
	\subfloat[]{\includegraphics[width=0.25\textwidth,trim={0cm 0cm 0cm 0cm},clip]{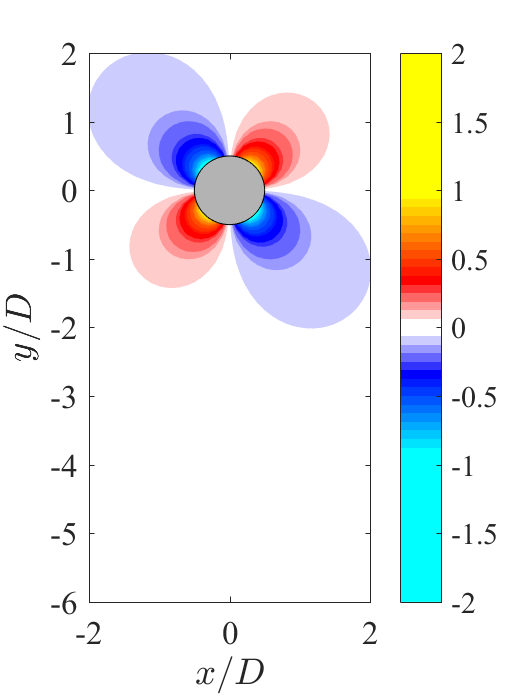}\label{Fig31d}}
	\caption{\color{black}An illustration of the non-dimensional planar velocity field of $\tilde{\kappa}$ in the inviscid non-lifting cylinder flow. $(a)$ $V_{\theta}=-{\partial \tilde{\kappa}}/{\partial r}$, $(b)$ $V_{r}={\partial \tilde{\kappa}}/{r\partial \theta}$, $(c)$ $V_{x}={\partial \tilde{\kappa}}/{\partial x}$ and $(d)$ $V_{y}={\partial \tilde{\kappa}}/{\partial y}$ \cite{Amoloye2021}. Reprinted from "A Closed-Form Analytical Solution to the Complete Three-Dimensional Unsteady Compressible Navier-Stokes Equations", by Taofiq Omoniyi Amoloye, 2022, Research Square (DOI:\url{https://doi.org/10.21203/rs.3.rs-1157529/v1}). CC-BY $4.0.$\color{black}}
	\label{Fig31}
\end{figure*}
\begin{figure}
	\centering
	\subfloat[]{\includegraphics[width=0.25\textwidth,trim={0cm 0cm 0cm 0cm},clip]{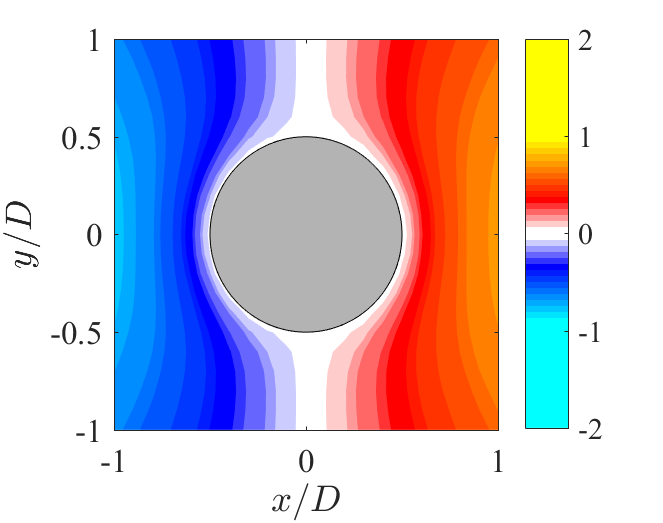}\label{Fig32a}}
	\subfloat[]{\includegraphics[width=0.25\textwidth,trim={0cm 0cm 0cm 0cm},clip]{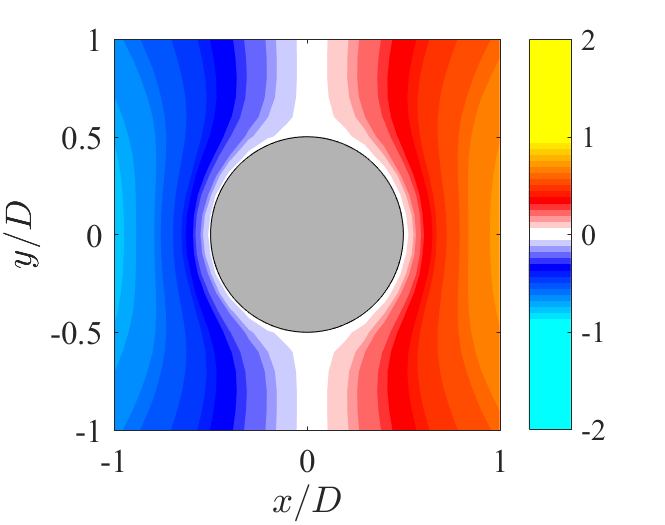}\label{Fig32b}}\\
	\subfloat[]{\includegraphics[width=0.25\textwidth,trim={0cm 0cm 0cm 0cm},clip]{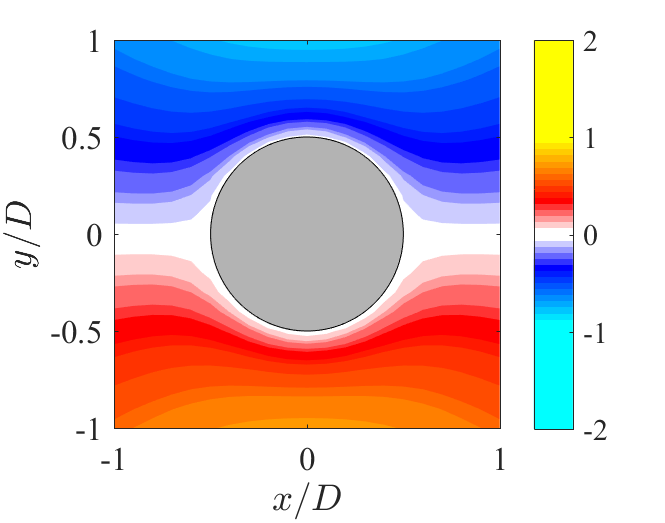}\label{Fig32c}}
	\caption{\color{black}A zoomed-in view of the radial velocity field. $(a)$ $V_{r}={\partial \psi}/{r\partial \theta}$, $(b)$ $V_{r}={\partial \phi}/{\partial r}$ and $(c)$ $V_{r}={\partial \tilde{\kappa}}/{r\partial \theta}$ \cite{Amoloye2021}. Reprinted from "A Closed-Form Analytical Solution to the Complete Three-Dimensional Unsteady Compressible Navier-Stokes Equations", by Taofiq Omoniyi Amoloye, 2022, Research Square (DOI:\url{https://doi.org/10.21203/rs.3.rs-1157529/v1}). CC-BY $4.0.$\color{black}}
	\label{Fig32}
\end{figure}
These are illustrated in figures\ref{Fig31c} and \ref{Fig31d} for the $z=0$ plane. The positive $z$-axis extends out of the page. Figure~\ref{Fig32} shows that the radial velocity vanishes at the surface because the surface is impermeable. \color{black}

\color{black}
The $\mathbf{\omega}$ components in the principal axes are computed as
\begin{equation}
	\label{dVxdz}
	\begin{array}{l}
		\dfrac{\partial V_{\tilde{x}}}{\partial z}=
		\dfrac{\partial^2 \tilde{\kappa}}{\partial z\partial x}\\[10pt]=\dfrac{\partial }{\partial z}\left(\dfrac{\partial \tilde{\kappa}}{\partial r}\dfrac{\partial r}{\partial x}+\dfrac{\partial \tilde{\kappa}}{r\partial \theta}\dfrac{r\partial\theta }{\partial x}\right)\\[10pt]
		=\left(\dfrac{\partial^2 \tilde{\kappa}}{\partial r^2}\dfrac{\partial r}{\partial x}+\dfrac{\partial^2 \tilde{\kappa}}{r\partial r\partial \theta}\dfrac{r\partial\theta }{\partial x}\right)\dfrac{\partial r}{\partial z}\\[10pt]=-\left(\dfrac{\partial V_{\theta}}{\partial r}\cos{\theta}+\dfrac{\partial V_r}{\partial r}\sin{\theta}\right)\cos{\varphi}\hspace{3pt},
	\end{array}
\end{equation}
\begin{equation}
	\label{dVydz}
	\begin{array}{l}
		\dfrac{\partial V_{\tilde{y}}}{\partial z}=
		\dfrac{\partial^2 \tilde{\kappa}}{\partial z\partial y}\\[10pt]=\dfrac{\partial }{\partial z}\left(\dfrac{\partial \tilde{\kappa}}{\partial r}\dfrac{\partial r}{\partial y}+\dfrac{\partial \tilde{\kappa}}{r\partial \theta}\dfrac{r\partial\theta }{\partial y}\right)\\[10pt]
		=\left(\dfrac{\partial^2 \tilde{\kappa}}{\partial r^2}\dfrac{\partial r}{\partial y}+\dfrac{\partial^2 \tilde{\kappa}}{r\partial r\partial \theta}\dfrac{r\partial\theta }{\partial y}\right)\dfrac{\partial r}{\partial z}\\[10pt]=-\left(\dfrac{\partial V_{\theta}}{\partial r}\sin{\theta}-\dfrac{\partial V_r}{\partial r}\cos{\theta}\right)\cos{\varphi}\hspace{3pt},
	\end{array}
\end{equation}
\begin{equation}
	\label{dVzdx}
	\begin{array}{l}
		\dfrac{\partial V_{z}}{\partial x}=
		\dfrac{\partial^2 \tilde{\kappa}}{\partial x\partial z}\\[10pt]=\dfrac{\partial }{\partial x}\left(\dfrac{\partial \tilde{\kappa}}{\partial r}\dfrac{\partial r}{\partial z}\right)\\[10pt]=\dfrac{\partial^2 \tilde{\kappa}}{\partial r^2}\dfrac{\partial r}{\partial x}\dfrac{\partial r}{\partial z}+\dfrac{\partial^2 \tilde{\kappa}}{r\partial \theta \partial r}\dfrac{r\partial \theta}{\partial x}\dfrac{\partial r}{\partial z}\\[10pt]=-\left(\dfrac{\partial V_{\theta}}{\partial r}\cos{\theta}+\dfrac{\partial V_{r}}{\partial r}\sin{\theta}\right)\cos{\varphi}\hspace{3pt},
	\end{array}
\end{equation}
\begin{equation}
	\label{dVzdy}
	\begin{array}{l}
		\dfrac{\partial V_{z}}{\partial y}=
		\dfrac{\partial^2 \tilde{\kappa}}{\partial y\partial z}\\[10pt]=\dfrac{\partial }{\partial y}\left(\dfrac{\partial \tilde{\kappa}}{\partial r}\dfrac{\partial r}{\partial z}\right)\\[10pt]=\dfrac{\partial^2 \tilde{\kappa}}{\partial r^2}\dfrac{\partial r}{\partial y}\dfrac{\partial r}{\partial z}+\dfrac{\partial^2 \tilde{\kappa}}{r\partial \theta \partial r}\dfrac{r\partial \theta}{\partial y}\dfrac{\partial r}{\partial z}\\[10pt]=-\left(\dfrac{\partial V_{\theta}}{\partial r}\sin{\theta}-\dfrac{\partial V_{r}}{\partial r}\cos{\theta}\right)\cos{\varphi}\hspace{3pt},
	\end{array}
\end{equation}
\begin{widetext}
\begin{equation}
	\label{dVxdy}
	\begin{array}{l}
		\dfrac{\partial V_{\tilde{x}}}{\partial y}=\dfrac{\partial^2 \tilde{\kappa}}{\partial y \partial x}=\dfrac{\partial }{\partial y}\left(\dfrac{\partial \tilde{\kappa}}{\partial r}\dfrac{\partial r}{\partial x}+\dfrac{\partial \tilde{\kappa}}{r\partial \theta}\dfrac{r\partial\theta }{\partial x}\right)\\[10pt] 
		=\left(\dfrac{\partial^2 \tilde{\kappa}}{\partial r^2}\dfrac{\partial r}{\partial y}+\dfrac{\partial^2 \tilde{\kappa}}{r\partial r\partial \theta}\dfrac{r\partial\theta }{\partial y}\right)\dfrac{\partial r}{\partial x}+\dfrac{\partial \tilde{\kappa}}{\partial r}\dfrac{\partial^2 r}{\partial y\partial x}+\left(\dfrac{\partial^2 \tilde{\kappa}}{r\partial r\partial \theta}\dfrac{\partial r}{\partial y}+\dfrac{\partial^2 \tilde{\kappa}}{r^2\partial \theta^2}\dfrac{r\partial\theta }{\partial y}\right)\dfrac{r\partial\theta }{\partial x}+\dfrac{\partial \tilde{\kappa}}{r^2\partial \theta}\dfrac{r^2\partial^2\theta}{\partial y\partial x}\\[10pt]
		=\left(-\dfrac{\partial V_{\theta}}{\partial r}\sin{\theta}\cos{\theta}+\dfrac{\partial V_r}{\partial r}\cos^2{\theta}+\dfrac{V_{\theta}}{r}\sin{\theta}\cos{\theta}\right)-\dfrac{\partial V_r}{\partial r}\sin^2{\theta}-\dfrac{V_r}{r}(1-2\cos^2{\theta})-\dfrac{\partial V_{r}}{r\partial \theta}\sin{\theta}\cos{\theta} \hspace{3pt}, 
	\end{array}
\end{equation}
\begin{equation}
	\label{dVydx}
	\begin{array}{l}
		\dfrac{\partial V_{\tilde{y}}}{\partial x}=\dfrac{\partial^2 \tilde{\kappa}}{\partial x \partial y}=\dfrac{\partial }{\partial x}\left(\dfrac{\partial \tilde{\kappa}}{\partial r}\dfrac{\partial r}{\partial y}+\dfrac{\partial \tilde{\kappa}}{r\partial \theta}\dfrac{r\partial\theta }{\partial y}\right)\\[10pt]
		=\left(\dfrac{\partial^2 \tilde{\kappa}}{\partial r^2}\dfrac{\partial r}{\partial x}+\dfrac{\partial^2 \tilde{\kappa}}{r\partial r\partial \theta}\dfrac{r\partial\theta }{\partial x}\right)\dfrac{\partial r}{\partial y}+\dfrac{\partial \tilde{\kappa}}{\partial r}\dfrac{\partial^2 r}{\partial x\partial y}+\left(\dfrac{\partial^2 \tilde{\kappa}}{r\partial r\partial \theta}\dfrac{\partial r}{\partial x}+\dfrac{\partial^2 \tilde{\kappa}}{r^2\partial \theta^2}\dfrac{r\partial\theta }{\partial x}\right)\dfrac{r\partial\theta }{\partial y}+\dfrac{\partial \tilde{\kappa}}{r^2\partial \theta}\dfrac{r^2\partial^2\theta}{\partial x\partial y}\\[10pt]
		=\left(-\dfrac{\partial V_{\theta}}{\partial r}\sin{\theta}\cos{\theta}-\dfrac{\partial V_r}{\partial r}\sin^2{\theta}+\dfrac{V_{\theta}}{r}\sin{\theta}\cos{\theta}\right)+\dfrac{\partial V_r}{\partial r}\cos^2{\theta}-\dfrac{V_r}{r}(1-2\cos^2{\theta})-\dfrac{\partial V_{r}}{r\partial \theta}\sin{\theta}\cos{\theta}\hspace{3pt},
	\end{array}
\end{equation}
\begin{equation}
	\label{omegax}
	\begin{array}{l}
		\omega_{x}=\dfrac{\partial V_{z}}{\partial y}-\dfrac{\partial V_{\tilde{y}}}{\partial z}\\[10pt]=
		-\left(\dfrac{\partial V_{\theta}}{\partial r}\sin{\theta}-\dfrac{\partial V_{r}}{\partial r}\cos{\theta}\right)\cos{\varphi}-\left(-\left(\dfrac{\partial V_{\theta}}{\partial r}\sin{\theta}-\dfrac{\partial V_r}{\partial r}\cos{\theta}\right)\cos{\varphi}\right)=0 \hspace{3pt},
	\end{array}
\end{equation}
\begin{equation}
	\label{omegay}
	\begin{array}{l}
		\omega_{y}=\dfrac{\partial V_{\tilde{x}}}{\partial z}-\dfrac{\partial V_{z}}{\partial x}
		\\[10pt]=-\left(\dfrac{\partial V_{\theta}}{\partial r}\cos{\theta}+\dfrac{\partial V_r}{\partial r}\sin{\theta}\right)\cos{\varphi}-\left(-\left(\dfrac{\partial V_{\theta}}{\partial r}\cos{\theta}+\dfrac{\partial V_{r}}{\partial r}\sin{\theta}\right)\cos{\varphi}\right)=0 \hspace{3pt}
	\end{array}
\end{equation}
\end{widetext}
and
\begin{widetext}
\begin{equation}
	\label{omegaztilde}
	\begin{array}{l}
		\omega_{z}=\dfrac{\partial V_{\tilde{y}}}{\partial x}-\dfrac{\partial V_{\tilde{x}}}{\partial y}\\[10pt]
		=\left(-\dfrac{\partial V_{\theta}}{\partial r}\sin{\theta}\cos{\theta}-\dfrac{\partial V_r}{\partial r}\sin^2{\theta}+\dfrac{V_{\theta}}{r}\sin{\theta}\cos{\theta}\right)+\dfrac{\partial V_r}{\partial r}\cos^2{\theta}-\dfrac{V_r}{r}(1-2\cos^2{\theta})-\dfrac{\partial V_{r}}{r\partial \theta}\sin{\theta}\cos{\theta}\\[10pt]
		-\left[\left(-\dfrac{\partial V_{\theta}}{\partial r}\sin{\theta}\cos{\theta}+\dfrac{\partial V_r}{\partial r}\cos^2{\theta}+\dfrac{V_{\theta}}{r}\sin{\theta}\cos{\theta}\right)-\dfrac{\partial V_r}{\partial r}\sin^2{\theta}-\dfrac{V_r}{r}(1-2\cos^2{\theta})-\dfrac{\partial V_{r}}{r\partial \theta}\sin{\theta}\cos{\theta}\right]\\[10pt]
		=0\hspace{3pt}.
	\end{array}
\end{equation}
\end{widetext}
However, the flow field remains rotational especially towards the surface and in the wake \cite{Schlichting1979,Amoloye2020}. In the polar coordinate system, $\tilde{\kappa}$ is a stream function, yielding these non-zero components of $\mathbf{\omega}$ \cite{Amoloye2020}.

The pressure is invariant to the rotation of the Cartesian axes \cite{Schlichting1979,Amoloye2020}. The Laplacian of $\tilde{\kappa}$ required to obtain the viscous term of the pressure field (Eq.~\ref{Velpol}) is computed as\cite{Amoloye2020}
\begin{widetext}
		\begin{equation}
			\label{dVxdx}
			\begin{array}{l}
				\dfrac{\partial V_{\tilde{x}}}{\partial x}=\dfrac{\partial^2 \tilde{\kappa}}{\partial x^2}=\dfrac{\partial }{\partial x}\left(\dfrac{\partial \tilde{\kappa}}{\partial r}\dfrac{\partial r}{\partial x}+\dfrac{\partial \tilde{\kappa}}{r\partial \theta}\dfrac{r\partial\theta }{\partial x}\right)\\[10pt]
				=\left(\dfrac{\partial^2 \tilde{\kappa}}{\partial r^2}\dfrac{\partial r}{\partial x}+\dfrac{\partial^2 \tilde{\kappa}}{r\partial r\partial \theta}\dfrac{r\partial\theta }{\partial x}\right)\dfrac{\partial r}{\partial x}+\dfrac{\partial \tilde{\kappa}}{\partial r}\dfrac{\partial^2 r}{\partial x^2}+\left(\dfrac{\partial^2 \tilde{\kappa}}{r\partial r\partial \theta}\dfrac{\partial r}{\partial x}+\dfrac{\partial^2 \tilde{\kappa}}{r^2\partial \theta^2}\dfrac{r\partial\theta }{\partial x}\right)\dfrac{r\partial\theta }{\partial x}+\dfrac{\partial \tilde{\kappa}}{r^2\partial \theta}\dfrac{r^2\partial^2\theta}{\partial x^2}\\[10pt]
				=-\left(\dfrac{\partial V_{\theta}}{\partial r}\cos^2{\theta}+\dfrac{\partial V_r}{\partial r}\sin{\theta}\cos{\theta}+\dfrac{V_{\theta}}{r}(1-\cos^2{\theta})\right)-\left(\dfrac{\partial V_r}{\partial r}-\dfrac{V_r}{r}\right)\sin{\theta}\cos{\theta} +\dfrac{\partial V_{r}}{r\partial \theta}\sin^2{\theta}\hspace{3pt},
			\end{array}
		\end{equation}
		\begin{equation}
			\label{dVydy}
			\begin{array}{l}
				\dfrac{\partial V_{\tilde{y}}}{\partial y}=\dfrac{\partial^2 \tilde{\kappa}}{\partial y^2}=\dfrac{\partial }{\partial y}\left(\dfrac{\partial \tilde{\kappa}}{\partial r}\dfrac{\partial r}{\partial y}+\dfrac{\partial \tilde{\kappa}}{r\partial \theta}\dfrac{r\partial\theta }{\partial y}\right)\\[10pt]
				=\left(\dfrac{\partial^2 \tilde{\kappa}}{\partial r^2}\dfrac{\partial r}{\partial y}+\dfrac{\partial^2 \tilde{\kappa}}{r\partial r\partial \theta}\dfrac{r\partial\theta }{\partial y}\right)\dfrac{\partial r}{\partial y}+\dfrac{\partial \tilde{\kappa}}{\partial r}\dfrac{\partial^2 r}{\partial y^2}+\left(\dfrac{\partial^2 \tilde{\kappa}}{r\partial r\partial \theta}\dfrac{\partial r}{\partial y}+\dfrac{\partial^2 \tilde{\kappa}}{r^2\partial \theta^2}\dfrac{r\partial\theta }{\partial y}\right)\dfrac{r\partial\theta }{\partial y}+\dfrac{\partial \tilde{\kappa}}{r^2\partial \theta}\dfrac{r^2\partial^2\theta}{\partial y^2}\\[10pt]
				=-\left(\dfrac{\partial V_{\theta}}{\partial r}\sin^2{\theta}-\dfrac{\partial V_r}{\partial r}\sin{\theta}\cos{\theta}+\dfrac{V_{\theta}}{r}(1-\sin^2{\theta})\right)+\left(\dfrac{\partial V_r}{\partial r}-\dfrac{V_r}{r}\right)\sin{\theta}\cos{\theta}+\dfrac{\partial V_{r}}{r\partial \theta}\cos^2{\theta}\hspace{3pt},
			\end{array}
		\end{equation}
        
		\begin{equation}
			\label{dVzdz}
			\begin{array}{l}
				\dfrac{\partial V_{z}}{\partial z}=\dfrac{\partial^2 \tilde{\kappa}}{\partial z^2}=\dfrac{\partial }{\partial z}\left(\dfrac{\partial \tilde{\kappa}}{\partial r}\dfrac{\partial r}{\partial z}\right)\\[10pt]=\dfrac{\partial^2 \tilde{\kappa}}{\partial r^2}\dfrac{\partial r}{\partial z}\dfrac{\partial r}{\partial z}+\dfrac{\partial \tilde{\kappa}}{\partial r}\dfrac{\partial^2 r}{\partial z^2}\\[10pt]=-\left(\dfrac{\partial V_{\theta}}{\partial r}\cos^2{\varphi}+\dfrac{V_{\theta}}{r}\sin^2{\varphi}\right)\hspace{3pt},
			\end{array}
		\end{equation}
        \end{widetext}
		and
\begin{widetext}
		\begin{equation}
			\label{crossderive}
			\begin{array}{l}
				{{\nabla{}}}^2\tilde{\kappa}=\dfrac{\partial V_{\tilde{x}}}{\partial x}+\dfrac{\partial V_{\tilde{y}}}{\partial y}+\dfrac{\partial V_{z}}{\partial z}=\dfrac{\partial^2 \tilde{\kappa}}{\partial x^2}+\dfrac{\partial^2 \tilde{\kappa}}{\partial y^2}+\dfrac{\partial^2 \tilde{\kappa}}{\partial z^2}\\[10pt]
				=-\left(\dfrac{\partial V_{\theta}}{\partial r}\cos^2{\theta}+\dfrac{\partial V_r}{\partial r}\sin{\theta}\cos{\theta}+\dfrac{V_{\theta}}{r}(1-\cos^2{\theta})\right)-\left(\dfrac{\partial V_r}{\partial r}-\dfrac{V_r}{r}\right)\sin{\theta}\cos{\theta}\\[10pt]+\dfrac{\partial V_{r}}{r\partial \theta}\sin^2{\theta}
				-\left(\dfrac{\partial V_{\theta}}{\partial r}\sin^2{\theta}-\dfrac{\partial V_r}{\partial r}\sin{\theta}\cos{\theta}+\dfrac{V_{\theta}}{r}(1-\sin^2{\theta})\right)\\[10pt]+\left(\dfrac{\partial V_r}{\partial r}-\dfrac{V_r}{r}\right)\sin{\theta}\cos{\theta}+\dfrac{\partial V_{r}}{r\partial \theta}\cos^2{\theta}
				-\left(\dfrac{\partial V_{\theta}}{\partial r}\cos^2{\varphi}+\dfrac{V_{\theta}}{r}\sin^2{\varphi}\right)\\[10pt]
				=-\dfrac{\partial V_{\theta}}{\partial r}-\dfrac{V_{\theta}}{r}+\dfrac{\partial V_{r}}{r\partial \theta}
				-\dfrac{\partial V_{\theta}}{\partial r}\cos^2{\varphi}-\dfrac{V_{\theta}}{r}\sin^2{\varphi}\hspace{3pt}.
			\end{array}
		\end{equation}
\end{widetext}
\section{Dimensional Consistency of the Pressure Field Equation} \label{DimAnal}
For physical credibility and mathematical rigor, the dimensional analysis of Eq.~\ref{MechPressure} in Section~\ref{RPT} is presented as follows.

The fundamental dimensions are defined as Mass $[\mathrm{M}]$, Length $[\mathrm{L}]$, and Time $[\mathrm{T}]$. In terms of these, the primary variables, operators and parameters that feature in Eq.~\ref{MechPressure} have the following dimensions:
\begin{itemize}
    \item $\tilde{\kappa}$: $[\mathrm{L}^{2}\mathrm{T}^{-1}]$
    \item $\rho$: $[\mathrm{M}\mathrm{L}^{-3}]$
    \item $V$: $[\mathrm{L}\mathrm{T}^{-1}]$
    \item $t$: $[\mathrm{T}]$
    \item $\tilde{p}$: $[\mathrm{M}\mathrm{L}^{-1}\mathrm{T}^{-2}]$
     \item ${p}_{0}$: $[\mathrm{M}\mathrm{L}^{-1}\mathrm{T}^{-2}]$
    \item $\mu$: $[\mathrm{M}\mathrm{L}^{-1}\mathrm{T}^{-1}]$
    \item $\lambda$: $[\mathrm{M}\mathrm{L}^{-1}\mathrm{T}^{-1}]$
    \item $\nabla \equiv \partial/\partial x$: $[\mathrm{L}^{-1}]$ 
\end{itemize}

The pressure field equation is given as
\begin{widetext}
\[\tilde{p}=\left[\dfrac{\partial{}\kappa{}}{\partial{}t}+\rho{}\left|\dfrac{\nabla \kappa}{\rho}\right|^2-\dfrac{4}{3}\mu{\nabla{}}\cdot\left(\dfrac{\nabla{}\kappa{}}{\rho}\right)-p_{0}\right]-\left(\lambda{}+\frac{2}{3}\mu{}\right){\nabla{}}\cdot\left(\dfrac{\nabla{}\kappa{}}{\rho}\right)\hspace{3pt}.\]
\end{widetext}
The principle of dimensional homogeneity requires that each additive term in this equation be evaluated to identical fundamental dimensions. Therefore, each term is analyzed below.
\begin{enumerate}
    \item Left-Hand Side (LHS) \newline
    The dimension of the mechanical pressure is
    \[[\tilde{p}]=[\mathrm{M}\mathrm{L}^{-1}\mathrm{T}^{-2}].\]
    \item Right-Hand Side (RHS)
    \begin{itemize}
        \item The time derivative of the scalar $\kappa$ field yields the following:
        \[ \left[\dfrac{\partial{}\kappa{}}{\partial{}t}\right]= \left[\dfrac{\partial{}(\rho \tilde{\kappa{}})}{\partial{}t}\right] =\dfrac{[\mathrm{M}\mathrm{L}^{-3}]\cdot [\mathrm{L}^{2}\mathrm{T}^{-1}] }{[\mathrm{T}]}=[\mathrm{M}\mathrm{L}^{-1}\mathrm{T}^{-2}]\]
        \item The product of density and the magnitude of the compressible velocity field yields:
        \[\left[\rho{}\left|\dfrac{\nabla \kappa}{\rho}\right|^2\right]=[\mathrm{M}\mathrm{L}^{-3}]\cdot \left[\dfrac{[L^{-1}][\mathrm{M}\mathrm{L}^{-3}]\cdot [\mathrm{L}^{2}\mathrm{T}^{-1}] }{[\mathrm{M}\mathrm{L}^{-3}]}\right]^2=[\mathrm{M}\mathrm{L}^{-1}\mathrm{T}^{-2}]\]
        \item The product of dynamic viscosity and the divergence operator applied to the compressible velocity field yields the following results:
\begin{widetext}
        \[\left[\mu{\nabla{}}\cdot\left(\dfrac{\nabla{}\kappa{}}{\rho}\right)\right]=[\mathrm{M}\mathrm{L}^{-1}\mathrm{T}^{-1}]\cdot[\mathrm{L}^{-1}]\cdot \dfrac{[\mathrm{L}^{-1}]\cdot[\mathrm{M}\mathrm{L}^{-3}]\cdot[\mathrm{L}^{2}\mathrm{T}^{-1}]}{[\mathrm{M}\mathrm{L}^{-3}]}=[\mathrm{M}\mathrm{L}^{-1}\mathrm{T}^{-2}] \] 
\end{widetext}
         \item The dimension of the stagnation pressure is
        \[[p_{0}]=[\mathrm{M}\mathrm{L}^{-1}\mathrm{T}^{-2}].\]
        \item The product of the viscosity coefficients and the divergence operator applied to the compressible velocity field yields the following:
\begin{widetext}
        \[\left[\left(\lambda +\dfrac{2}{3}\mu\right){\nabla{}}\cdot\left(\dfrac{\nabla{}\kappa{}}{\rho}\right)\right]=[\mathrm{M}\mathrm{L}^{-1}\mathrm{T}^{-1}]\cdot[\mathrm{L}^{-1}]\cdot \dfrac{[\mathrm{L}^{-1}]\cdot[\mathrm{M}\mathrm{L}^{-3}]\cdot[\mathrm{L}^{2}\mathrm{T}^{-1}]}{[\mathrm{M}\mathrm{L}^{-3}]}=[\mathrm{M}\mathrm{L}^{-1}\mathrm{T}^{-2}] \] 
\end{widetext}
    \end{itemize}
\end{enumerate}
The reduction in dimensions of all individual terms of both LHS and RHS yields the identical unit profile of \([\text{M L}^{-1}\text{T}^{-2}]\), which physically corresponds to pressure (\(\text{N/m}^{2}\)). This uniform reduction rigorously demonstrates the dimensional homogeneity and mathematical consistency of the equation.
\color{black}

\bibliography{references}

@PREAMBLE{
 "\providecommand{\noopsort}[1]{}" 
 # "\providecommand{\singleletter}[1]{#1}%" 
}

@book{Anderson2011,
    author    = {Anderson, John D.},
    title     = {{Fundamentals of Aerodynamics}},
    edition={5th},
    chapter   = {},
    year      = {2011},
    publisher = {McGraw-Hill},
    address   = {{New York}},
    doi={},
    ISBN={978-0073398105 },
}

@book{Anderson2003,
    author    = {Anderson, John D.},
    title     = {{Mordern Compressible Flows with Historical Perspectives}},
    edition={3rd},
    chapter   = {},
    year      = {2003},
    publisher = {McGraw-Hill},
    address   = {{New York}},
    doi={},
    ISBN={O-01-242443-5 },
}

@book{Anderson2006,
    author    = {Anderson, John D.},
    title     = {{Hypersonics and High-Temperature Gas Dynamics}},
    edition={2nd},
    chapter   = {},
    year      = {2006},
    publisher = {{American Institute of Aeronautics and Astronautics, Inc.}},
    address   = {Reston, Virginia},
    doi={},
    ISBN={978-1-56347-780-5},
}

@book{White2006,
    author    = {White, Frank M.},
    title     = {{Viscous Fluid Flow}},
    edition = {3rd},
    chapter   = {},
    year      = {2006},
    publisher = {McGraw-Hill},
    address   = {{New York}},
    doi={},
    ISBN={978-0-07-240231-5},
}

@book{CMI,
	author = {Feffermann,Charles L.},
	title = {{The Millennium Prize Problems: Existence and Smoothness of the Navier-Stokes}},
	chapter={},
	year = {2006},
           month={},
           day={},
           publisher = {{American Mathematical Society}},
	address   = {{Providence, RI}},
           doi={},
           ISBN={0-8218-3679-X},
           note={Available at \url{https://www.ams.org/publications/authors/books/postpub/mprize} - Last accessed: August 25, 2020}
}

@article{Amoloye2018,
	author = {Amoloye, Taofiq O.},
	journal = {{2018 {AIAA} Aerospace Sciences Meeting, {AIAA} {SciTech} Forum}},
	title = {{Kwasu Function: A Closed-Form Analytical Solution to the Complete Three-Dimensional Unsteady Compressible Navier-Stokes Equation}},
          number={{AIAA 2018-1288}},
          volume={},
           pages={},
	year = {2018},
	note={Available at \url{https://doi.org/10.2514/6.2018-1288 } - Last accessed: August 25, 2020},
}

@article{Joseph2006,
	author = {Joseph, Daniel D.},
	journal = {{International Journal of Multiphase Flow}},
           publisher={Elsevier},
	title = {{Potential Flow of Viscous Fluids: Historical Notes}},
	volume = {32},
	issue = {3},
	pages = {285-310},
           month = {},
	year = {2006},
          doi = {10.1016/j.ijmultiphaseflow.2005.09.004},
          note={Available at \url{https://doi.org/10.1016/j.ijmultiphaseflow.2005.09.004} - Last accessed: August 26, 2020},
}

@article{Parnaudeauetal2008,
	author = {Parnaudeau, Philippe and  Johan,Carlier and Heitz, Dominique  and  Lamballais, Eric},
	journal = {{Physics of Fluids}},
	title = {{Experimental and Numerical Studies of the Flow Over a Circular Cylinder at Reynolds Number $3900$}},
	volume = {20},
	issue = {8},
	pages = {1-14},
           month = {},
	year = {2008},
           doi = {10.1063/1.2957018},
           note={Available at \url{https://doi.org/10.1063/1.2957018} - Last accessed: August 26, 2020},
}

@phdthesis{Amoloye2020,
author = {Amoloye, Taofiq O.},
title = {{A Refined Potential Theory for the Incompressible Unsteady Subcritical-Reynolds number Flows on Canonical Bluff Bodies}},
school = {{Georgia Institute of Technology}},
year = {2020},
note={Available at \url{http://hdl.handle.net/1853/64150} - Last accessed: January 13, 2021},
}

@article{KravchenkoandMoin2000,
	author = {Kravchenko, A. G. and Moin, P.},
	journal = {Physics of Fluids},
           publisher={American Institute of Physics},
	title = {{Numerical Studies of Flow Over a Circular Cylinder at  $ Re_D= 3900$}},
	volume = {12},
	number = {2},
	pages = {403-417},
           month = {},
	year = {2000},
           doi = {},
           note={Available at \url{https://doi.org/10.1063/1.870318}  - Last accessed: August 26, 2020},
}

@article{Matlab,
	author = {Matlab},
	journal = {{MATLAB R2022b - Home Use, The MathWorks, Inc., Natick, Massachusetts, United States.}},
           publisher={},
	title = {Matlab},
	volume = {},
	number = {},
	pages = {},
           day={},
           month = {},
	year = {2022},
           doi = {},
}

@article{Roshko1961,
	author = {Roshko, Anatol},
	journal = {Journal of Fluid Mechanics},
           publisher={Cambridge University Press},
	title = {{Experiments on the Flow Past a Circular Cylinder at Very High Reynolds Number}},
	volume = {10},
	number = {3},
	pages = {345-356},
           day={},
           month = {},
	year = {1961},
           note ={Available at \url{http://dx.doi.org/10.1017/S0022112061000950} - Last accessed: August 26, 2020},
}

@article{GadelHak1995,
	author = {Gad-el-Hak, Mohamed},
	journal = {{Journal of Fluids Engineering}},
           publisher={{The American Society of Mechanical Engineers}},
	title = {{Questions in Fluid Mechanics: Stokes' Hypothesis for a Newtonian, Isotropic Fluid}},
	volume = {117},
	number = {1},
	pages = {3-5},
           day={},
           month = {},
	year = {1995},
           note ={Available at \url{https://doi.org/10.1115/1.2816816} - Last accessed: August 26, 2020},
}

@book{VanDyke1982,
    author    = {Van Dyke, Milton},
    title     = {{An Album of Fluid Motion}},
    edition={4th},
    chapter   = {},
    year      = {1988},
    publisher = {{The Parabolic Press}},
    address   = {Stanford, California 94305-0030},
   note={Available at \url{http://courses.washington.edu/me431/handouts/Album-Fluid-Motion-Van-Dyke.pdf} - Last accessed: August 26, 2020},
    ISBN={ },
}

@book{Schlichting1979,
	author = {Schlichting, Hermann},
	title = {{Boundary-Layer Theory}},
           edition={7th},
           chapter   = {},
	year = {1979},
           month={},
           day={},
           publisher = {{McGraw-Hill Book Company}},
	 address   = {New York},
           doi={},
           ISBN={},
           url={ }
}

@article{Amoloye2021,
	author = {Amoloye, Taofiq O.},
	journal = {{}},
        publisher={{}},
	title = {{A Closed-Form Analytical Solution to the Complete Three-Dimensional Unsteady Compressible Navier-Stokes Equations}},
	volume = {},
	number = {},
	pages = {},
        day={},
        month = {},
	year = {2021},
        note= {preprint (Version 1) available at Research Square \url{https://doi.org/10.21203/rs.3.rs-1157529/v1} - Last accessed: January 3, 2022},
}

@article{Farwig2021,
  title = {{From Jean Leray to The Millennium Problem: The Navier–Stokes Equations}},
  author = {Farwig, Reinhard},
  journal = {{Journal of Evolution Equations}},
  volume = {21},
  issue = {},
  pages = {3243–3263},
  numpages = {21},
  year = {2021},
  month = {},
  publisher = {Springer Nature},
  doi = {10.1007/s00028-020-00645-3},
  note={Available at \url{https://link.springer.com/article/10.1007/s00028-020-00645-3}- Last accessed: March 19, 2022},
}

@article{CoutandBouard1977a,
	author = {Coutanceau, M. and Bouard, R. },
	journal = {{Journal of Fluid Mechanics}},
	publisher = {{Cambridge University Press}},
	title = {{Experimental Determination of the Main Features of the
Viscous Flow in the Wake of a Circular Cylinder in Uniform Translation. Part 1. Steady flow}},
    volume = {79},
    number = {2},
    pages = {231-256},
    year = {1977},
    doi = {10.1017/S0022112077000135},
	note={Available at \url{https://doi.org/10.1017/S0022112077000135} - Last accessed: February 27, 2023},
}

@article{CoutandBouard1977b,
	author = {Coutanceau, M. and Bouard, R. },
	journal = {{Journal of Fluid Mechanics}},
	publisher = {{Cambridge University Press}},
	title = {{Experimental Determination of the Main Features of the
Viscous Flow in the Wake of a Circular Cylinder in Uniform Translation. Part 2. Unsteady flow}},
    volume = {79},
    number = {2},
    pages = {257-272},
    year = {1977},
    doi = {10.1017/S0022112077000147},
	note={Available at \url{https://doi.org/10.1017/S0022112077000147} - Last accessed: February 27, 2023},
}

@article{Amoloye2022,
	author = {Amoloye, Taofiq O.},
	journal = {{}},
        publisher={{}},
	title = {{Analysis of an Unsteady Incompressible Crossflow on a Stationary Circular Cylinder at Reynolds number $3~900$ Using Refined Potential Flow Theory}},
	volume = {},
	number = {},
	pages = {},
        day={},
        month = {},
	year = {2022},
        note= {Preprint (Version 1) available at Research Square \url{ https://doi.org/10.21203/rs.3.rs-1723482/v1} - Last accessed: February 4, 2023},
}

@article{Amoloye2024,
	author = {Amoloye, Taofiq O.},
	journal = {{Phys. Scr.}},
        publisher = {{IOP Publishing Ltd}},
	title = {{Modeling the unsteady wake of an impulsively started circular cylinder using refined potential flow theory}},
          number={8},
          volume={99},
           pages={},
	year = {2024},
        doi={10.1088/1402-4896/ad4ace},
	note={Available at \url{https://iopscience.iop.org/article/10.1088/1402-4896/ad4ace} - Last accessed: February 01, 2025},
}

@article{Ackermanetal2008,
	author = {Ackerman,~J. and Gostelow,~J. and Rona,~A. and Carscallen,~W.},
	journal = {{38th Fluid Dynamics Conference and Exhibit}},
	publisher = {{AIAA}},
	title = {{Base Pressure Measurements on a Circular Cylinder in Subsonic Cross Flow.}},
        volume = {},
        number = {},
        pages = {1-14},
        year = {2008},
        doi = {https://doi.org/10.2514/6.2008-4305},
	note={Available at \url{https://arc.aiaa.org/doi/10.2514/6.2008-4305} - Last accessed: April 05, 2025},
}

@article{Anyojietal2011,
	author = {Anyoji,~M. and Nose,~K. and Ida,~S. and Numata,~D. and Nagai,~H. and Asai,~K.},
	journal = {{Transactions of the Japan Society for Aeronautical and Space Sciences, Aerospace Technology Japan}},
	publisher = {{The Japan Society for Aeronautical and Space Sciences}},
	title = {{Development of a Low-Density Wind Tunnel for Simulating Martian Atmospheric Flight}},
        volume = {9},
        number = {},
        pages = {21-27},
        year = {2011},
        doi = {https://doi.org/10.2322/tastj.9.21},
	note={Available at \url{https://www.jstage.jst.go.jp/article/tastj/9/0/9_0_21/_article} - Last accessed: April 05, 2025},
}

@article{BobenriethMiserdaandLeal2006,
	author = {Bobenrieth Miserda,~R.~F. and Leal,~R.~G.},
	journal = {{44th AIAA Aerospace Sciences Meeting}},
	publisher = {{AIAA}},
	title = {{Numerical simulation of the unsteady aerodynamic forces over a circular cylinder in transonic flow}},
        volume = {22},
        number = {},
        pages = {16941-16957},
        year = {2006},
        doi = {https://doi.org/10.2514/6.2006-1408},
	note={Available at \url{https://doi.org/10.2514/6.2006-1408} - Last accessed: April 05, 2025},
}

@article{Botta1995,
	author = {Botta,~N.},
	journal = {{Journal of Fluid Mechanics}},
	publisher = {{Cambridge University Press}},
	title = {{The inviscid transonic flow about a cylinder}},
        volume = {301},
        number = {},
        pages = {225–250},
        year = {1995},
        doi = {https://doi.org/10.1017/S0022112095003879},
	note={Available at \url{https://doi.org/10.1017/S0022112095003879} - Last accessed: April 05, 2025},
}

@article{BurbeauandSagaut2002,
	author = {Burbeau,~A. and Sagaut,~P.},
	journal = {{Computers and Fluids}},
	publisher = {{Elsevier}},
	title = {{Simulation of a viscous compressible flow past a circular cylinder with high-order discontinuous Galerkin methods}},
        volume = {31},
        number = {8},
        pages = {867–889},
        year = {2002},
        doi = {https://doi.org/10.1016/S0045-7930(01)00055-X},
	note={Available at \url{https://doi.org/10.1016/S0045-7930(01)00055-X} - Last accessed: April 05, 2025},
}

@article{CanutoandTaira2015,
	author = {Canuto,~D. and Taira,~K.},
	journal = {{Journal of Fluid Mechanics}},
	publisher = {{Cambridge University Press}},
	title = {{Two-dimensional compressible viscous flow around a circular cylinder}},
        volume = {785},
        number = {},
        pages = {349-371},
        year = {2015},
        doi = {https://doi.org/10.1017/jfm.2015.635},
	note={Available at \url{ https://doi.org/10.1017/jfm.2015.635} - Last accessed: April 05, 2025},
}

@article{CherryandTemple1947,
	author = {Cherry,~T.,~M. and Temple,~G.~F.~J.},
	journal = {{Proceedings of the Royal Society A:Mathematical and Physical Sciences}},
	publisher = {{The Royal Society Publishing}},
	title = {{Flow of a compressible fluid about a cylinder}},
        volume = {192},
        number = {1028},
        pages = {45-79},
        year = {1947},
        doi = {https://doi.org/10.1098/rspa.1947.0138},
	note={Available at \url{ https://royalsocietypublishing.org/doi/pdf/10.1098/rspa.1947.0138} - Last accessed: April 05, 2025},
}

@article{DennisandChang1970,
	author = {Dennis,~S.~C.~R. and Chang,~G.~Z.},
	journal = {{Journal of Fluid Mechanics}},
	publisher = {{Cambridge University Press}},
	title = {{Numerical solutions for steady flow past a circular cylinder at Reynolds numbers up to 100}},
        volume = {42},
        number = {3},
        pages = {471-489},
        year = {1970},
        doi = {https://doi.org/10.1017/S0022112070001428},
	note={Available at \url{https://doi.org/10.1017/S0022112070001428} - Last accessed: April 05, 2025},
}

@article{GowenandPerkins1952,
	author = {Gowen,~F.~E. and Perkins,~E.~W.},
	journal = {{NACA Technical Note}},
	publisher = {{NASA}},
	title = {{Drag of circular cylinders for a wide range of Reynolds numbers and Mach numbers}},
        volume = {},
        number = {NACA-RM-A52C20},
        pages = {1-26},
        year = {1952},
        doi = {},
	note={Available at \url{https://ntrs.nasa.gov/citations/19930087134} - Last accessed: April 05, 2025},
}

@article{HoffmannandWeiss2023,
	author = {Hoffmann,~J. and Weiss,~D.~A.},
	journal = {{Fluids}},
	publisher = {{MDPI}},
	title = {{Compressible and Viscous Effects in Transonic Planar Flow around a Circular Cylinder—A Numerical Analysis Based on a Commercially Available CFD Tool}},
        volume = {8},
        number = {6},
        pages = {1-28},
        year = {2023},
        doi = {https://doi.org/10.3390/fluids8060182},
	note={Available at \url{https://doi.org/10.3390/fluids8060182} - Last accessed: April 05, 2025},
}

@article{IshiiandKuwahara1984,
	author = {Ishii,~K. and Kuwahara,~K.},
	journal = {{17th Fluid Dynamics, Plasma Dynamics, and Lasers Conference}},
	publisher = {{AIAA}},
	title = {{Computation of Compressible Flow Around a Circular Cylinder}},
        volume = {},
        number = {},
        pages = {1-11},
        year = {1984},
        doi = {https://doi.org/10.2514/6.1984-1631},
	note={Available at \url{https://doi.org/10.2514/6.1984-1631} - Last accessed: April 05, 2025},
}

@article{LiSetal2024,
	author = {Li,~S. and Yang,~J. and Teng,~P.},
	journal = {{Applied Ocean Research}},
	publisher = {{Elsevier}},
	title = {{Data-driven prediction of cylinder-induced unsteady wake flow}},
        volume = {150},
        number = {},
        pages = {1-19},
        year = {2024},
        doi = {https://doi.org/10.1016/j.apor.2024.104114},
	note={Available at \url{https://doi.org/10.1016/j.apor.2024.104114} - Last accessed: April 05, 2025},
}

@article{Lindsey1938,
	author = {Lindsey,~W.~F.},
	journal = {{NACA Technical Report}},
	publisher = {{NASA}},
	title = {{Drag of cylinders of simple shapes}},
        volume = {},
        number = {NACA-TR-619},
        pages = {169-176},
        year = {1938},
        doi = {},
	note={Available at \url{https://ntrs.nasa.gov/citations/19930091694} - Last accessed: April 05, 2025},
}

@article{Liuetal2023b,
	author = {Liu,~Y. and Ding,~Z. and Tao,~Y. and Qu,~J. and Xie,~X. and Qiu,~X.},
	journal = {{Physics of Fluids}},
	publisher = {{AIP Publishing}},
	title = {{Numerical study on compressible flow around a circular cylinder in proximity to the wall}},
        volume = {35},
        number = {6},
        pages = {1-17},
        year = {2023},
        doi = {https://doi.org/10.1063/5.0148846},
	note={Available at \url{https://doi.org/10.1063/5.0148846} - Last accessed: April 05, 2025},
}

@article{Lysenkoetal2012,
	author = {Lysenko,~D.~A. and Ertesvåg,~I.~S. and Rian,~K.~E.},
	journal = {{Flow, Turbulence and Combustion}},
	publisher = {{Springer Nature Link}},
	title = {{Large-Eddy Simulation of the Flow Over a Circular Cylinder at Reynolds Number 3900 Using the OpenFOAM Toolbox}},
        volume = {89},
        number = {4},
        pages = {491-518},
        year = {2012},
        doi = {https://doi.org/10.1007/s10494-012-9405-0},
	note={Available at \url{https://doi.org/10.1007/s10494-012-9405-0} - Last accessed: April 05, 2025},
}

@article{Mataretal2023,
	author = {Matar,~C. and Cinnella,~P. and Gloerfelt,~X. and Reinker,~F. and aus der Wiesche,~S.},
	journal = {{Energy}},
	publisher = {{Elsevier}},
	title = {{Investigation of non-ideal gas flows around a circular cylinder}},
        volume = {268},
        number = {},
        pages = {375-395},
        year = {2023},
        doi = {https://doi.org/10.1016/j.energy.2022.126563},
	note={Available at \url{https://doi.org/10.1016/j.energy.2022.126563} - Last accessed: April 05, 2025},
}

@article{McCarrhyandKubota1964,
	author = {McCarrhy JR,~J.~F. and Kubota,~T.},
	journal = {{AIAA Journa}},
	publisher = {{AIAA}},
	title = {{A study of wakes behind a circular cylinder at M equal 5.7}},
        volume = {2},
        number = {4},
        pages = {629-636},
        year = {1964},
        doi = {https://doi.org/10.2514/3.2399},
	note={Available at \url{https://doi.org/10.2514/3.2399} - Last accessed: April 05, 2025},
}

@article{MurthyandRose1977,
	author = {Murthy,~V. and Rose,~W.},
	journal = {{10th Fluid and Plasmadynamics Conference}},
	publisher = {{AIAA}},
	title = {{Form drag, skin friction, and vortex shedding frequencies for subsonic and transonic crossflows on circular cylinder}},
        volume = {},
        number = {},
        pages = {},
        year = {1977},
        doi = {https://doi.org/10.2514/6.1977-687},
	note={Available at \url{https://doi.org/10.2514/6.1977-687} - Last accessed: April 05, 2025},
}

@article{Nagataetal2020,
	author = {Nagata,~T. and Noguchi,~A. and Nonomura,~T. and Ohtani,~K. and Asai,~K.},
	journal = {{Shock waves}},
	publisher = {{Springer Nature Link}},
	title = {{Experimental investigation of transonic and supersonic flow over a sphere for Reynolds numbers of $10^3–10^5$ by free-flight tests with schlieren visualization}},
        volume = {30},
        number = {2},
        pages = {139-151},
        year = {2020},
        doi = {https://doi.org/10.1007/s00193-019-00924-0},
	note={Available at \url{https://doi.org/10.1007/s00193-019-00924-0} - Last accessed: April 05, 2025},
}

@article{Nagataetal2020b,
	author = {Nagata,~T. and Noguchi,~A. and Kusama,~K. and Nonomura,~T. and Komuro,~A. and Ando,~A. and Asai,~K.},
	journal = {{Journal of Fluid Mechanics}},
	publisher = {{Cambridge University Press}},
	title = {{Experimental investigation on compressible flow over a circular cylinder at Reynolds number of between 1000 and 5000}},
        volume = {893},
        number = {A13},
        pages = {},
        year = {2020},
        doi = {10.1017/jfm.2020.221},
	note={Available at \url{https://doi.org/10.1017/jfm.2020.221} - Last accessed: April 05, 2025},
}

@article{PandolfiandLarocca1989,
	author = {Pandolfi,~M. and Larocca,~F.},
	journal = {{Computers and Fluids}},
	publisher = {{Elsevier}},
	title = {{Transonic flow about a circular cylinder}},
        volume = {17},
        number = {1},
        pages = {205-220},
        year = {1989},
        doi = {https://doi.org/10.1016/0045-7930(89)90017-0},
	note={Available at \url{https://doi.org/10.1016/0045-7930(89)90017-0} - Last accessed: April 05, 2025},
}

@article{Rodriguezetal2023,
	author = {Rodriguez,~I. and Eiximeno,~B. and Gasparino,~L. and Tur-Mongè,~C. and Muela,~J. and Lehmkuhl,~O.},
	journal = {{}},
	publisher = {{Universitat Politecnica De Catalunya Barcelonatech}},
	title = {{Compressibility effects on the wake dynamics of a circular cylinder}},
        volume = {},
        number = {},
        pages = {1-6},
        year = {2023},
        doi = {},
	note={Available at \url{chrome-extension://efaidnbmnnnibpcajpcglclefindmkaj/https://upcommons.upc.edu/bitstream/2117/394832/1/ETMM14_Rodriguez.pdf} - Last accessed: April 05, 2025},
}

@article{Rodriguez1984,
	author = {Rodriguez,~O.},
	journal = {{AIAA Journal}},
	publisher = {{AIAA}},
	title = {{The circular cylinder in subsonic and transonic flow}},
        volume = {22},
        number = {12},
        pages = {1713–1718},
        year = {1984},
        doi = {https://doi.org/10.2514/3.8842},
	note={Available at \url{https://doi.org/10.2514/3.8842} - Last accessed: April 05, 2025},
}

@article{Rolandietal2023,
	author = {Rolandi,~L.~V. and Fontane,~J. and Jardin,~T. and Gressier,~J. and Joly,~L.},
	journal = {{Journal of Fluid Mechanics}},
	publisher = {{Cambridge University Press}},
	title = {{Compressibility effects on the secondary instabilities of the circular cylinder wake}},
        volume = {966},
        number = {},
        pages = {1-25},
        year = {2023},
        doi = {https://doi.org/10.1017/jfm.2023.430},
	note={Available at \url{https://doi.org/10.1017/jfm.2023.430} - Last accessed: April 05, 2025},
}

@article{Roshko1953,
	author = {Roshko,~A.},
	journal = {{National Advisory Committee for Aeronautics}},
	publisher = {{}},
	title = {{On the Development of Turbulent Wakes from Vortex Streets}},
        volume = {},
        number = {NACA TN 2913},
        pages = {1-78},
        year = {1953},
        doi = {},
	note={Available at \url{https://authors.library.caltech.edu/records/jpc6y-qvc49} - Last accessed: April 05, 2025},
}

@article{Sansicaetal2018,
	author = {Sansica,~A. and Robinet,~J.~C. and Alizard,~F. and Goncalves,~E.},
	journal = {{Journal of Fluid Mechanics}},
	publisher = {{Cambridge University Press}},
	title = {{Three-dimensional instability of a flow past a sphere: Mach evolution of the regular and Hopf bifurcations}},
        volume = {855},
        number = {},
        pages = {1088-1115},
        year = {2018},
        doi = {doi:10.1017/jfm.2018.664},
	note={Available at \url{https://doi.org/10.1017/jfm.2018.664} - Last accessed: April 05, 2025},
}

@article{Shang1982,
	author = {Shang,~J.},
	journal = {{20th Aerospace Sciences Meeting}},
	publisher = {{}},
	title = {{Oscillatory compressible flow around a cylinder}},
        volume = {},
        number = {},
        pages = {},
        year = {1982},
        doi = {https://doi.org/10.2514/6.1982-98},
	note={Available at \url{https://doi.org/10.2514/6.1982-98} - Last accessed: April 05, 2025},
}

@article{Shirani2001,
	author = {Shirani,~E.},
	journal = {{Journal of Applied Sciences}},
	publisher = {{Science Alert}},
	title = {{Compressible Flow Around a Circular Cylinder}},
        volume = {1},
        number = {4},
        pages = {472-476},
        year = {2001},
        doi = {10.3923/jas.2001.472.476},
	note={Available at \url{https://doi.org/10.3923/jas.2001.472.476} - Last accessed: April 05, 2025},
}

@article{Taneda1956,
	author = {Taneda,~S.},
	journal = {{Journal of the Physical Society of Japan}},
	publisher = {{The Physical Society of Japan}},
	title = {{Experimental Investigation of the Wakes behind Cylinders and Plates at Low Reynolds Numbers}},
        volume = {11},
        number = {3},
        pages = {302-307},
        year = {1956},
        doi = {https://doi.org/10.1143/JPSJ.11.302},
	note={Available at \url{https://doi.org/10.1143/JPSJ.11.302} - Last accessed: April 05, 2025},
}

@article{Xiaetal2016,
	author = {Xia,~Z. and Xiao,~Z. and Shi,~Y. and Chen,~S.},
	journal = {{AIAA Journal}},
	publisher = {{AIAA}},
	title = {{Mach Number Effect of Compressible Flow Around a Circular Cylinder}},
        volume = {54},
        number = {6},
        pages = {2004-2009},
        year = {2016},
        doi = {https://doi.org/10.2514/1.J054420},
	note={Available at \url{https://doi.org/10.2514/1.J054420} - Last accessed: April 05, 2025},
}

@article{Xuetal2009,
	author = {Xu,~C. and Chen,~L. and Lu,~X.},
	journal = {{Chinese Science Bulletin}},
	publisher = {{Springer Nature Link}},
	title = {{Effect of Mach number on transonic flow past a circular cylinder}},
        volume = {54},
        number = {11},
        pages = {1886-1893},
        year = {2009},
        doi = {https://doi.org/10.1007/s11434-009-0325-x},
	note={Available at \url{https://doi.org/10.1007/s11434-009-0325-x} - Last accessed: April 05, 2025},
}

@article{Xuetal2020,
	author = {Xu,~C.~-Y. and Hou,~B. and Wang,~Z. and Zhang,~Y.~-T. and Sun,~J.~-H.},
	journal = {{Aerospace Science and Technology}},
	publisher = {{Elsevier}},
	title = {{Effect of Mach number on the compressible flow past a wavy-axis cylinder}},
        volume = {104},
        number = {},
        pages = {},
        year = {2020},
        doi = {https://doi.org/10.1016/j.ast.2020.105943},
	note={Available at \url{https://doi.org/10.1016/j.ast.2020.105943} - Last accessed: April 05, 2025},
}

@article{Xueetal2024,
	author = {Xue,~K. and Li,~Q. and Zhao,~L.},
	journal = {{Physics of Fluids}},
	publisher = {{AIP Publishing}},
	title = {{Compressibility effect on flow characteristics over a circular cylinder at Reynolds number of 3900}},
        volume = {36},
        number = {8},
        pages = {},
        year = {2004},
        doi = {https://doi.org/10.1063/5.0217452},
	note={Available at \url{https://doi.org/10.1063/5.0217452} - Last accessed: April 05, 2025},
}

@article{WallersteinandKeshet2022,
	author = {Wallerstein,~I.~S. and Keshet, U.},
	journal = {{Journal of Fluid Mechanics}},
	publisher = {{Cambridge University Press}},
	title = {{Compressible potential flows around round
bodies: Janzen–Rayleigh expansion inferences}},
    volume = {932},
    number = {A6},
    pages = {1-22},
    year = {2022},
    doi = {10.1017/jfm.2021.965},
	note={Available at \url{https://doi.org/10.1017/jfm.2021.965} - Last accessed: November 09, 2025},
}

@article{KeshetandNaor2016,
	author = {Keshet,~U. and Naor,~Y.},
	journal = {{The Astrophysical Journal}},
	publisher = {{The American Astronomical Society}},
	title = {{Compressible Flow in Front of an Axisymmetric Blunt Object: Analytic Approximation and Astrophysical Implications}},
    volume = {830},
    number = {2},
    pages = {1-9},
    year = {2016},
    doi = {10.3847/0004-637X/830/2/147},
	note={Available at \url{https://iopscience.iop.org/article/10.3847/0004-637X/830/2/147} - Last accessed: November 09, 2025},
}

@article{Jelineketal2022,
	author = {Jelínek,~P. Belov,~S. and Karlický,~M.},
	journal = {{The Astrophysical Journal}},
	publisher = {{The American Astronomical Society}},
	title = {{Numerical Simulations of Oscillations in Solar Corona Excited by Vortex Shedding}},
    volume = {941},
    number = {2},
    pages = {1-8},
    year = {2022},
    doi = {10.3847/1538-4357/aca40d},
	note={Available at \url{https://doi.org/10.3847/1538-4357/aca40d} - Last accessed: November 09, 2025},
}

@article{Taoetal2025,
	author = {Tao,~Y. and Qu,~J. and Gao,~S. and Qiu,~X. and Zhang, ~X. and Liu,~Y.},
	journal = {{Physics of Fluids}},
	publisher = {{AIP Publishing}},
	title = {{Energy evolution in the Kármán vortex street of compressible flow around a circular cylinder}},
    volume = {37},
    number = {026129},
    pages = {1–17},
    year = {2025},
    doi = {10.1063/5.0254923},
	note={Available at \url{https://doi.org/10.1063/5.0254923} - Last accessed: January 10, 2026},
}

@article{Li2025,
	author = {Li,~S.},
	journal = {{Physical Review Fluids}},
	publisher = {{American Physical Society}},
	title = {{Sound emission in a quasi-steady transonic turbulent flow
past a circular cylinder}},
    volume = {10},
    number = {034603},
    pages = {1–28},
    year = {2025},
    doi = {https://doi.org/10.1103/PhysRevFluids.10.034603},
	note={Available at \url{https://doi.org/10.1103/PhysRevFluids.10.034603} - Last accessed: February 15, 2026},
}

@article{SinclairandCui2017,
	author = {Sinclair,~J. and Cui,~X.},
	journal = {{Physics of Fluids}},
	publisher = {{American Institute of Physics}},
	title = {{A theoretical approximation of the shock standoff distance for supersonic flows
around a circular cylinder}},
    volume = {29},
    number = {026102},
    pages = {1–13},
    year = {2017},
    doi = {https://doi.org/10.1063/1.4975983},
	note={Available at \url{https://doi.org/10.1063/1.4975983} - Last accessed: February 25, 2026},
}

@article{Yildizetal2023,
	author = {Yildiz,~U. and Vatansever,~D. and Celik,~B.},
	journal = {{Theoretical and Computational Fluid Dynamics}},
	publisher = {{Springer Nature}},
	title = {{Shock stand-off distances over sharp wedges for thermally non-equilibrium dissociating nitrogen flows}},
    volume = {37},
    number = {},
    pages = {799-821},
    year = {2023},
    doi = {https://doi.org/10.1007/s00162-023-00669-8},
	note={Available at \url{https://doi.org/10.1007/s00162-023-00669-8} - Last accessed: February 26, 2026},
}

@article{Awasthietal2025,
	author = {Awasthi,~M. and McCreton,~S. and Moreau,~D.~J. and Doolan,~C.~J.},
	journal = {{Journal of Fluid Mechanics}},
	publisher = {{Cambridge University Press}},
	title = {{Coherent oscillations and acoustic waves in a supersonic cylinder wake}},
    volume = {1012},
    number = {A27},
    pages = {1-31},
    year = {2025},
    doi = {10.1017/jfm.2025.10190},
	note={Available at \url{https://doi.org/10.1017/jfm.2025.10190} - Last accessed: February 27, 2026},
}

@article{MatheswaranandMiller2024a,
	author = {Matheswaran,~V. and Miller,~L.~S.},
	journal = {{Journal of Engineering Mathematics}},
	publisher = {{Springer}},
	title = {{A hybrid potential flow model for shedding flow around a
circular cylinder}},
    volume = {147},
    number = {15},
    pages = {1–20},
    year = {2024},
    doi = {https://doi.org/10.1007/s10665-024-10386-8},
	note={Available at \url{https://doi.org/10.1007/s10665-024-10386-8} - Last accessed: March 07, 2026},
}

@article{MatheswaranandMiller2024b,
	author = {Matheswaran,~V. and DeLillo,~T.~K. and Miller,~L.~S.},
	journal = {{Journal of Engineering Mathematics}},
	publisher = {{Springer}},
	title = {{Vortex shedding from bluff bodies: a conformal mapping
approach}},
    volume = {146},
    number = {17},
    pages = {1–20},
    year = {2024},
    doi = {https://doi.org/10.1007/s10665-024-10367-x},
	note={Available at \url{https://doi.org/10.1007/s10665-024-10367-x} - Last accessed: March 07, 2026},
}

@article{Joseph2003,
	author = {Joseph, Daniel D.},
	journal = {{Journal of Fluid Mechanics}},
    publisher={Cambridge University Press},
	title = {{Viscous Potential Flow}},
	volume = {479},
	issue = {},
	pages = {191-197},
	year = {2003},
    doi = {10.1017/S0022112002003634},
    note={Available at \url{https://doi.org/10.1017/S0022112002003634} - Last accessed: May 15, 2026},
}

@incollection{Arfenetal2013,
  author		= {Arfken, G.~B. and Weber, H.~J. and Harris, F.~E.},
  title		= {{Chapter 3 - Vector Analysis}},
  editor		= {},
  booktitle	= {{Mathematical Methods for Physicists (Seventh Edition)}},
  pages		= {123-203},
  address		= {},
  publisher	= {Academic Press},
  year		= {2013},
  ISBN        = {9780123846549},
  note        = {Available at \url{https://doi.org/10.1016/B978-0-12-384654-9.00003-7} - Last accessed: May 17, 2026},
}

@book{Josephetal2007,
	author = {Joseph,~D. and Funada,~T. and Wang,~J.},
    title = {{Potential Flows of Viscous and Viscoelastic Liquids}},
    address ={},
    publisher={Cambridge University Press},
	year = {2007},
    doi = {https://doi.org/10.1017/CBO9780511550928.012},
    note={Available at \url{https://doi.org/10.1017/CBO9780511550928.012} - Last accessed: May 17, 2026},
}

@misc{Amoloyeetal2026,
  author		= {Amoloye,~T.~O. and Oladimeji, ~L.~T. and Hayajnh,~M.~A. and Olayemi,~O.~A.},
  title		= {An exposition on some viscosity dependent issues in refined potential flow
theory},
  journal     = {{AIP Advances}},
  note		= {Unpublished},
}

@article{Economon2020,
	author = {Economon,~T.~D.},
	journal = {AIAA Journal },
	publisher = {AIAA},
	title = {{Simulation and Adjoint-Based Design for Variable Density Incompressible Flows with Heat Transfer}},
	volume={58},
    number={2},
    pages={757-769},
	month = {},
	year = {2020},
    doi={https://doi.org/10.2514/1.J058222},
    note={Available at \url{https://doi.org/10.2514/1.J058222} - Last accessed: May 18, 2026},
}

@article{Shivamoggi2025,
	author = {Shivamoggi,~B.~K.},
	journal = {The Astrophysical Journal},
	publisher = {American Astronomical Society},
	title = {{Potential Flow Theory Formulation of Parker’s Unsteady Solar Wind Model and
Nonlinear Stability of Parker’s Steady Solar Wind Solution}},
	volume={985},
    number={1},
    pages={1-8},
	month = {},
	year = {2025},
    doi={https://doi.org/10.3847/1538-4357/adcd64},
    note={Available at \url{https://doi.org/10.3847/1538-4357/adcd64} - Last accessed: May 24, 2026},
}

@article{HadjadjandKudryavtsev2005,
	author = {Hadjadj,~A. and Kudryavtsev,~A.},
	journal = {Journal of Turbulence},
	publisher = {American Astronomical Society},
	title = {{Computation and Flow Visualization in High-Speed Aerodynamics}},
	volume={6},
    number={16},
    pages={33-81},
	month = {},
	year = {2005},
    doi={https://doi.org/10.1080/14685240500209775},
    note={Available at \url{https://doi.org/10.1080/14685240500209775} - Last accessed: May 24, 2026},
}

@article{KleimannandRoken2024,
	author = {Kleimann,~J. and Röken,~C.},
	journal = {{Physics of Fluids}},
	publisher = {{American Institute of Physics}},
	title = {{An Exact Analytical Solution for the Weakly Magnetized Flow around an Axially Symmetric Paraboloid, With Application to Magnetosphere Models}},
    volume = {36},
    number = {093625},
    pages = {1–17},
    year = {2024},
    doi = {https://doi.org/10.1063/5.0215849},
	note={Available at \url{https://doi.org/10.1063/5.0215849} - Last accessed: July 02, 2026},
}

@article{Schereretal2016,
	author = {Scherer,~K. and Fichtner,~H. and Fahr,~H.~J. and Röken,~C. and Kleimann,~J.},
	journal = {{The Astrophysical Journal}},
	publisher = {{The American Astronomical Society}},
	title = {{Generalized Multi-polytropic Rankine Hugoniot Relations and The Entropy Condition}},
    volume = {833},
    number = {38},
    pages = {1-10},
    year = {2016},
    doi = {http://dx.doi.org/10.3847/1538-4357/833/1/38},
	note={Available at \url{http://dx.doi.org/10.3847/1538-4357/833/1/38} - Last accessed: July 02, 2026},
}

@article{Livadiotis2015,
	author = {Livadiotis,~G.},
	journal = {{The Astrophysical Journal}},
	publisher = {{The American Astronomical Society}},
	title = {{Shock Strength in Space and Astrophysical Plasmas}},
    volume = {809},
    number = {2},
    pages = {1-21},
    year = {2015},
    doi = {10.1088/0004-637X/809/2/111},
	note={Available at \url{http://dx.doi.org/10.1088/0004-637X/809/2/111} - Last accessed: July 03, 2026},
}

\end{document}